\documentclass[final,3p,times]{elsarticle}

\usepackage{graphicx}
\usepackage{multicol,multirow}
\usepackage{amsmath,amssymb,amsfonts}
\usepackage{mathrsfs}
\usepackage{amsthm}
\usepackage{rotating}
\usepackage{appendix}
\usepackage{ifpdf}
\usepackage[T1]{fontenc}
\usepackage{newtxtext}
\usepackage{newtxmath}
\usepackage{textcomp}
\usepackage{lipsum}
\usepackage[colorlinks,allcolors=blue]{hyperref}
\usepackage[export]{adjustbox}

\usepackage{amssymb}
\usepackage{amsmath}
\usepackage{graphicx}
\usepackage{algorithm2e}
\RestyleAlgo{ruled}
\usepackage{caption}
\usepackage{subcaption}
\usepackage{pdflscape}
\usepackage{tabularx}
\usepackage{booktabs}
\usepackage{multirow}
\usepackage{cases}
\DeclareMathOperator*{\argmin}{\arg\!\min}

\allowdisplaybreaks
\usepackage{enumitem}

\SetKwInput{KwData}{Inputs}
\SetKwInput{KwResult}{Outputs}

\usepackage[dvipsnames,table]{xcolor}
\usepackage{soul}
\usepackage{ulem}
\usepackage{dutchcal}
\usepackage{float}
\usepackage{multirow}
\usepackage{fancyhdr}
\usepackage[pscoord]{eso-pic}
\usepackage{nicematrix} 
\usepackage{afterpage} 
\usepackage{pdflscape}
\usepackage{natbib}
\usepackage{epstopdf}
\usepackage[english]{babel}
\usepackage{amsthm}

\theoremstyle{thmstyletwo}%
\newtheorem{example}{Example}%

\theoremstyle{remark}
\newtheorem{remark}{Remark}[section]

\theoremstyle{definition}

\usepackage{lineno}

\usepackage{pifont}

\newcolumntype{C}[1]{>{\centering\arraybackslash}m{#1}} 

\newcommand{\placetextbox}[3]{
  \setbox0=\hbox{#3}
  \AddToShipoutPictureFG*{
    \put(\LenToUnit{#1\paperwidth},\LenToUnit{#2\paperheight}){\vtop{{\null}\makebox[0pt][c]{#3}}}%
  }%
}%
\usepackage[english]{babel}
\addto\captionsenglish{\renewcommand{\appendixname}{Appendix}}

\journal{xxxx}

\begin{document}
\pagestyle{fancy}
\fancyhf{}
\renewcommand{\headrulewidth}{0pt}
\cfoot{{\textsf{Distribution Statement A. Approved for public release. Distribution is unlimited.}}}
\rfoot{\thepage}

\begin{frontmatter}

\title{Hyper-reduction-free reduced-order Newton solvers for projection-based model-order reduction of nonlinear dynamical systems}

\author[1]{Liam K. Magargal}
\author[1]{Parisa Khodabakhshi}
\ead{pak322@lehigh.edu}
\author[2]{Steven N. Rodriguez}

\affiliation[1]{organization={Department of Mechanical Engineering and Mechanics, Lehigh University},
            city={Bethlehem},
            state={PA},
            country={United States}}

\affiliation[2]{organization={Optical Sciences Division, U. S. Naval Research Laboratory},
            city={Washington},
            state={DC},
            country={United States}}

\begin{keyword}
Projection-based model-order reduction \sep Hyper-reduction-free approach \sep Galerkin projection
\sep Least-squares Petrov-Galerkin projection \sep Lifting transformations

\end{keyword}

\begin{abstract} 
This study proposes an intrusive projection-based model-order reduction framework for nonlinear problems with a polynomial structure, solved iteratively using a Newton solver in the reduced space. It is demonstrated that for the targeted class of polynomial nonlinearities, all operators appearing in the projected approximate residual and Jacobian can be precomputed in the offline phase, eliminating the need for hyper-reduction. Additionally, the evaluation of both the projected approximate residual and its Jacobian scales only with the dimension of the reduced space, and does not depend on the dimension of the full-order model, enabling effective offline-online decomposition. The proposed hyper-reduction-free (HRF) framework is applied to both Galerkin (HRF-G) and least-squares Petrov-Galerkin (HRF-LSPG) projection schemes. The accuracy and computational efficiency of the proposed HRF schemes are evaluated in two numerical experiments and compared with a commonly used hyper-reduction scheme, namely the energy-conserving sampling and weighting method, for both the Galerkin and LSPG schemes. In the first numerical example, a parametric Burgers' equation is used to assess the predictive capabilities of the considered model reduction approaches on parameter sets not seen in the training snapshots. In the second example, a parametric heat equation with a cubic reaction term is studied, for which a lifting transformation is employed to expose the desired structure. The efficacy of the HRF methods in accurately reducing the dimensionality of the lifted formulation is investigated. For the studied problems, the results show that HRF-G and HRF-LSPG achieve two and one order of magnitude speedup, respectively, with respect to the full-order model while resulting in state prediction errors below $O(10^{-2})$.
\end{abstract}

\end{frontmatter}

\placetextbox{0.5}{0.1}{\textsf{Distribution Statement A. Approved for public release. Distribution is unlimited.}}



\section{Introduction}

Projection-based model-order reduction (PMOR) seeks to reduce the computational cost associated with many-query applications of large scale time-dependent numerical simulations of dynamical models by projecting them onto a low-dimensional subspace \cite{benner2015Survey}. The resulting reduced-order model (ROM) is less expensive to evaluate, yet is relatively accurate in approximating the high-dimensional model. In PMOR, the ROM is typically constructed during an offline phase. This phase involves collecting snapshot solutions from the full-order model (FOM), extracting a reduced basis from the snapshot data, and projecting the FOM onto the low-dimensional manifold spanned by this basis. The resulting ROM is then employed in the online phase to make rapid approximations of the FOM behavior. This general PMOR framework has been shown to be effective in many large-scale engineering applications \cite{lieu2006aircraft,amsallem2015optimization}.

Dimensionality reduction for PMOR is commonly achieved using linear manifolds, such as bases recovered from the proper orthogonal decomposition (POD) \cite{sirovich1987turbulence} and its variants \cite{willcox2002balancedpod,schmid2020spod}. For linear dynamical systems, the use of linear manifolds has been shown to significantly reduce dimensionality, yielding ROMs that achieve significant computational savings \cite{antoulas2009approximation}. However, a naive application of linear manifolds in PMORs of nonlinear dynamical systems can result in a ROM that lacks offline-online decomposition, with the evaluation of the reduced nonlinear terms continuing to scale with the dimension of the FOM \cite{chaturantabut2010deim,bhattacharyya2025hyperreduction}. To overcome this limitation, various hyper-reduction methods have been introduced in the literature that approximate the reduced nonlinear terms by sparsely sampling selected indices from the FOM solution, thereby enabling an efficient low-dimensional representation of the reduced nonlinear term. Examples of hyper-reduction schemes include the empirical interpolation method (EIM) \cite{barrault2004eim}, discrete empirical interpolation (DEIM) \cite{chaturantabut2010deim,drmac2016deimerror}, Gauss-Newton with approximated tensors (GNAT) \cite{carlberg2013gnat}, and energy-conserving sampling and weighting (ECSW) \cite{farhat2015ecsw,grimberg2021ecsw}. It is important to note that, due to their sparse sampling approach, hyper-reduction schemes introduce an additional layer of approximation. Furthermore, for highly nonlinear problems, the required number of sampled points to maintain sufficient accuracy can become very high \cite{nasika2023deim,chmiel2024unified}, which can impede potential cost savings.

\pagestyle{fancy}

Alternatively, some studies address the complexity associated with model reduction of nonlinear systems by adopting operator learning strategies such as dynamic mode decomposition (DMD) \cite{schmid2010dmd}, and operator inference (OpInf) \cite{peherstorfer2016data}. In particular, OpInf employs a data-driven approach and seeks to infer low-dimensional operators for systems with polynomial nonlinearity directly from state solution data by exploiting the fact that the linear Galerkin projection preserves the polynomial structure \cite{peherstorfer2016data}. While many dynamical systems do not naturally exhibit a polynomial form, numerous studies have shown that through variable transformation and the possible introduction of auxiliary variables, commonly referred to as a lifting transformation, a system with general nonlinearity can be recast into a polynomial form, enabling the application of OpInf \cite{gu2011qlmor,kramer2019lifting,qian2020lift,mcquarrie2021data,khodabakhshi2022opinf}. To improve the predictive capabilities of ROMs for nonlinear systems, researchers have explored the use of nonlinear manifolds, such as quadratic manifolds \cite{barnett2022quadratic,geelen2023operator}, autoencoders \cite{lee2020deeplspg,magargal2025projection}, probabilistic manifold decompositions \cite{guo2026probabilistic}, adaptive bases \cite{geelen2022localized,mohaghegh2025adaptive}, and closure models \cite{Barnett2023NNAugmentedPROM,ardesdeparga2026pmor}. Among studies leveraging autoencoders to construct nonlinear manifolds, some aim to approximate low-dimensional dynamics without direct knowledge of the high-dimensional system. Instead, some of these studies employ long-short term memory modules \cite{wiewel2019latent}, self-attention networks \cite{fu2023aerom,fu2025paerom}, or feed-forward neural networks \cite{fresca2021comprehensive}, while others, inspired by sparse identification of nonlinear dynamics (SINDy) \cite{brunton2016discovering} and OpInf, infer latent space operators directly from data \cite{fries2022lasdi,he2023glasdi}. Collectively, these approaches enable the practical implementation of nonlinear manifolds for model reduction while maintaining computational efficiency. However, much like hyper-reduction schemes, these methods still require some notion of an \textit{approximation} of the projection scheme to achieve a low-dimensional system. 
 
The focus of this study is on model reduction of dynamical systems with desired polynomial nonlinearity solved using implicit time-integration schemes with Newton solvers, and efficiently constructed in an intrusive manner that avoids the additional approximation introduced by hyper-reduction. The use of Newton solvers in projection-based ROMs commonly involves projecting the reconstructed residual and Jacobian onto a test basis \cite{carlberg2011lspg,lee2020deeplspg,barnett2022quadratic,Barnett2023NNAugmentedPROM,chmiel2024unified,magargal2025projection}, which again scales with the dimension of the FOM. While hyper-reduction schemes have been successfully integrated into PMOR schemes on nonlinear models using Newton solvers \cite{carlberg2013gnat,farhat2015ecsw}, they inherently introduce an additional approximation of the projected nonlinear terms. This approximation not only contributes to error, but also introduces hyperparameters that must be carefully tuned by the user. In this paper, a hyper-reduction-free (HRF) reduced Newton solver is introduced for the class of polynomial dynamical systems. Specifically, for these systems, the projected residual and Jacobian can be expressed explicitly and intrusively, bypassing the need for hyper-reduction while still achieving computational savings. The versatility and efficacy of HRF reduced Newton solvers are demonstrated by developing hyper-reduction-free Galerkin (HRF-G) and hyper-reduction-free least-squares Petrov-Galerkin (HRF-LSPG) Newton solvers, corresponding to Galerkin and least-squares Petrov-Galerkin (LSPG) projections, respectively.

The paper is organized as follows. Section \ref{sec:problemFormulation} introduces the polynomial form of the dynamical system and the corresponding Newton solvers considered in this study, provides background on the residual minimizing projection schemes deployed by the HRF approaches, and briefly discusses hyper-reduction schemes and some of their limitations. Section \ref{sec:HRF} presents the framework for obtaining low-dimensional operators for the HRF schemes. Section \ref{sec:experiments} presents two numerical examples that demonstrate the application of HRF schemes in different applications. The first application is a one-dimensional (1D) Burgers' equation model, used to benchmark the HRF approaches with hyper-reduction schemes and to demonstrate that they solve the Galerkin and LSPG projection schemes exactly, without approximation. The second application is a 1D unsteady heat equation with a cubic reaction term. This example illustrates how a lifting transformation can be used to transform the cubic reaction term into a quadratic system and, once again, compare the HRF approaches to hyper-reduction schemes. Finally, Section \ref{sec:conclusions} discusses conclusions and future work.

\section{Projection-based model-order reduction} \label{sec:problemFormulation}

In Section \ref{ssec:FOM}, the polynomial form of the semi-discretized FOM and the general linear multistep time-integration scheme used in this study to solve it are introduced. Section \ref{ssec:projectionSchemes} summarizes the key features of the Galerkin and LSPG projection schemes, along with the implementation of the linear multistep time-integration scheme. Finally, Section \ref{ssec:hyperreduction} discusses the role of hyper-reduction in enabling efficient implementation of implicit time-integration schemes.

\subsection{Full-order model} \label{ssec:FOM}

In this study, problems governed by a partial differential equation (PDE) with affine parametric dependence are considered. The spatially semi-discretized PDE yields a system of $N$ ordinary differential equations of the form \eqref{eq:quadraticForm}, with $N$ representing the dimension of the FOM
\begingroup
\begin{align} \label{eq:quadraticForm}
    &\frac{\mathrm{d}}{\mathrm{dt}}\mathbf{x}(t;\boldsymbol{\mu}) = \mathbf{C}(\boldsymbol{\mu}) + \mathbf{A}(\boldsymbol{\mu})\mathbf{x}(t;\boldsymbol{\mu}) + \mathbf{F}(\boldsymbol{\mu})(\mathbf{x}(t;\boldsymbol{\mu})\otimes\mathbf{x}(t;\boldsymbol{\mu})) + \mathbf{B}(\boldsymbol{\mu})\mathbf{u}(t;\boldsymbol{\mu}) + \mathbf{N}(\boldsymbol{\mu})(\mathbf{u}(t;\boldsymbol{\mu}) \otimes \mathbf{x}(t;\boldsymbol{\mu})),\notag \\
    &\mathbf{x}(0;\boldsymbol{\mu}) = \mathbf{x}_0(\boldsymbol{\mu}), 
\end{align}
\endgroup
where $\mathbf{x}: [0,T_f] \times \mathcal{D} \rightarrow \mathbb{R}^N$ denotes the semi-discrete state vector and $\mathbf{u}: [0, T_f] \times \mathcal{D} \rightarrow \mathbb{R}^{N_u}$ denotes the semi-discrete input vector with $N_u$ denoting the number of inputs, $T_f$ denotes the final time and $\mathcal{D}\subset \mathbb{R}^{N_p}$ is the parameter domain with $N_p$ parameters. The operators $\mathbf{C}(\boldsymbol{\mu}) \in \mathbb{R}^{N}$, $\mathbf{A}(\boldsymbol{\mu}) \in\mathbb{R}^{N \times N}$, and $\mathbf{F}(\boldsymbol{\mu}) \in \mathbb{R}^{N \times N^2}$ correspond to the constant, linear, and quadratic (in the state) terms, respectively. Furthermore, 
$\mathbf{B}(\boldsymbol{\mu}) \in \mathbb{R}^{N \times N_u}$ represents the operator associated with terms that are linear in the input, and $\mathbf{N}(\boldsymbol{\mu}) \in\mathbb{R}^{N \times N_u N}$ denotes the operator corresponding to terms that are bilinear in the input and state vectors. All operators are assumed to have affine parametric dependence on $\boldsymbol{\mu} \in \mathcal{D}$. Finally, $\mathbf{x}_0(\boldsymbol{\mu}) \in \mathbb{R}^{N}$ denotes the parameterized initial conditions. A general residual minimizing linear multistep scheme is used to solve \eqref{eq:quadraticForm}, in which the residual is defined as,
\begingroup
\begin{align} 
    \mathbf{r}^{m}:(\mathbf{x}^{m},\mathbf{x}^{m-1},\ldots,\mathbf{x}^{m-\tau}; \boldsymbol{\mu}) \mapsto \sum_{j=0}^{\tau} \alpha_j \mathbf{x}^{m-j} - &\Delta t \sum_{j=0}^{\tau} \beta_j \Big[\mathbf{C}(\boldsymbol{\mu}) + \mathbf{A}(\boldsymbol{\mu})\mathbf{x}^{m-j} + \mathbf{F}(\boldsymbol{\mu})(\mathbf{x}^{m-j} \otimes \mathbf{x}^{m-j}) \notag \\
    & \qquad \qquad   + \mathbf{B}(\boldsymbol{\mu}) \mathbf{u}^{m-j} + \mathbf{N}(\boldsymbol{\mu})(\mathbf{u}^{m-j} \otimes \mathbf{x}^{m-j})\Big], \label{eq:linearMultistep}
\end{align}
\endgroup
where $\mathbf{r}^m\in \mathbb{R}^N$ is the residual associated with the time-integration scheme at time step $m$, and $\mathbf{x}^{m-j}\in \mathbb{R}^N$ for $j=0,\cdots,\tau$ represents the state vector at time step $m-j$ (i.e., $\mathbf{x}((m-j) \Delta t; \boldsymbol{\mu})$). Here, $\Delta t\in\mathbb{R}_+$ denotes the time step size (chosen to be fixed for simplicity). The scalars $\alpha_j,\,\beta_j\in\mathbb{R}$ are the coefficients used to define the time-integration scheme, and $\tau\in\mathbb{N}$ is the total number of steps used in the linear multistep scheme. The residual minimizing scheme used to obtain $\mathbf{x}^m$ can be formally written as
\begin{equation} \label{eq:minResFOM}
    \mathbf{x}^{m} = \underset{\boldsymbol{\xi} \in \mathbb{R}^N}{\argmin} \Big\vert \Big\vert \mathbf{r}^{m}(\boldsymbol{\xi},\mathbf{x}^{m-1},\ldots,\mathbf{x}^{m-\tau}; \boldsymbol{\mu}) \Big\vert \Big\vert _2,
\end{equation}
where $\boldsymbol{\xi}$ denotes the unknown solution at time step $m$ (i.e., time $t_m = m\Delta t$), and $\vert\vert\, \cdot \,\vert \vert_2$ denotes the $L^2$-norm. For explicit time-integration schemes, where $\beta_0=0$, solving \eqref{eq:minResFOM} is straightforward. However, implicit time-integration schemes often require an iterative Newton solver. As an example of how the coefficients $\alpha_j$ and $\beta_j$ are determined for a given time-integration scheme, the residual expression for the Crank-Nicolson method, an implicit scheme, is derived.

\begin{example}
    The update for the Crank-Nicolson scheme \cite{crank1947crankNicolson} is given by
    \begin{equation*}
        \mathbf{x}(t^m; \boldsymbol{\mu}) = \mathbf{x}(t^{m-1}; \boldsymbol{\mu}) + \frac{\Delta t}{2} \left( \frac{\mathrm{d}\mathbf{x}}{\mathrm{dt}}\Bigg\vert_{(t^{m};\boldsymbol{\mu})} + \frac{\mathrm{d}\mathbf{x}}{\mathrm{dt}}\Bigg\vert_{(t^{m-1};\boldsymbol{\mu})}\right),
    \end{equation*}
    which can be rearranged to match the form of \eqref{eq:linearMultistep},
    \begin{equation*}
        \mathbf{0} = \mathbf{x}(t^m; \boldsymbol{\mu}) - \mathbf{x}(t^{m-1}; \boldsymbol{\mu}) - \Delta t\left[ \frac{1}{2} \left(\frac{\mathrm{d}\mathbf{x}}{\mathrm{dt}}\Bigg \vert_{(t^{m};\boldsymbol{\mu})}\right) + \frac{1}{2}\left(\frac{\mathrm{d}\mathbf{x}}{\mathrm{dt}}\Bigg \vert_{(t^{m-1};\boldsymbol{\mu})}\right)\right],
    \end{equation*}
    where it becomes apparent that for this scheme $\tau=1$, $\alpha_0=1$, $\alpha_1 = -1$, and $\beta_0 = \beta_1 = \frac{1}{2}$. 
    
\end{example}

\subsection{Projection schemes} \label{ssec:projectionSchemes}

The main focus of this study is on Galerkin and LSPG projection schemes. This section provides a brief overview of PMOR, the implementation of a ROM in the time-integration scheme to approximate the evolution of the full-order state, and the difference between the two projection schemes considered. The goal of PMOR is to project the FOM onto a low-dimensional ROM. The FOM state vector is approximated as $\mathbf{x}^m \approx \tilde{\mathbf{x}}^m = \mathbf{\Phi}\hat{\mathbf{x}}^m$, where $\mathbf{\Phi}\in\mathbb{R}^{N\times n}$ (with $n \ll N$ denoting the dimension of the ROM) is the trial basis over the span of which the full-order state vector evolves. Additionally, $\hat{\mathbf{x}}^m \in \mathbb{R}^n$ is the low-dimensional representation of the approximated full-order state solution, $\tilde{\mathbf{x}}^m \in \mathbb{R}^N$. By direct substitution of $\tilde{\mathbf{x}}^m$ into \eqref{eq:linearMultistep} and applying the mixed Kronecker product identity \cite{magnus1988kronecker} on the quadratic and bilinear terms, the \textit{approximate} residual is obtained,
\begin{align} 
    \tilde{\mathbf{r}}^{m}:(\mathbf{\hat{x}}^{m},\mathbf{\hat{x}}^{m-1},\ldots,\mathbf{\hat{x}}^{m-\tau},\boldsymbol{\Phi}; \boldsymbol{\mu}) \mapsto \sum_{j=0}^{\tau} \alpha_j \boldsymbol{\Phi} \mathbf{\hat{x}}^{m-j} - \Delta t \sum_{j=0}^{\tau}  &\beta_j \Big[\mathbf{C}(\boldsymbol{\mu}) + \mathbf{A}(\boldsymbol{\mu})\boldsymbol{\Phi}\mathbf{\hat{x}}^{m-j}   +\mathbf{F}(\boldsymbol{\mu})(\boldsymbol{\Phi} \otimes \boldsymbol{\Phi})(\mathbf{\hat{x}}^{m-j} \otimes \mathbf{\hat{x}}^{m-j}) 
    \notag \\ 
    & + \mathbf{B}(\boldsymbol{\mu}) \mathbf{u}^{m-j} + \mathbf{N}(\boldsymbol{\mu})(\mathbf{I}_{N_u} \otimes \boldsymbol{\Phi})(\mathbf{u}^{m-j} \otimes \mathbf{\hat{x}}^{m-j})\Big], \label{eq:approxRes}
\end{align}
where $\mathbf{I}_{N_u}$ denotes the identity matrix of dimension $N_u$. Substituting the approximate residual \eqref{eq:approxRes} into \eqref{eq:minResFOM} and projecting onto the test basis $\boldsymbol{\Psi}\in\mathbb{R}^{N\times n}$ yields
\begin{equation} \label{eq:lowDimRes}
    \hat{\mathbf{x}}^{m} = \underset{\hat{\boldsymbol{\xi}} \in \mathbb{R}^n}{\argmin} \Big\vert \Big\vert \mathbf{\Psi}(\hat{\boldsymbol{\xi}}; \boldsymbol{\mu})^\top \tilde{\mathbf{r}}(\hat{\boldsymbol{\xi}},\hat{\mathbf{x}}^{m-1},\ldots,\hat{\mathbf{x}}^{m-\tau},\boldsymbol{\Phi}; \boldsymbol{\mu}) \Big\vert \Big\vert _2,
\end{equation}
where $\hat{\boldsymbol{\xi}}$ denotes the unknown reduced state solution at $t_m$. Depending on the projection scheme, the test basis $\boldsymbol{\Psi}$ may or may not depend on the sought-after solution. To solve \eqref{eq:lowDimRes}, a Newton solver is employed based on the procedure presented in \cite{carlberg2011lspg,Barnett2023NNAugmentedPROM}, in which the root of the first-order Taylor expansion of the projected residual centered about the reduced state solution is iteratively sought; that is,
\begingroup
\begin{align*} 
    \mathbf{\Psi}(\hat{\mathbf{x}}^{m(k)}; \boldsymbol{\mu})^\top \tilde{\mathbf{r}} (\hat{\mathbf{x}}^{m(k)},&\hat{\mathbf{x}}^{m-1},\ldots,\hat{\mathbf{x}}^{m-\tau},\boldsymbol{\Phi}; \boldsymbol{\mu}) \approx \mathbf{\Psi}(\hat{\mathbf{x}}^{m(k)}; \boldsymbol{\mu})^\top \left(\tilde{\mathbf{r}}(\hat{\mathbf{x}}^{m(k)},\hat{\mathbf{x}}^{m-1},\ldots,\hat{\mathbf{x}}^{m-\tau},\boldsymbol{\Phi}; \boldsymbol{\mu}) + \frac{\partial{\tilde{\mathbf{r}}}}{\partial \hat{\mathbf{x}} } \Big \vert_{\hat{\mathbf{x}}^{m(k)}} \hat{\mathbf{p}}^k \right) = \mathbf{0},
\end{align*}
\endgroup
where $\hat{\mathbf{p}}^k \in \mathbb{R}^n$ is the step direction at the $k^{\mathrm{th}}$ iteration, and the superscript in $\hat{\mathbf{x}}^{m(k)}$ denotes the reduced state vector at the $k^{\mathrm{th}}$ iteration for the $m^{\mathrm{th}}$ time step. Dropping the function arguments for notational convenience, this class of projection schemes is succinctly written in the form, 
\begin{align} 
    &\mathbf{\Psi}^\top \left(\frac{\partial{\tilde{\mathbf{r}}}}{\partial {\hat{\mathbf{x}}}} \Big \vert_{\hat{\mathbf{x}}^{m(k)}} \right) \hat{\mathbf{p}}^k = - \mathbf{\Psi}^\top \tilde{\mathbf{r}}^{m(k)}, \quad \text{with}\;\, \tilde{\mathbf{r}}^{m(k)}=\tilde{\mathbf{r}}(\hat{\mathbf{x}}^{m(k)},\hat{\mathbf{x}}^{m-1},\ldots,\hat{\mathbf{x}}^{m-\tau},\boldsymbol{\Phi}; \boldsymbol{\mu}) \label{eq:genProjRes}\\
    &\hat{\mathbf{x}}^{m(k+1)} = \hat{\mathbf{x}}^{m(k)} + \gamma \hat{\mathbf{p}}^k, \label{eq:genUpdate}
\end{align}
where $\gamma \in \mathbb{R}_+$ is the step length. The solver \eqref{eq:genProjRes}-\eqref{eq:genUpdate} is applied iteratively until a convergence criterion such as the one in \eqref{eq:convergence} is met,
\begin{equation} \label{eq:convergence}
    \Big\vert \Big\vert \mathbf{\Psi}(\hat{\mathbf{x}}^{m(k)}; \boldsymbol{\mu})^\top \tilde{\mathbf{r}}(\hat{\mathbf{x}}^{m(k)},\hat{\mathbf{x}}^{m-1},\ldots,\hat{\mathbf{x}}^{m-\tau},\boldsymbol{\Phi}; \boldsymbol{\mu}) \Big\vert \Big\vert _2 < \kappa_{\mathrm{res}},
\end{equation}
where $\kappa_{\mathrm{res}} \in \mathbb{R}_+$ is a user-defined tolerance. 

In a Galerkin projection, the test basis $\boldsymbol{\Psi}$ is chosen to be identical to the trial basis $\boldsymbol{\Phi}$, whereas in a Petrov-Galerkin projection, the two bases differ, i.e., $\boldsymbol{\Psi} \neq \boldsymbol{\Phi}$. In this work, the LSPG projection is employed, in which the test basis, $\mathbf{\Psi}$ incorporates the Jacobian of the residual and is updated at each Newton iteration, often leading to improved stability \cite{carlberg2011lspg,grimberg2020pmor}.

\subsection{Hyper-reduction schemes} \label{ssec:hyperreduction}
While dimensionality reduction is often attainable using the reduced Newton solver in \eqref{eq:genProjRes}-\eqref{eq:genUpdate}, the operation count for nonlinear problems still scales with the dimension of the FOM, $N$, due to the repeated evaluation of $\mathbf{\Psi}^\top \left(\frac{\partial{\tilde{\mathbf{r}}}}{\partial {\hat{\mathbf{x}}}} \Big \vert_{\hat{\mathbf{x}}^{m(k)}} \right)$ and $\mathbf{\Psi}^\top \tilde{\mathbf{r}}^{m(k)}$ in \eqref{eq:genProjRes}. To alleviate this issue, hyper-reduction schemes are commonly employed, performing sparse weighted sampling of the left- and right-hand sides of \eqref{eq:genProjRes}, thereby enabling efficient \textit{approximate} evaluations with an operation count that does not scale directly with $N$. In particular, GNAT \cite{carlberg2013gnat} and ECSW \cite{farhat2015ecsw,grimberg2021ecsw} were specifically developed for Newton solvers like the one described in Section \ref{ssec:projectionSchemes}. While effective in reducing operation count for general nonlinearities, all hyper-reduction schemes are ultimately \textit{approximations} of the underlying projection scheme.

\section{Hyper-reduction-free reduced-order Newton solvers for nonlinear operators} \label{sec:HRF}
In this section, a framework is presented for efficient implementation of Newton solvers for Galerkin and LSPG projections applied to polynomial nonlinear problems in the form \eqref{eq:quadraticForm}, without the need for a hyper-reduction scheme. The framework builds upon two key observations discussed in the following.

The first observation is that, when a dynamical system is expressed in the form \eqref{eq:quadraticForm}, both the approximated residual and its Jacobian, denoted by  $\tilde{\mathbf{r}}^{m(k)}$ and $\frac{\partial{\tilde{\mathbf{r}}}}{\partial {\hat{\mathbf{x}}}} \Big \vert_{\hat{\mathbf{x}}^{m(k)}}$, respectively, can be written out explicitly. This structure enables the direct derivation of low-dimensional representations of both the Galerkin and LSPG solvers. To illustrate, the Jacobian of \eqref{eq:approxRes} with respect to $\hat{\mathbf{x}}^{m(k)}$ can be expressed as follows,
\begin{align} 
    \frac{\partial \tilde{\mathbf{r}}}{\partial \hat{\mathbf{x}}^{m(k)}} =  \alpha_0 \mathbf{\Phi} - \Delta t \beta_0 \Big[\mathbf{A}(\boldsymbol{\mu})\boldsymbol{\Phi} + \mathbf{F}(\boldsymbol{\mu})(\boldsymbol{\Phi} \otimes \boldsymbol{\Phi})&\left((\mathbf{\hat{x}}^{m(k)} \otimes \mathbf{I}_n) + (\mathbf{I}_n \otimes \mathbf{\hat{x}}^{m(k)} \right) + \mathbf{N}(\boldsymbol{\mu})(\mathbf{I}_{N_u} \otimes \boldsymbol{\Phi})(\mathbf{u}^{m} \otimes \mathbf{I}_n)\Big], \label{eq:approxJacobian}
\end{align}
where $\mathbf{I}_n$ denotes the identity matrix of dimension $n$.

The second observation is that the Kronecker terms in \eqref{eq:approxJacobian} can be written as a summation over the reduced state vector, $\hat{\mathbf{x}}$, and substituted into \eqref{eq:genProjRes}-\eqref{eq:genUpdate}. To illustrate, explicitly writing out $\hat{\mathbf{x}}^{m(k)} \otimes \mathbf{I}_n$ yields the result, 
\begingroup
\begin{equation} \label{eq:xkronI}
    \hat{\mathbf{x}}^{m(k)} \otimes \mathbf{I}_n =  \begin{bmatrix}
        \hat{x}_1^{m(k)}\mathbf{I}_n \\[3pt]
        \hat{x}_2^{m(k)}\mathbf{I}_n \\
        \vdots \\[3pt]
        \hat{x}_n^{m(k)}\mathbf{I}_n
        \end{bmatrix}
        = \sum_{\ell=1}^{n} \mathbf{G}^{\ell} \hat{x}_{\ell}^{m(k)} \in \mathbb{R}^{n^2 \times n}, \quad \text{ with }\;\; \mathbf{G}^{\ell} =  \begin{bmatrix}
        \mathbf{G}_1^{\ell} \\[3pt]
        \mathbf{G}_2^{\ell} \\
        \vdots \\[2pt]
        \mathbf{G}_n^{\ell} \\
        \end{bmatrix}\in \mathbb{R}^{n^2 \times n},
\end{equation}
\endgroup
where the $i^{\mathrm{th}}$ block matrix, $\mathbf{G}_i^{\ell} \in \mathbb{R}^{n\times n}$, is 
\begin{equation}
   \mathbf{G}^{\ell}_i =  
   \begin{cases}
       \mathbf{I}_n, \text{\quad if } i=\ell, \\
       \mathbf{0}, \text{ \quad otherwise.}
   \end{cases}
\end{equation}

Likewise, explicitly writing out $\mathbf{I}_n \otimes \hat{\mathbf{x}}^{m(k)}$ yields the result,
\begingroup
\begin{equation} \label{eq:Ikronx}
    \mathbf{I}_n \otimes \hat{\mathbf{x}}^{m(k)} = \begin{bmatrix}
        \hat{\mathbf{x}}^{m(k)} & \mathbf{0} & \cdots & \mathbf{0} \\
        \mathbf{0} & \hat{\mathbf{x}}^{m(k)} & \cdots & \mathbf{0}\\
        \vdots & \vdots & \ddots & \vdots\\
        \mathbf{0} & \mathbf{0} & \cdots & \hat{\mathbf{x}}^{m(k)}
        \end{bmatrix}
        = \sum_{\ell=1}^{n} \mathbf{H}^{\ell} \hat{x}_{\ell}^{m(k)} \in \mathbb{R}^{n^2 \times n}, \; \text{ with }\;\; \mathbf{H}^{\ell} =  \begin{bmatrix}
        \boldsymbol{\eta}^{\ell} & \mathbf{0}  & \cdots & \mathbf{0} \\
        \mathbf{0} & \boldsymbol{\eta}^{\ell} & \cdots & \mathbf{0} \\
        \vdots & \vdots & \ddots & \vdots \\
        \mathbf{0} & \mathbf{0} & \cdots & \boldsymbol{\eta}^{\ell}
        \end{bmatrix}\in \mathbb{R}^{n^2 \times n},
\end{equation}
\endgroup
where the $i^{\mathrm{th}}$ element in the vector, $\boldsymbol{\eta}^{\ell}\in\mathbb{R}^n$, is
\begin{equation}
   \eta^{\ell}_i =  
   \begin{cases}
       1, \text{\quad if } i=\ell, \\
       0, \text{ \quad otherwise.}
   \end{cases}
\end{equation}

Although only a dynamical system of the quadratic form \eqref{eq:quadraticForm} is presented here, cubic and higher-order terms exhibit identities similar to those presented above. However, rather than summing over the indices of $\hat{\mathbf{x}}^{m(k)}$, a cubic term would require summation over the indices of $\left(\hat{\mathbf{x}}^{m(k)} \otimes \hat{\mathbf{x}}^{m(k)}\right)$, a quartic term would require summation over the indices of $\left(\hat{\mathbf{x}}^{m(k)} \otimes \hat{\mathbf{x}}^{m(k)} \otimes \hat{\mathbf{x}}^{m(k)}\right)$, and higher-order terms follow the same pattern. 

The usefulness of these identities is demonstrated next by first considering a Galerkin projection. Substituting \eqref{eq:approxJacobian} into \eqref{eq:genProjRes}, setting $\boldsymbol{\Psi}=\boldsymbol{\Phi}$ for the Galerkin projection, and applying the identities \eqref{eq:xkronI} and \eqref{eq:Ikronx}, the left- and right-hand side operators of \eqref{eq:genProjRes} can be written, 
\begin{equation} \label{eq:HFGalerkinLHS}
    \begin{split}
    \boldsymbol{\Phi}^\top \left(\frac{\partial \tilde{\mathbf{r}}}{\partial \hat{\mathbf{x}}^{m(k)}}\right) =  \alpha_0 {\underbrace{\boldsymbol{\Phi}^\top \boldsymbol{\Phi}}_{\in \mathbb{R}^{n\times n}}} - \Delta t \beta_0 \Bigg\{ \underbrace{\boldsymbol{\Phi}^\top \mathbf{A}(\boldsymbol{\mu})\boldsymbol{\Phi}}_{\in \mathbb{R}^{n \times n}} &+ \sum_{\ell=1}^{n} \Big[\underbrace{\boldsymbol{\Phi}^\top \mathbf{F}(\boldsymbol{\mu})(\boldsymbol{\Phi} \otimes \boldsymbol{\Phi}) \mathbf{G}^{\ell}}_{\in \mathbb{R}^{n \times n}} + \underbrace{\boldsymbol{\Phi}^\top  \mathbf{F}(\boldsymbol{\mu})(\boldsymbol{\Phi} \otimes \boldsymbol{\Phi})\mathbf{H}^{\ell} }_{\in\mathbb{R}^{n\times n}}\Big]\hat{x}_{\ell}^{m(k)} \\ 
    & \quad + \underbrace{\boldsymbol{\Phi}^\top \mathbf{N}(\boldsymbol{\mu})(\mathbf{I}_{N_u} \otimes \boldsymbol{\Phi})}_{\in \mathbb{R}^{n \times N_u n}}(\mathbf{u}^{m} \otimes \mathbf{I}_n)\Bigg\},
    \end{split}
\end{equation}
\begin{equation} \label{eq:HFGalerkinRHS}
    \begin{split}
    \boldsymbol{\Phi}^\top \tilde{\mathbf{r}}^{m(k)} = \sum_{j=0}^{\tau} \alpha_j {\underbrace{\boldsymbol{\Phi}^\top \boldsymbol{\Phi}}_{\in \mathbb{R}^{n\times n}}} \mathbf{\hat{x}}^{m-j} - \Delta t \sum_{j=0}^{\tau} \beta_j \bigg[\underbrace{\boldsymbol{\Phi}^\top \mathbf{C}(\boldsymbol{\mu})}_{\in \mathbb{R}^n} &+ \underbrace{\boldsymbol{\Phi}^\top \mathbf{A}(\boldsymbol{\mu})\boldsymbol{\Phi}}_{\in\mathbb{R}^{n\times n}}\mathbf{\hat{x}}^{m-j} + \underbrace{\boldsymbol{\Phi}^\top \mathbf{F}(\boldsymbol{\mu})(\boldsymbol{\Phi} \otimes \boldsymbol{\Phi})}_{\in \mathbb{R}^{n \times n^2}}(\mathbf{\hat{x}}^{m-j} \otimes \mathbf{\hat{x}}^{m-j})
    \\ 
    &+ \underbrace{\boldsymbol{\Phi}^\top \mathbf{B}(\boldsymbol{\mu})}_{\in \mathbb{R}^{n \times N_u}} \mathbf{u}^{m-j} + \underbrace{\boldsymbol{\Phi}^\top \mathbf{N}(\boldsymbol{\mu})(\mathbf{I}_{N_u} \otimes \boldsymbol{\Phi})}_{\in \mathbb{R}^{n \times N_u n}}(\mathbf{u}^{m-j} \otimes \mathbf{\hat{x}}^{m-j})\bigg],
    \end{split}
\end{equation}
where in \eqref{eq:HFGalerkinRHS}, the superscript $(k)$ from $\mathbf{x}^{m(k)}$ has been dropped for notational convenience. Note that the underbraced reduced operators in \eqref{eq:HFGalerkinLHS}-\eqref{eq:HFGalerkinRHS} depend only on the dimension of the reduced space, $n$, and have no explicit dependence on the full-order dimension $N$. Moreover, all of these reduced operators can be \textit{precomputed} during the offline phase. Consequently, this approach is expected to yield computational savings without requiring a hyper-reduction scheme. 

A similar result holds for LSPG projections \cite{carlberg2011lspg}, where the test basis is taken to be, $\boldsymbol{\Psi} = \frac{\partial{\tilde{\mathbf{r}}}}{\partial {\hat{\mathbf{x}}}} \Big \vert_{\hat{\mathbf{x}}^{m(k)}}$ \cite{carlberg2011lspg,barnett2022quadratic,Barnett2023NNAugmentedPROM}. \ref{appendix:HRFlspg} presents the left- and right-hand side operators of \eqref{eq:genProjRes} for the HRF-LSPG scheme. Additionally, the interested reader is directed to \cite{carlberg2011lspg,lee2020deeplspg,grimberg2020pmor} for discussion on some of the advantages of using LSPG schemes.

\begin{remark}\label{rem:vanillaROM}
Hereafter, ``Galerkin-ROM'' and ``LSPG-ROM'' refer to the ROMs in the form \eqref{eq:genProjRes}-\eqref{eq:genUpdate}, where an offline-online decomposition is not leveraged. Instead, the projected form of the residual, $\boldsymbol{\Psi}^{\top}\tilde{\mathbf{r}}^{m}(\hat{\mathbf{x}}^{m(k)},\hat{\mathbf{x}}^{m-1},\ldots,\hat{\mathbf{x}}^{m-\tau}, \boldsymbol{\Phi};\boldsymbol{\mu})$, and its Jacobian, $\boldsymbol{\Psi}^{\top} \left( \frac{\partial \tilde{\mathbf{r}}}{\partial \mathbf{x}} \vert_{\boldsymbol{\Phi} \hat{\mathbf{x}}^{m(k)}} \right)\boldsymbol{\Phi}$, are computed via reconstruction of the high-dimensional state solution, full-state evaluation of the residual and its Jacobian, then projection onto the test basis, where the identity $\left( \frac{\partial \tilde{\mathbf{r}}}{\partial \hat{\mathbf{x}}} \vert_{\hat{\mathbf{x}}^{m(k)}} \right) = \left( \frac{\partial \tilde{\mathbf{r}}}{\partial \mathbf{x}} \vert_{\boldsymbol{\Phi} \hat{\mathbf{x}}^{m(k)}} \right)\boldsymbol{\Phi}$ is obtained from chain rule. Because these solvers lack an offline-online decomposition, and therefore have an operation count complexity that scales with the FOM dimension, $N$, they are not anticipated to achieve computational savings. Computational cost savings for such PMOR methods are typically achieved through hyper-reduction techniques. In this study, the ECSW hyper-reduction scheme is employed using both Galerkin (ECSW-G) and LSPG (ECSW-LSPG) projection schemes, following the procedure outlined in \cite{farhat2015ecsw,grimberg2021ecsw}. A brief overview of ECSW is provided in \ref{appendix:ECSW}. 
\end{remark}

A summary of the methods used in this study is presented in Table \ref{table:methodsSummary}. The HRF-G and HRF-LSPG methods (the primary contributions of this manuscript) efficiently compute the reduced Newton update using the precomputed reduced operators presented in \eqref{eq:HFGalerkinLHS}-\eqref{eq:HFGalerkinRHS} and \eqref{eq:HRFLSPGLHS3}-\eqref{eq:HRFLSPGRHS2}, respectively. In contrast, the baseline Galerkin-ROM and LSPG-ROM approaches reconstruct the approximate high-dimensional solution, evaluate the corresponding full-order residual and its Jacobian, and subsequently project onto the test basis. Finally, ECSW-G and ECSW-LSPG follow a similar procedure to Galerkin-ROM and LSPG-ROM but operate on a sparsely sampled representation of the high-dimensional solution, evaluating a subset of the residual and Jacobian entries before projecting onto the corresponding sampled test basis.

\begin{table}[h]
\centering
\caption{Summary of the methods used in this study. The bold-faced methods in the first column (HRF-G and HRF-LSPG) represent the main contributions of this manuscript. The second column reports the test basis $\boldsymbol{\Psi}$ (the trial basis is $\boldsymbol{\Phi}$ for all methods). The third and fourth columns provide the left- and right-hand sides of the Newton solver \eqref{eq:genProjRes} used in each method. The last column summarizes the computational procedure, distinguishing between complete offline-online decomposition (HRF methods), full-order reconstruction and projection (Galerkin-ROM and LSPG-ROM), and sparse sampling with projection (ECSW-based methods).}
\rowcolors{1}{gray!25}{white}
\begin{tabular}{C{2.2cm} C{2.2cm} C{4.8cm} C{3.0cm} C{3cm}}
    \hline
    method & test basis, $\boldsymbol{\Psi}$ & left-hand side & right-hand side & computational procedure\\
    \hline
    \textbf{HRF-G} & $\boldsymbol{\Phi}$ & \eqref{eq:HFGalerkinLHS} & \eqref{eq:HFGalerkinRHS} & offline-online decomposition  \\
    
    \textbf{HRF-LSPG} & $\dfrac{\partial \tilde{\mathbf{r}}}{\partial \hat{\mathbf{x}}} \big|_{\hat{\mathbf{x}}^{m(k)}}$ & \eqref{eq:HRFLSPGLHS3} & \eqref{eq:HRFLSPGRHS2} & offline-online decomposition \\
    
    Galerkin-ROM & $\boldsymbol{\Phi}$ & $\boldsymbol{\Phi}^{\top}\left( \dfrac{\partial \tilde{\mathbf{r}}}{\partial \mathbf{x}} \big|_{\boldsymbol{\Phi} \hat{\mathbf{x}}^{m(k)}} \right)\boldsymbol{\Phi}$ & $-\boldsymbol{\Phi}^{\top}\tilde{\mathbf{r}}^{m}$ & full-order reconstruction \& projection \\
    
    LSPG-ROM & $\left(\dfrac{\partial \tilde{\mathbf{r}}}{\partial \mathbf{x}} \big|_{\boldsymbol{\Phi} \hat{\mathbf{x}}^{m(k)}}\right)\boldsymbol{\Phi}$ & $\boldsymbol{\Phi}^{\top}\left( \dfrac{\partial \tilde{\mathbf{r}}}{\partial \mathbf{x}} \big|_{\boldsymbol{\Phi} \hat{\mathbf{x}}^{m(k)}} \right)^{\top}\left( \dfrac{\partial \tilde{\mathbf{r}}}{\partial \mathbf{x}} \big|_{\boldsymbol{\Phi} \hat{\mathbf{x}}^{m(k)}} \right)\boldsymbol{\Phi}$ & $-\boldsymbol{\Phi}^{\top}\left( \dfrac{\partial \tilde{\mathbf{r}}}{\partial \mathbf{x}} \big|_{\boldsymbol{\Phi} \hat{\mathbf{x}}^{m(k)}} \right)^{\top} \tilde{\mathbf{r}}^{m}$ & full-order reconstruction \& projection \\
    
    ECSW-G & $\boldsymbol{\Phi}$ & \eqref{eq:ecswLHS} & \eqref{eq:ecswRHS} & sparse sampling with projection \\
    
    ECSW-LSPG & $\left(\dfrac{\partial \tilde{\mathbf{r}}}{\partial \mathbf{x}} \big|_{\boldsymbol{\Phi} \hat{\mathbf{x}}^{m(k)}}\right)\boldsymbol{\Phi}$ & \eqref{eq:ecswLHS} & \eqref{eq:ecswRHS} & sparse sampling with projection\\
    \hline
\end{tabular}
\label{table:methodsSummary}
\end{table}

\section{Numerical experiments} \label{sec:experiments}
In this study, the accuracy and computational efficiency of the proposed HRF-G and HRF-LSPG model reduction methods are assessed through two numerical experiments. The HRF ROM results are compared with the FOM results, as well as with those obtained from Galerkin-ROM and LSPG-ROM, which, as already discussed in Remark \ref{rem:vanillaROM}, are obtained by reconstructing the reduced state, evaluating the residual, and projecting the reconstructed residual and its Jacobian onto a test basis in \eqref{eq:genProjRes}-\eqref{eq:genUpdate} without an offline-online decomposition. Additionally, both numerical experiments include results obtained by ECSW, a hyper-reduction method specifically tailored for Newton solvers. As mentioned before, a brief description of ECSW is provided in \ref{appendix:ECSW}. The ECSW results include a tolerance parameter, $\epsilon_{\mathrm{ecsw}}$, which ranges between $(0,\,1]$. Smaller values of $\epsilon_{\mathrm{ecsw}}$  correspond to more sampled points for the hyper-reduction scheme, and hence more accurate predictions. \ref{appendix:ECSW} provides information on the number of points sampled by ECSW in the numerical examples considered in this study.

Two numerical experiments are presented in this section. For illustrative purposes, both experiments are conducted on 1D spatial domains. However, the HRF-G and HRF-LSPG frameworks are readily extendable to systems with 2D and 3D spatial domains, provided that the semi-discrete dynamical system can be expressed in the form of \eqref{eq:quadraticForm}. In the first numerical experiment, a parametric study is performed on the 1D viscous Burgers' equation, a benchmark problem commonly used in the literature for nonlinear model reduction. The goal of this numerical experiment is to demonstrate that HRF-G and HRF-LSPG achieve results with accuracy comparable to Galerkin-ROM and LSPG-ROM, without the additional error associated with hyper-reduction schemes. Furthermore, it is demonstrated that HRF-G exhibits a favorable computational efficiency in comparison with ECSW-G. In the second numerical experiment, a 1D unsteady heat equation with a cubic reaction term is considered to evaluate the predictive capabilities of the ROMs over time and under varying parameters. Since the highest order polynomial in \eqref{eq:quadraticForm} is quadratic, a lifting transformation \cite{kramer2019lifting,qian2020lift} is applied to convert the cubic PDE into a system of two quadratic PDEs, each conforming to \eqref{eq:quadraticForm}. This numerical experiment aims to assess the performance of HRF-G and HRF-LSPG for problems with general nonlinearities (or polynomial nonlinearities of order higher than quadratic) when using the lifting transformation to recast the system into the desired form (e.g., quadratic at most). For both numerical experiments, the reduced Newton solver applied to all ROMs utilizes a step length of $\mathrm{\gamma}=1$ in \eqref{eq:genUpdate} and a convergence tolerance of $\kappa_{\mathrm{res}}=10^{-6}$ in \eqref{eq:convergence}.

In the remainder of the manuscript, the trial basis $\boldsymbol{\Phi}$ is constructed using POD via the method of snapshots \cite{sirovich1987turbulence}. 
Let $\mathbf{X}\in \mathbb{R}^{N\times N_s}$ denote the snapshot matrix formed by the state solutions of dimension $N$ at $N_s$ sampled time steps. Performing a (thin) singular value decomposition (SVD) on the snapshot matrix $\mathbf{X}$ gives $\mathbf{X}=\boldsymbol{\Theta}\boldsymbol{\Sigma}\boldsymbol{\Xi}^\top$ where $\boldsymbol{\Sigma}$ is a diagonal matrix containing the singular values $\sigma_i$ of the snapshot matrix in non-increasing order, and $\boldsymbol{\Theta}$ contains the corresponding left singular vectors. The number of POD modes $n$ in the basis is chosen based on the truncated modal energy $\epsilon_{\mathrm{POD}}$ by selecting the smallest number of modes $n$ such that
\begin{equation} \label{eq:truncatedModal}
     1 - \dfrac{\sum_{i=1}^n \sigma_i^2}{\sum_{i=1}^N \sigma_i^2}<\epsilon_{\mathrm{POD}},
\end{equation}
where $\sigma_i \in \mathbb{R}_+$ is the $i^{\mathrm{th}}$ singular value of the snapshot matrix used to generate the POD basis.

In this study, three performance metrics are considered for the ROMs, two measuring error and one measuring computational efficiency:
\begin{enumerate}[leftmargin=*]
    \item \textit{Relative state prediction error} quantifies the difference between the ROM and FOM solutions. This metric is evaluated for all model reduction approaches considered in this study, including Galerkin-ROM, LSPG-ROM, HRF-G, HRF-LSPG, ECSW-G, and ECSW-LSPG and is defined as follows
    \begin{equation}\label{eq:state_err}
        \text{state prediction error} = \frac{\Big\vert\Big\vert \mathbf{X}(\boldsymbol{\mu}) - \tilde{\mathbf{X}}(\boldsymbol{\mu}) \Big\vert\Big\vert_F^2}{\Big\vert\Big\vert \mathbf{X}(\boldsymbol{\mu}) \Big\vert\Big\vert_F^2},
    \end{equation}
    where $\mathbf{X}(\boldsymbol{\mu}): \mathcal{D} \rightarrow \mathbb{R}^{N \times N_t}$ and $\tilde{\mathbf{X}}(\boldsymbol{\mu}): \mathcal{D} \rightarrow \mathbb{R}^{N \times N_t}$ denote the matrices containing the ground truth and ROM state solution trajectories, respectively, for a given parameter realization $\boldsymbol{\mu}$, and $\vert\vert \cdot \vert\vert_F$ denotes the Frobenius norm \cite{strang2006linearalgebra}. 
    Additionally, the \textit{relative projection error} is reported which serves as the lower bound for the state prediction error in \eqref{eq:state_err},
    \begin{equation}\label{eq:reconstruc_err}
        \text{projection error} = \frac{\Big\vert\Big\vert \left(\mathbf{I}_N-\boldsymbol{\Phi}\boldsymbol{\Phi}^\top\right)\mathbf{X}(\boldsymbol{\mu}) \Big\vert\Big\vert_F^2}{\Big\vert\Big\vert \mathbf{X}(\boldsymbol{\mu}) \Big\vert\Big\vert_F^2},
    \end{equation}
    where $\mathbf{I}_N$ denotes the identity matrix of dimension $N$.
    \item \textit{ROM evaluation error} measures the discrepancy between the HRF-G, HRF-LSPG, ECSW-G, and ECSW-LSPG solutions with respect to their corresponding Galerkin-ROM and LSPG-ROM counterparts. In essence, a lower ROM evaluation error indicates that the HRF or ECSW methods closely approximate the solution obtained by the conventional Galerkin-ROM or LSPG-ROM. This error is given by
    \begin{equation}\label{eq:eval_err}
        \text{ROM evaluation error} = \frac{\Big\vert\Big\vert \mathbf{X}_{\mathrm{ROM}}(\boldsymbol{\mu}) - \tilde{\mathbf{X}}(\boldsymbol{\mu}) \Big\vert\Big\vert_F^2}{\Big\vert\Big\vert \mathbf{X}_{\mathrm{ROM}}(\boldsymbol{\mu}) \Big\vert\Big\vert_F^2},
    \end{equation}
    where $\mathbf{X}_{\mathrm{ROM}}(\boldsymbol{\mu}): \mathcal{D} \rightarrow \mathbb{R}^{N \times N_t}$ denotes the matrix containing the state solution trajectory predicted by either Galerkin-ROM or LSPG-ROM for the parameter vector $\boldsymbol{\mu}$.
    \item \textit{Speedup factor} evaluates the computational cost savings obtained with the ROM relative to the FOM by comparing their respective online wall-clock times,
    \begin{equation}\label{eq:speedUpFactor}
        \text{speedup factor} = \frac{\text{wall-clock time for FOM}}{\text{wall-clock time for ROM}}.
    \end{equation}
    To mitigate minor fluctuations in the wall-clock times of the FOM and ROM, each time measurement is averaged over $10$ independent runs. All computations were carried out in MATLAB R2024b on an Intel\textsuperscript{\textregistered} Xeon\textsuperscript{\textregistered} Gold 6230R @ 2.10GHz 10 core CPU.
\end{enumerate}

\subsection{One-dimensional viscous Burgers' equation}

In the first numerical experiment, HRF-G and HRF-LSPG are applied to a parameterized 1D viscous Burgers' equation test case from \cite{peherstorfer2016data}, which is similar to the formulations found across the literature \cite{rewienski2003phd,carlberg2013gnat,lee2020deeplspg,Barnett2023NNAugmentedPROM}. The performance of HRF schemes is benchmarked against their ECSW-G and ECSW-LSPG counterparts, as well as Galerkin-ROM and LSPG-ROM. The ECSW results are shown for two tolerance values: $\epsilon_{\mathrm{ecsw}}=10^{-5}$ and $10^{-9}$. The governing equation is a parameterized first-order PDE subject to appropriate boundary and initial conditions
\begin{align} \label{eq:burgers}
    &\begin{cases}
        \dfrac{\partial (w(x,t;\boldsymbol{\mu}))}{\partial t} + w(x,t;\boldsymbol{\mu})\dfrac{\partial (w(x,t;\boldsymbol{\mu}))}{\partial x} = \mu_2 \dfrac{\partial^2 (w(x,t;\boldsymbol{\mu}))}{\partial x^2}, & \forall x\in (0,L), \;\forall t \in (0,T_f], \\[3pt]
        w(0,t;\boldsymbol{\mu}) = \mu_1, \;\;w(L,t;\boldsymbol{\mu}) = 0, & \forall t \in (0,T_f], \\
        w(x,0; \boldsymbol{\mu}) = w_0(x)= 0, & \forall x \in (0,L), 
    \end{cases}
\end{align}
where $L=1$ and $T_f=0.5$. The equation is parametrized by the Dirichlet boundary condition on the left boundary, $\mu_1$, and the viscosity parameter, $\mu_2$, giving $\boldsymbol{\mu} = (\mu_1,\mu_2)$. Following spatial discretization with a second-order central difference scheme on a uniform grid of $N=1024$ points, and taking the state solution vector as the discrete representation of $w$ (i.e., $\mathbf{x}(t;\boldsymbol{\mu}) = \mathbf{w}(t; \boldsymbol{\mu})$), the semi-discretized form of \eqref{eq:burgers} can be written as
\begin{equation} \label{eq:burgersQuadratic}
    \frac{\mathrm{d}}{\mathrm{d}t}\mathbf{x}(t;\boldsymbol{\mu}) = \mathbf{A}(\mu_2)\mathbf{x}(t;\boldsymbol{\mu}) + \mathbf{F} (\mathbf{x}(t;\boldsymbol{\mu}) \otimes \mathbf{x}(t;\boldsymbol{\mu})) + \mathbf{B}(\mu_2) \mathbf{u}(t;\boldsymbol{\mu}) + \mathbf{N}(\mathbf{u}(t;\boldsymbol{\mu}) \otimes \mathbf{x}(t;\boldsymbol{\mu}) ),
\end{equation}
which is of the desired form \eqref{eq:quadraticForm}. A brief overview of the construction of the operators of \eqref{eq:burgersQuadratic}, obtained via spatial discretization, is provided in \ref{appendix:fd_burgers}.

\begin{figure}[ht!]
    \centering
     \begin{tabular}{ccc}
        & {\footnotesize{$\mu_1=1.0$}} & {\footnotesize{$\mu_1=3.25$}} \\[-1pt]

        \raisebox{2.9em}{\rotatebox[origin=lb]{90}{\parbox{2.5cm}{\centering \footnotesize{$\mu_2 = 0.01 $}}}}& \includegraphics[height=0.3\textwidth]{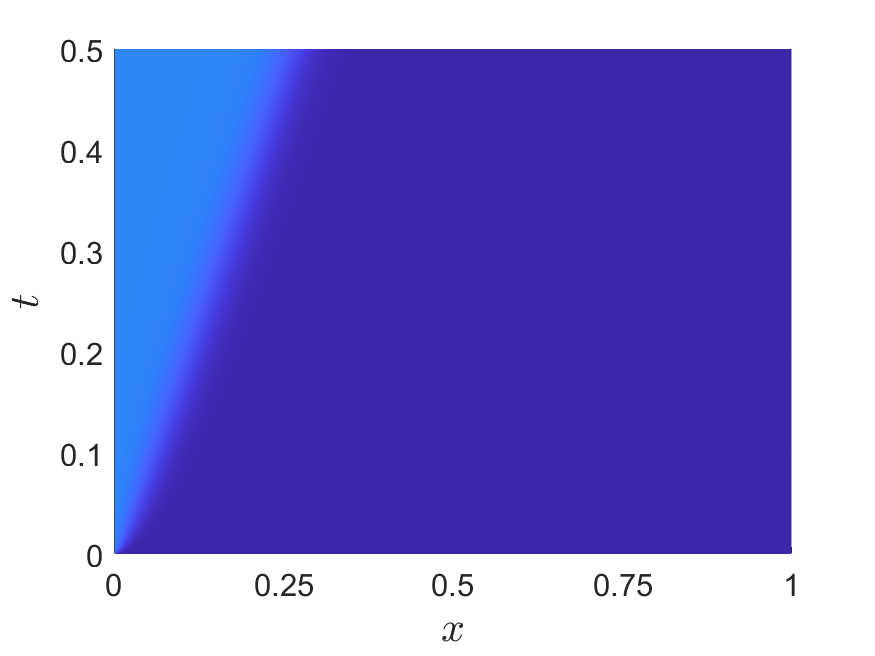}
        &
        \includegraphics[height=0.3\textwidth]{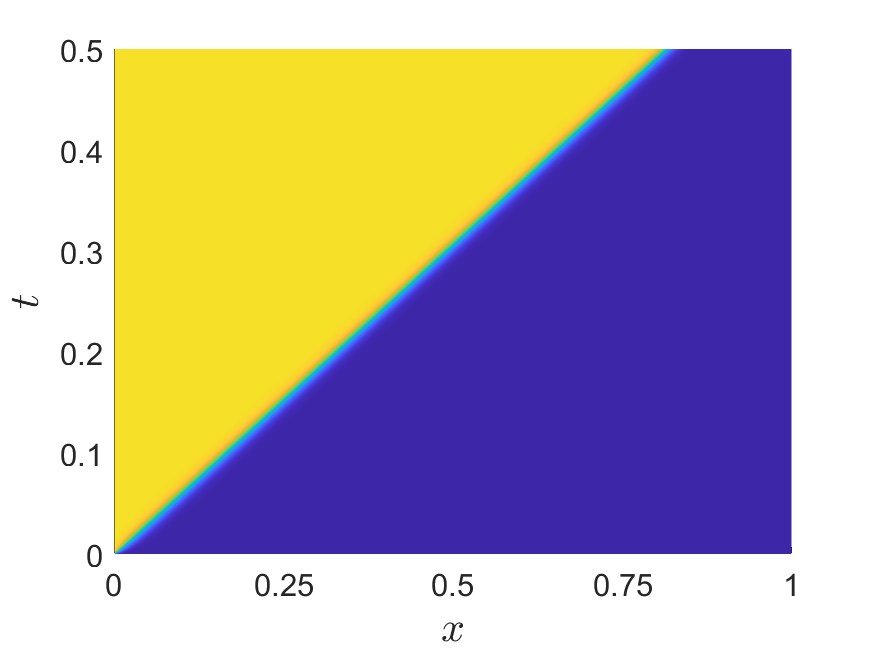}\\
        \raisebox{3.9em}{\rotatebox[origin=lb]{90}{\parbox{2cm}{\centering \footnotesize{$\mu_2 = 0.1$}}}} &
        \includegraphics[height=0.3\textwidth]{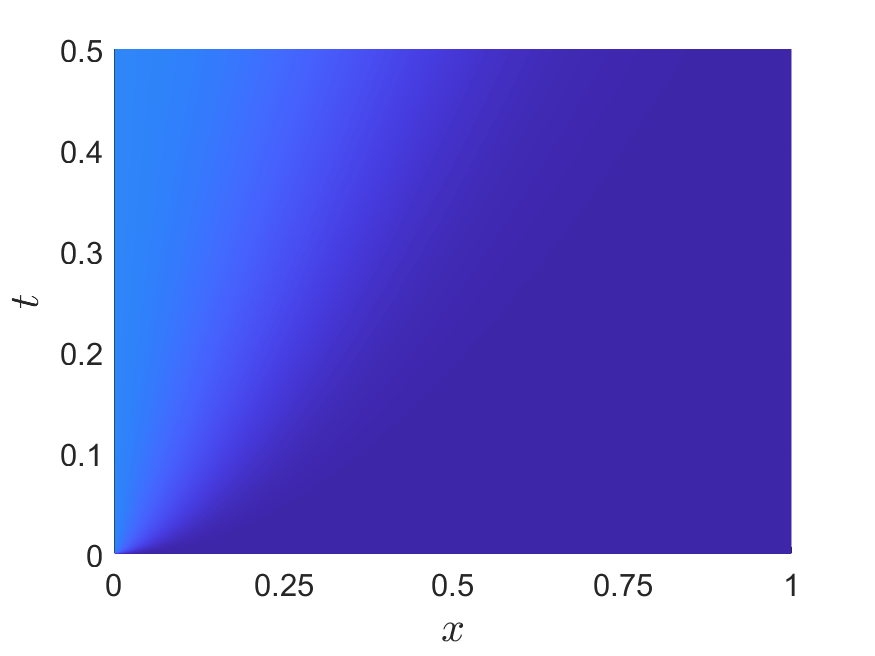} & \includegraphics[height=0.3\textwidth]{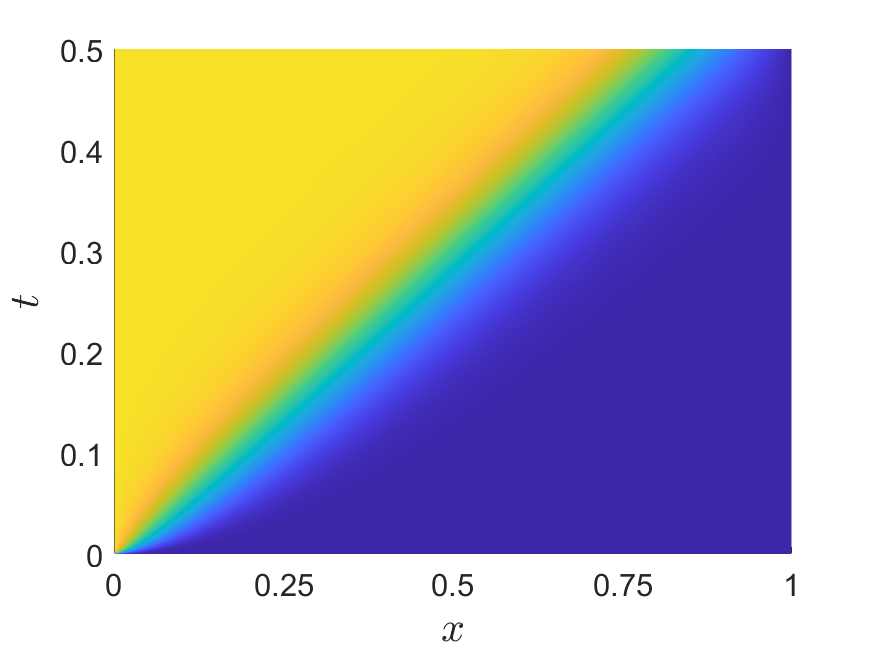} \\
        \multicolumn{3}{c}{\includegraphics[scale=.28]{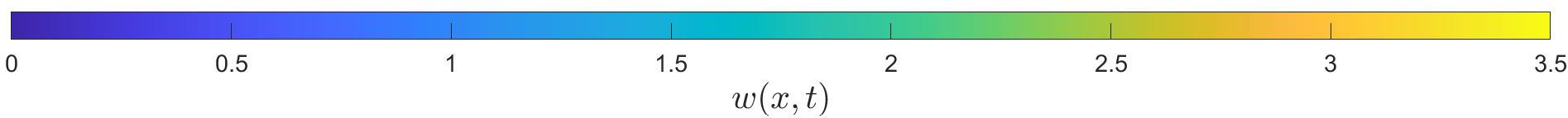}}
    \end{tabular}
    \captionsetup{justification=centering}
    \caption{Sample FOM solutions demonstrating the variation of the solution with respect to the parameters. The left and right columns present solutions with $\mu_1=1.0$, and $\mu_1 = 3.25$, respectively, while the top and bottom rows present solutions with $\mu_2 = 0.01$, and $\mu_2 = 0.1$, respectively. Higher values of the left Dirichlet boundary condition, $\mu_1$, result in the shock advecting more quickly through the domain, while smaller values of the diffusion coefficient, $\mu_2$, result in a sharper shock front. (Online version in color.)
    }
    \label{fig:burgers_samples}
\end{figure}

Using a constant time step size of $\Delta t=0.001$ over $N_t=501$ time steps (including initial conditions), temporal discretization is performed with the backward Euler scheme. Accordingly, the coefficients in \eqref{eq:linearMultistep} are $\tau=1$, $\alpha_0=1$, $\alpha_1=-1$, $\beta_0=1$, and $\beta_1=0$. To construct the POD basis, high-dimensional state solution trajectories are first generated by solving the FOM for 100 parameter realizations on a grid in the parameter space. A snapshot matrix is then formed by concatenating the solution trajectories, 
\begin{equation}
    \mathbf{X} = \begin{bmatrix}
    \mathbf{X}_{0,0} & \mathbf{X}_{1,0} & \cdots & \mathbf{X}_{9,0} & \mathbf{X}_{0,1} & \mathbf{X}_{1,1} & \cdots & \mathbf{X}_{9,1} & \mathbf{X}_{0,2} & \mathbf{X}_{1,2} & \cdots & \mathbf{X}_{9,9}
    \end{bmatrix} \in \mathbb{R}^{N \times 100N_t},
\end{equation}
where $\mathbf{X}_{i,j} \in \mathbb{R}^{N \times N_t}$ denotes the high-dimensional state trajectory corresponding to the parameter set $\boldsymbol{\mu} = (\mu_1, \mu_2) = (1 + 0.25 \mathrm{i}, \mu_2 = 0.01 + 0.01\mathrm{j})$. Finally, the POD basis is obtained by performing an SVD on $\mathbf{X}$ (in this case, $\mathbf{X} \in \mathbb{R}^{1024 \times 50100}$). Figure \ref{fig:burgers_samples} presents four solutions to training parameter sets that appear in the snapshot matrix. Varying the left Dirichlet boundary causes the shock wave to advect through the domain at different speeds, while varying the diffusion coefficient varies the sharpness of the shock front.

\begin{figure}[ht!]
    \centering
     \begin{tabular}{cc|cc}
        {\footnotesize{ground truth (FOM)}} & & {\footnotesize{HRF schemes}} & {\footnotesize{\, ECSW schemes ($\epsilon_{\mathrm{ecsw}}=10^{-9}$)}} \\[-1pt] 
        \includegraphics[height=0.225\textwidth]{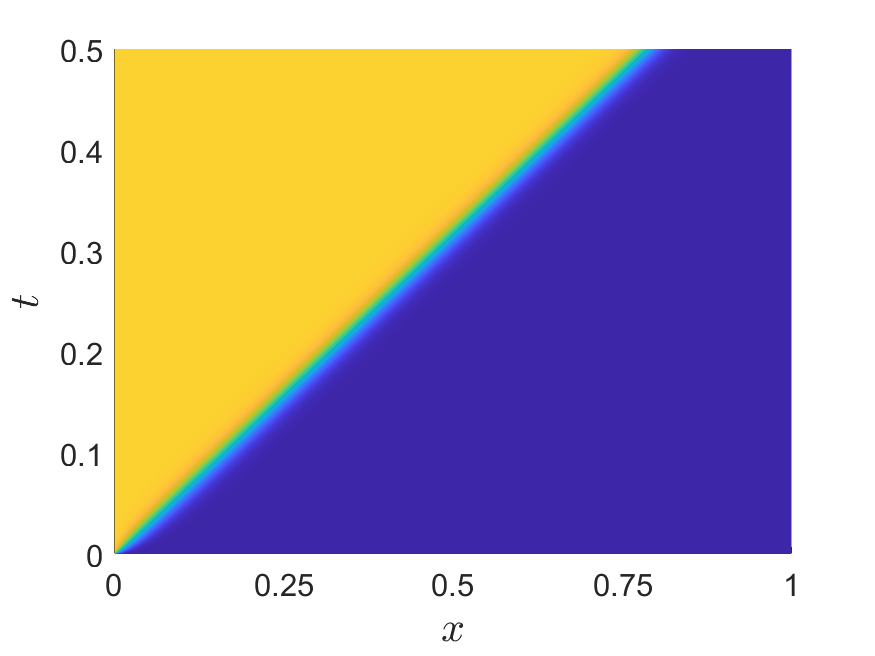}
        &
        \raisebox{1.1em}{\rotatebox[origin=lb]{90}{\parbox{2.5cm}{\centering \footnotesize{Galerkin projection}}}}&
        \includegraphics[height=0.225\textwidth]{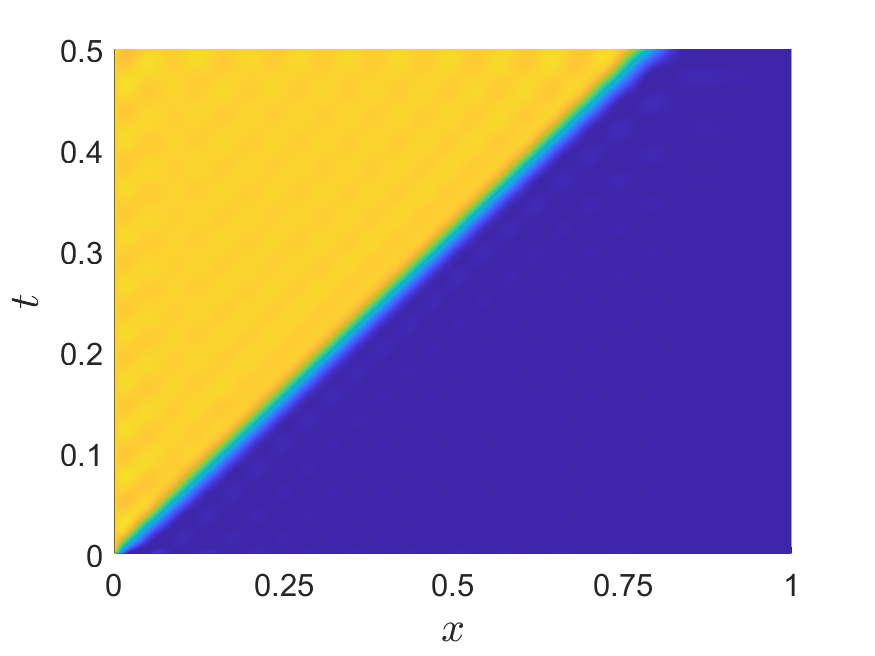}&
        \includegraphics[height=0.225\textwidth]{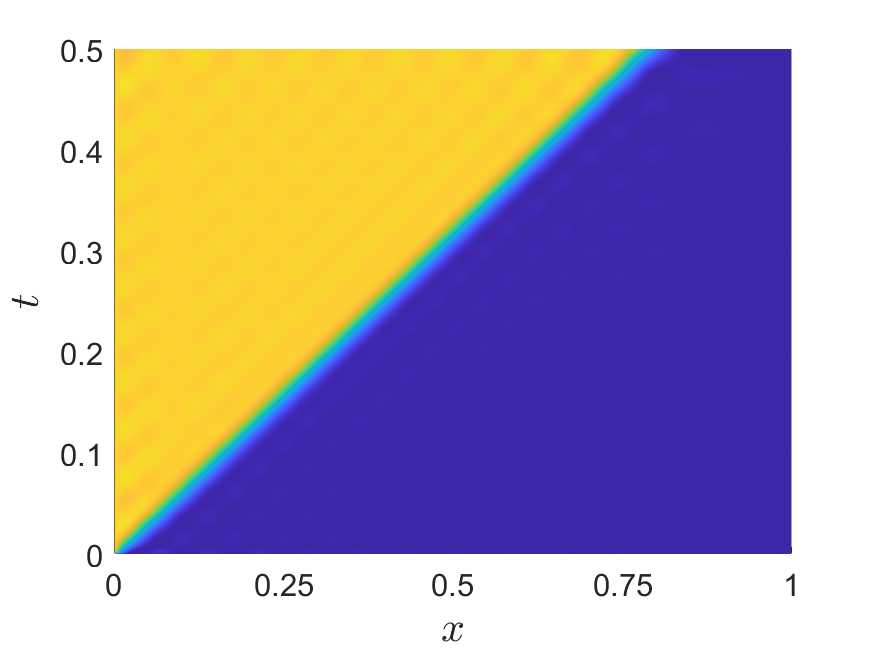}\\ \includegraphics[height=0.225\textwidth]{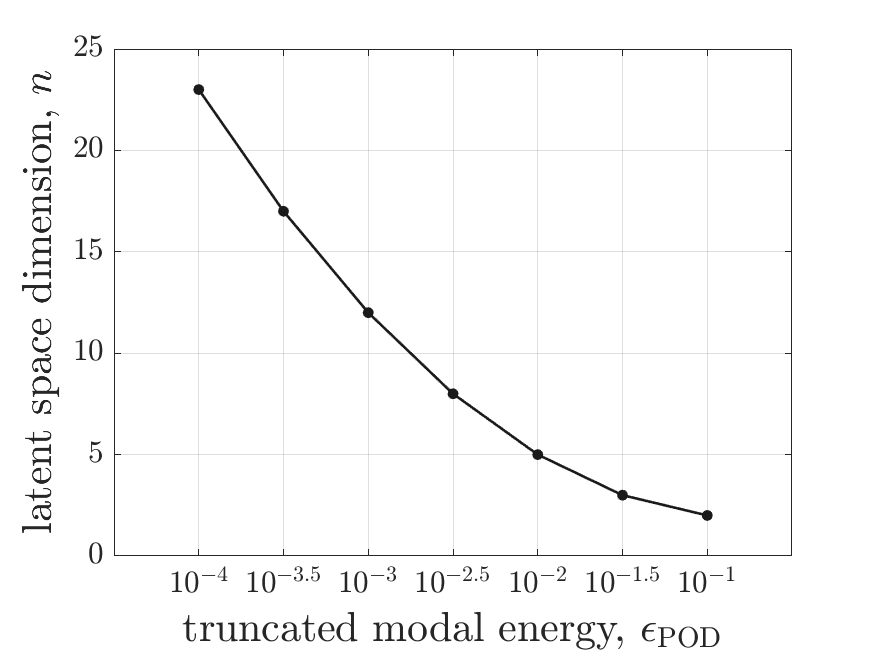}&
        \raisebox{2.0em}{\rotatebox[origin=lb]{90}{\parbox{2cm}{\centering \footnotesize{LSPG projection}}}} &
        \includegraphics[height=0.225\textwidth]{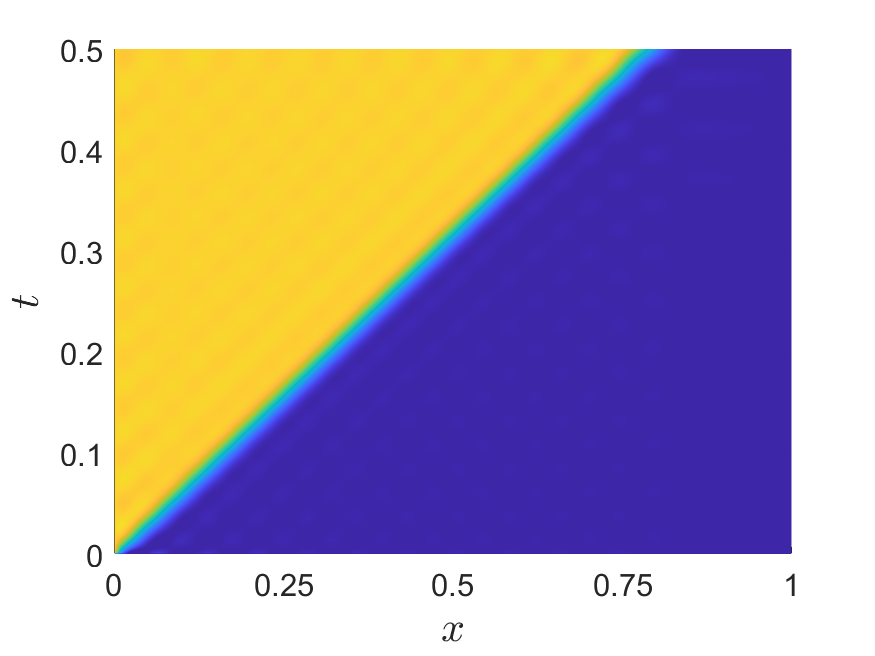}&
        \includegraphics[height=0.225\textwidth]{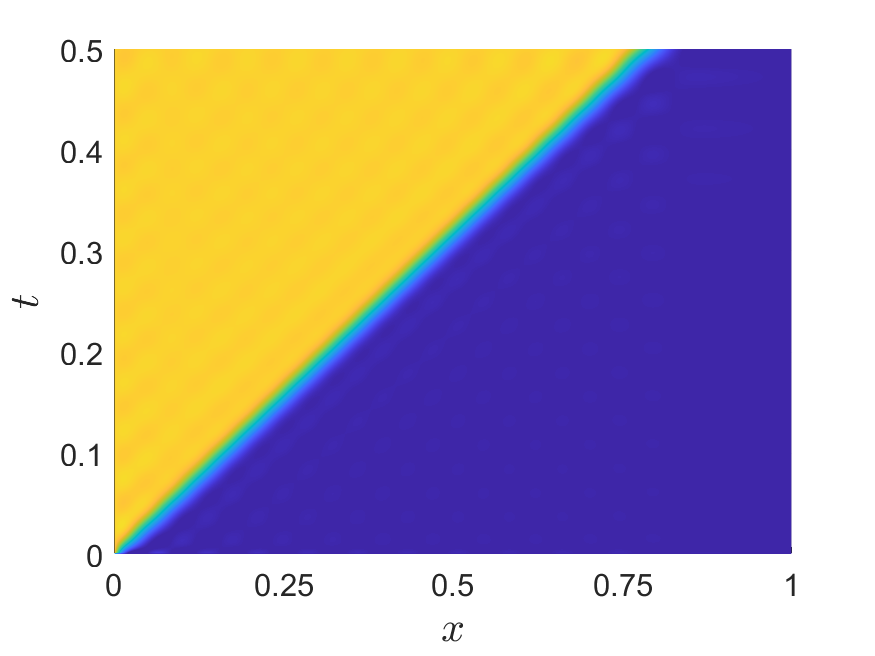}\\
        \multicolumn{4}{c}{\includegraphics[scale=.32]{BurgersFigures/Burgers_colorbar.png}}
    \end{tabular}
    \captionsetup{justification=centering}
    \caption{State solutions for the test parameter set $\boldsymbol{\mu} = (\mu_1, \mu_2)= (3.125, 0.0175)$ are shown for the FOM (top left figure), HRF (middle column), and ECSW (right column) schemes using Galerkin (top row) and LSPG (bottom row) projections. The solution exhibits a moving shock that starts at the left boundary and advects through the domain. The ROM results correspond to a truncated modal energy of $\epsilon_{\mathrm{POD}}=10^{-4}$, and the ECSW results are provided for tolerance $\epsilon_{\mathrm{ecsw}}=10^{-9}$. Qualitatively, all presented ROM solutions accurately predict the FOM solution. The bottom left figure shows the latent space dimension, $n$, plotted with respect to the truncated modal energy, $\epsilon_{\mathrm{POD}}$. As expected, $n$ increases as $\epsilon_{\mathrm{POD}}$ decreases. (Online version in color.)
    }
    \label{fig:burgers_solutions}
\end{figure}

Figure \ref{fig:burgers_solutions} presents the FOM solutions alongside those of HRF-G, HRF-LSPG, ECSW-G, and ECSW-LSPG for the test parameter set $\boldsymbol{\mu} = (\mu_1, \mu_2)= (3.125, 0.0175)$ using a truncated modal energy of $\epsilon_{\mathrm{POD}}=10^{-4}$ (i.e., $n=23$). ECSW solutions are only shown for $\epsilon_{\mathrm{ecsw}}=10^{-9}$, as the ECSW solutions for $\epsilon_{\mathrm{ecsw}}=10^{-5}$ using $\epsilon_{\mathrm{POD}}=10^{-4}$ failed to converge. Compared with the FOM, it is shown that HRF-G, HRF-LSPG, ECSW-G, and ECSW-LSPG produce qualitatively similar, accurate solutions for the test parameter set not seen in the snapshot matrix. Additionally, the figure shows the sensitivity of the ROM dimension $n$ to the truncated modal energy, $\epsilon_{\mathrm{POD}}$. As anticipated, lower values of $\epsilon_{\mathrm{POD}}$ correspond to higher values of $n$. The interested reader is referred to \ref{appendix:ECSW} for a breakdown of the number of sampled points in the ECSW schemes for each truncated modal energy, $\epsilon_{\mathrm{POD}}$, and ECSW tolerance value, $\epsilon_{\mathrm{ecsw}}$, for this example. Tighter ECSW tolerances and lower truncated modal energies generally require a larger number of sampled points in the ECSW scheme.

\begin{figure}[ht!]
    \centering
     \begin{tabular}{cc}
        {\footnotesize $\boldsymbol{\mu} = (3.125, 0.0175)$} & {\footnotesize $\boldsymbol{\mu} = (1.375, 0.0825)$} \\
        \includegraphics[height=0.3\textwidth]{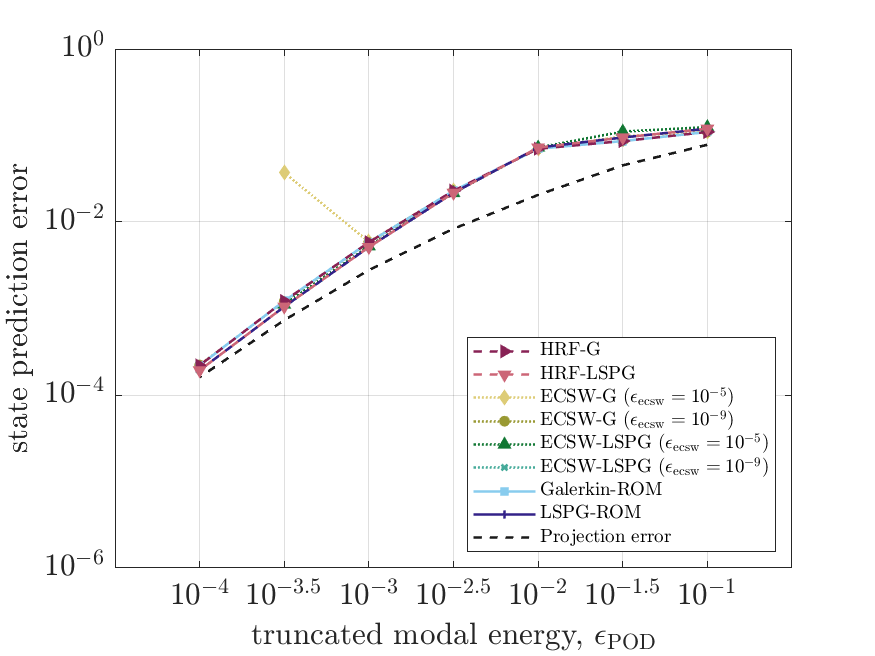}
        & \includegraphics[height=0.3\textwidth]{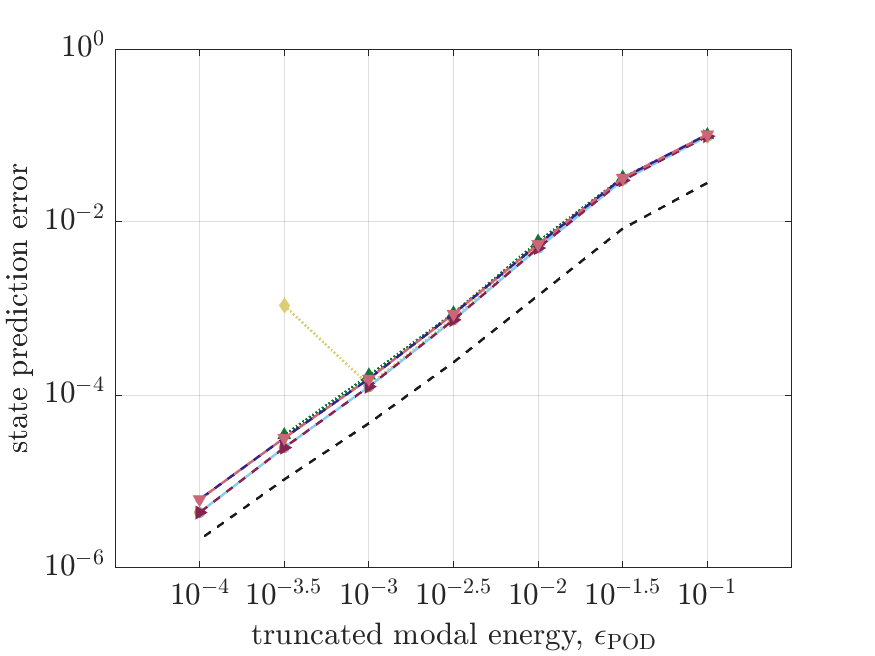}\\ \includegraphics[height=0.3\textwidth]{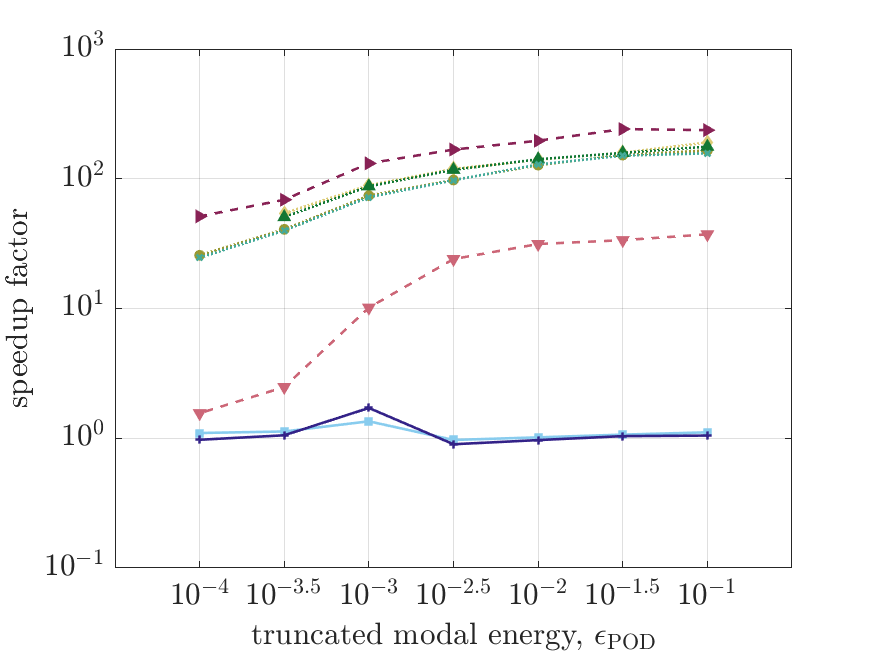}
        & \includegraphics[height=0.3\textwidth]{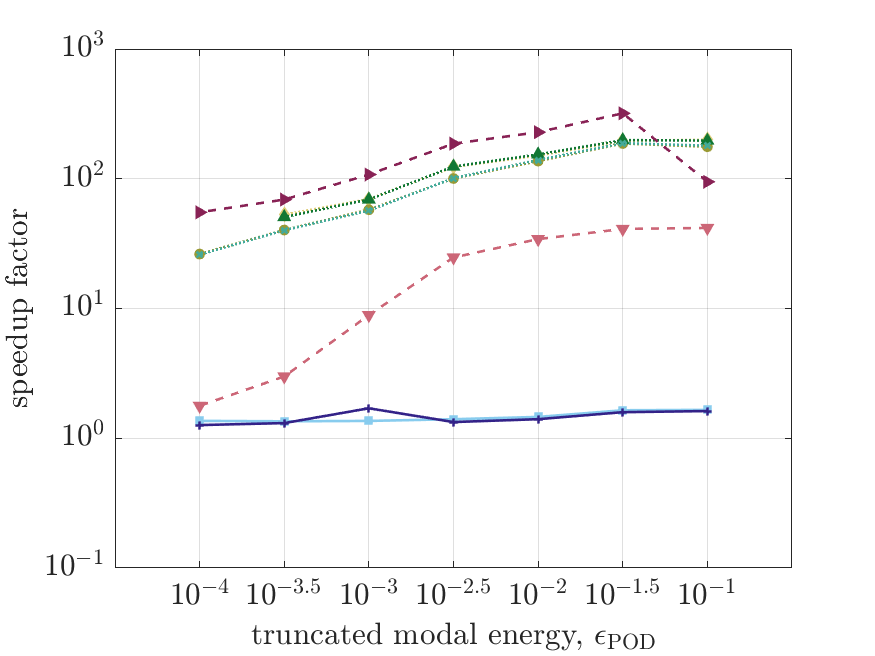}
    \end{tabular}
    \captionsetup{justification=centering}
    \caption{State prediction errors (top row) and speedup factors (bottom row) versus truncated modal energy, $\epsilon_{\mathrm{POD}}$, for HRF-G, HRF-LSPG, ECSW-G, ECSW-LSPG, Galerkin-ROM, and LSPG-ROM. ECSW results are presented for tolerances $\epsilon_{\mathrm{ecsw}}=10^{-5}$ and $10^{-9}$. The left and right columns correspond to the test parameter sets $\boldsymbol{\mu} = \left(\mu_1 ,\, \mu_2 \right)=\left( 3.125,\, 0.0175 \right)$ and $\left( 1.375,\, 0.0825 \right)$, respectively. In the figures showing the state prediction errors (top row), the projection error is also reported. As shown, the HRF-G and HRF-LSPG solutions consistently achieve the same accuracy as the Galerkin-ROM and LSPG-ROM solutions, indicating that the HRF schemes do not introduce the additional errors typically associated with hyper-reduction. In contrast, the accuracy of the ECSW results improves as the tolerance value decreases. HRF-G provides higher speedup factors in comparison with the ECSW solutions used in this study. The FOM wall-clock time for test parameter sets $\boldsymbol{\mu} =\left( 3.125,\, 0.0175 \right)$ and $\left( 1.375,\, 0.0825 \right)$ was 1.62 and 1.13 seconds, respectively. Note that the legend is the same for all figures and that the missing ECSW data points indicate solutions that failed to converge. (Online version in color.)}
    \label{fig:burgers_latTol_stateSpeedup}
\end{figure}

The accuracy and computational cost savings of Galerkin-ROM, LSPG-ROM, HRF-G, HRF-LSPG, ECSW-G, and ECSW-LSPG relative to the FOM are evaluated next. Each ROM is used to generate solutions to two test parameter sets not included in the trajectories generating the snapshot matrix: $\boldsymbol{\mu} = \left(\mu_1, \mu_2 \right) = \left(3.125, 0.0175 \right)$ and $\left(1.375, 0.0825 \right)$. Figure \ref{fig:burgers_latTol_stateSpeedup} presents the corresponding state prediction errors (along with the projection error) and speedup factors for truncated modal energies ranging from $10^{-1}$ to $10^{-4}$. As expected, the state prediction errors generally decrease as the truncated modal energy decreases. Across all truncated modal energies considered, HRF-G and HRF-LSPG provide state prediction errors that overlap perfectly with the Galerkin-ROM and LSPG-ROM counterparts. Additionally, HRF-G and HRF-LSPG provide state prediction errors comparable to their ECSW-G and ECSW-LSPG counterparts for the tolerance $\epsilon_{\mathrm{ecsw}}=10^{-9}$ for all studied $\epsilon_{\mathrm{POD}}$ and provide considerably lower state prediction errors than their ECSW-G and ECSW-LSPG counterparts for the tolerance $\epsilon_{\mathrm{ecsw}}=10^{-5}$ for $\epsilon_{\mathrm{POD}} < 10^{-3}$. The accuracy of the ECSW schemes depends on the tolerance parameter, with smaller values of $\epsilon_{\mathrm{ecsw}}$ leading to improved accuracy. For the larger tolerance (i.e., $\epsilon_{\mathrm{ecsw}}=10^{-5}$), both ECSW schemes struggle to converge at smaller truncated modal energies. Although HRF-LSPG achieves accuracy comparable to HRF-G, it achieves lower speedup factors and becomes roughly as expensive as the FOM at $\epsilon_{\mathrm{POD}}=10^{-4}$. Alternatively, HRF-G demonstrates a higher speedup than the ECSW-G solutions for both studied ECSW tolerances and for all truncated modal energies. Unlike HRF-G and HRF-LSPG, Galerkin-ROM and LSPG-ROM do not achieve any meaningful cost savings due to the lack of an offline-online decomposition. Note that the parameter set $\boldsymbol{\mu} = \left(3.125, 0.0175 \right)$ yields a solution with a sharper shock front compared to $\boldsymbol{\mu}=\left(1.375, 0.0825 \right)$, resulting in higher state prediction errors for the corresponding ROMs.

\begin{figure}[ht!]
    \centering
     \begin{tabular}{cc}
        {\footnotesize Galerkin projection} & {\footnotesize LSPG projection} \\
        \includegraphics[height=0.3\textwidth]{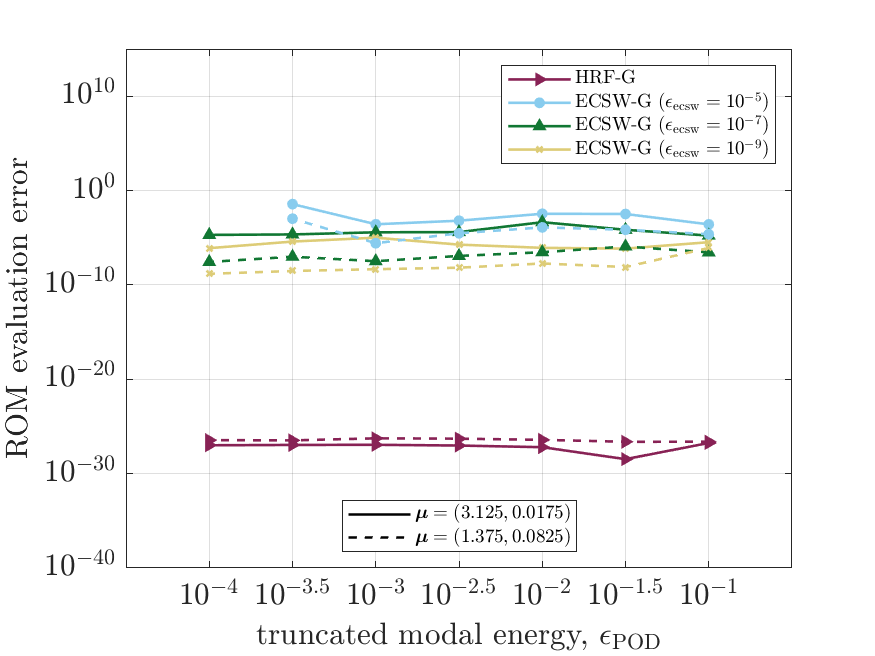} &
        \includegraphics[height=0.3\textwidth]{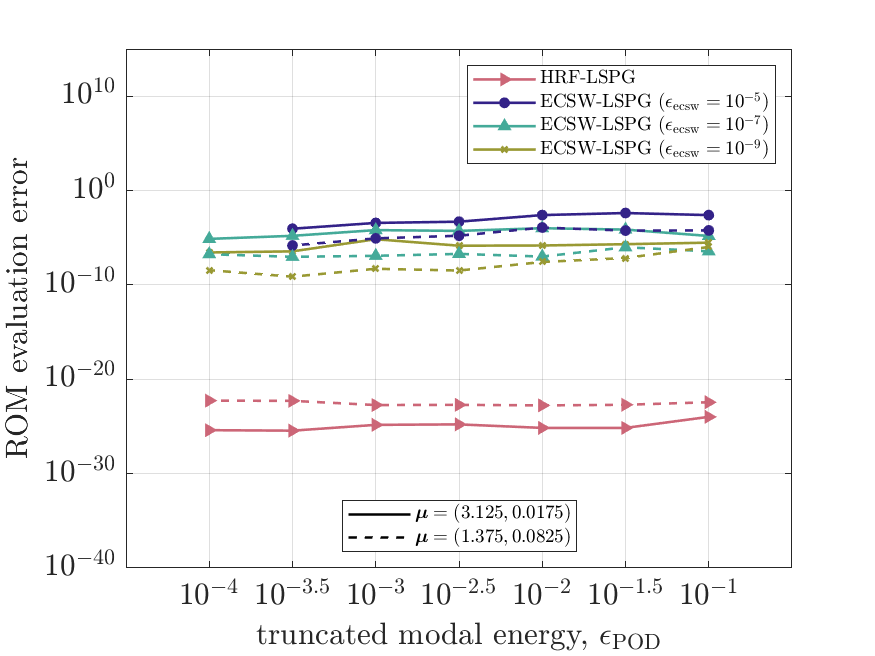}
    \end{tabular}
    \captionsetup{justification=centering}
    \caption{ROM evaluation errors for methods employing Galerkin (left) and LSPG (right) projections, plotted with respect to truncated modal energy, $\epsilon_{\mathrm{POD}}$, for HRF and ECSW schemes. ECSW results are shown for $\epsilon_{\mathrm{ecsw}}=10^{-5}$, $10^{-7}$, and $10^{-9}$. Solid and dashed lines correspond to the results for test parameter sets $\boldsymbol{\mu} = \left(3.125, 0.0175 \right)$ and $\left(1.375, 0.0825 \right)$, respectively. Since HRF-G and HRF-LSPG do not introduce the additional layer of approximation introduced by hyper-reduction, their ROM evaluation errors are several orders of magnitude lower than those of the corresponding ECSW schemes. Missing ECSW data points indicate cases where the solver failed to converge. (Online version in color.)}
    \label{fig:burgers_latTol_ROMeval}
\end{figure}

In Figure \ref{fig:burgers_latTol_ROMeval}, the ROM evaluation errors are presented, computed from \eqref{eq:eval_err}, plotted against the truncated modal energy, $\epsilon_{\mathrm{POD}}$, for HRF-G, HRF-LSPG, ECSW-G, and ECSW-LSPG, and for test parameter sets $\boldsymbol{\mu} = \left(\mu_1, \mu_2 \right) = \left(3.125, 0.0175 \right)$ and $\left(1.375, 0.0825 \right)$. To better illustrate the effect of the tolerance $\epsilon_{\mathrm{ecsw}}$ on the ROM evaluation error, Figure \ref{fig:burgers_latTol_ROMeval} presents ECSW results for $\epsilon_{\mathrm{ecsw}}=10^{-5}$, $10^{-7}$, and $10^{-9}$. Since the HRF schemes do not introduce the additional layer of approximation inherent to hyper-reduction, they achieve ROM evaluation errors that are orders of magnitude lower than those of the corresponding ECSW schemes, indicating a more accurate representation of their Galerkin-ROM and LSPG-ROM counterparts. As expected, decreasing the ECSW tolerance, $\epsilon_{\mathrm{ecsw}}$, reduces the ROM evaluation errors, showing that tighter tolerances yield ECSW-G and ECSW-LSPG solutions that more closely match the Galerkin-ROM and LSPG-ROM results, respectively. As observed in Figure \ref{fig:burgers_latTol_stateSpeedup}, both ECSW methods failed to converge for the combination $\epsilon_{\mathrm{POD}}=10^{-4}$ and $\epsilon_{\mathrm{ecsw}}=10^{-5}$.

\begin{figure}[ht!]
    \centering
     \begin{tabular}{cc}
        {\footnotesize $\boldsymbol{\mu} = (3.125, 0.0175)$} & {\footnotesize $\boldsymbol{\mu} = (1.375, 0.0825)$} \\
        \includegraphics[height=0.3\textwidth]{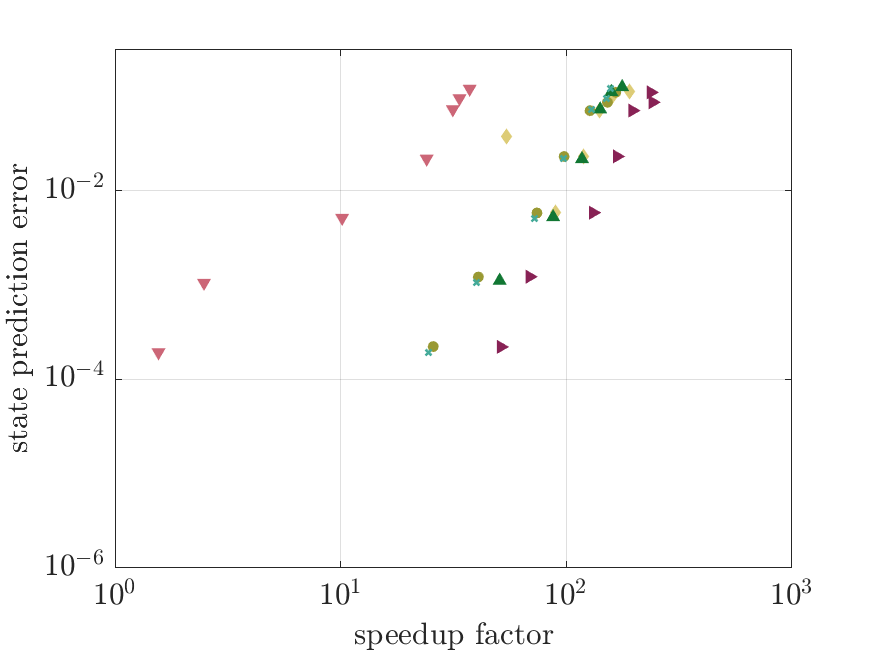} &
        \includegraphics[height=0.3\textwidth]{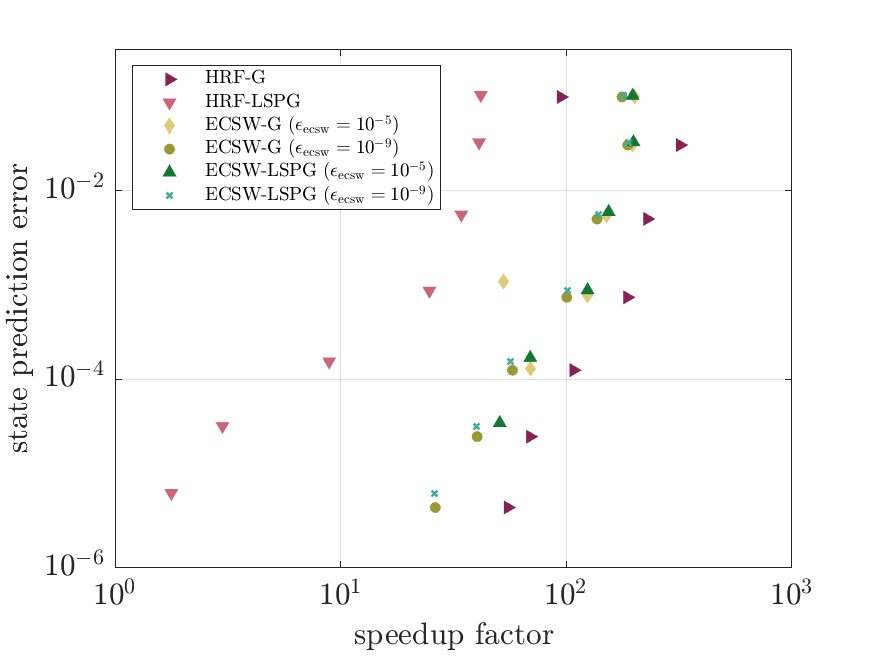}
    \end{tabular}
    \captionsetup{justification=centering}
    \caption{State prediction errors versus the speedup factor for HRF-G, HRF-LSPG, ECSW-G, and ECSW-LSPG schemes, where ECSW results are shown for tolerances $\epsilon_{\mathrm{ecsw}}=10^{-5}$ and $10^{-9}$. The left and right figures present results for test parameter sets $\boldsymbol{\mu} = \left(3.125, 0.0175 \right)$ and $\left(1.375, 0.0825 \right)$, respectively. Each data point in the scatter plots corresponds to a ROM with a specific truncated modal energy, $\epsilon_{\mathrm{POD}}$. While HRF-G provides the highest speedup factor for all state prediction errors, HRF-LSPG performs worse than ECSW-LSPG in terms of computational efficiency. The FOM wall-clock time for test parameter sets $\boldsymbol{\mu} =\left( 3.125,\, 0.0175 \right)$ and $\left( 1.375,\, 0.0825 \right)$ was 1.62 and 1.13 seconds, respectively. Note that the legend is the same for both figures. (Online version in color.)}
    \label{fig:burgers_tradeOff}
\end{figure}

 To investigate the trade-off between ROM accuracy and ROM computational cost, Figure \ref{fig:burgers_tradeOff} presents scatter plots of state prediction error versus the corresponding speedup factor for two test parameter sets. Data points located toward the bottom-right corner of the plot, indicating lower state prediction error and higher speedup, represent better overall performance. For a given state prediction error, the speedup factors achieved by HRF-G are comparable to those of ECSW-G and ECSW-LSPG, with HRF-G consistently performing slightly better. On the other hand, HRF-LSPG generally attains the smallest speedup factor. Therefore, for this numerical example, the most efficient implementation of the LSPG projection is through the ECSW hyper-reduction schemes, whereas the most efficient implementation of a Galerkin projection is achieved by HRF-G.

\begin{figure}[ht!]
    \centering
     \begin{tabular}{c|ccc}
        & {\footnotesize{HRF}} & {\footnotesize{ECSW, $\epsilon_{\mathrm{ecsw}}=10^{-5}$}} & {\footnotesize{ECSW, $\epsilon_{\mathrm{ecsw}}=10^{-9}$}} \\
        \raisebox{2.5em}{\rotatebox[origin=lb]{90}{\parbox{2cm}{\centering \footnotesize{training set}}}}&
        \includegraphics[height=0.25\textwidth]{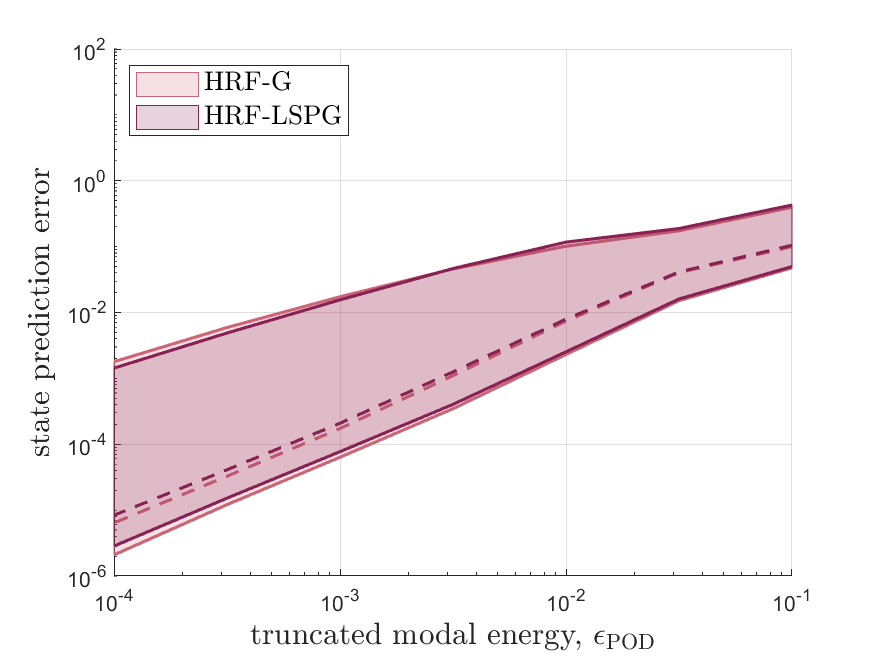}&\hspace{-.5cm}
        \includegraphics[height=0.25\textwidth]{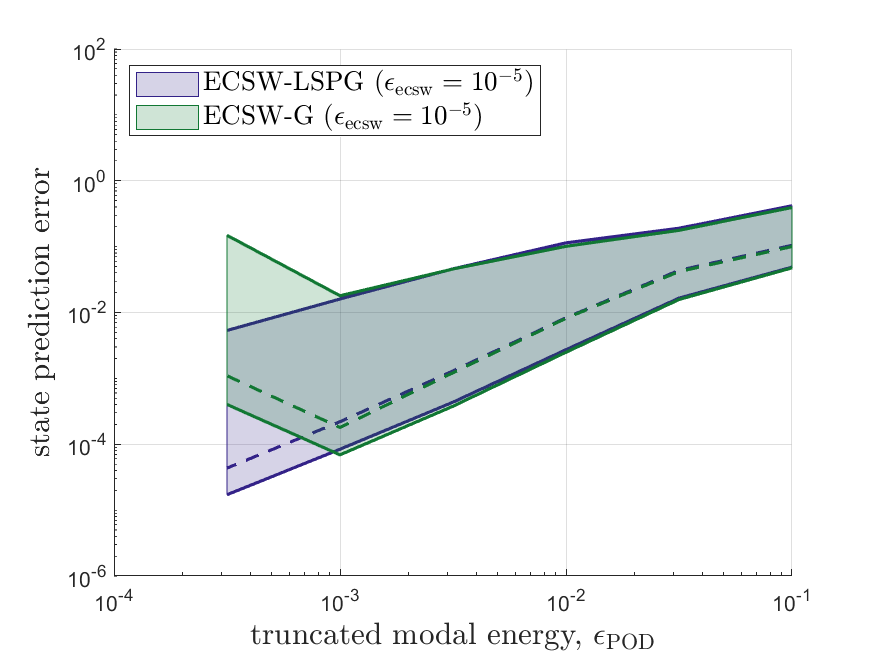}&\hspace{-.5cm}
        \includegraphics[height=0.25\textwidth]{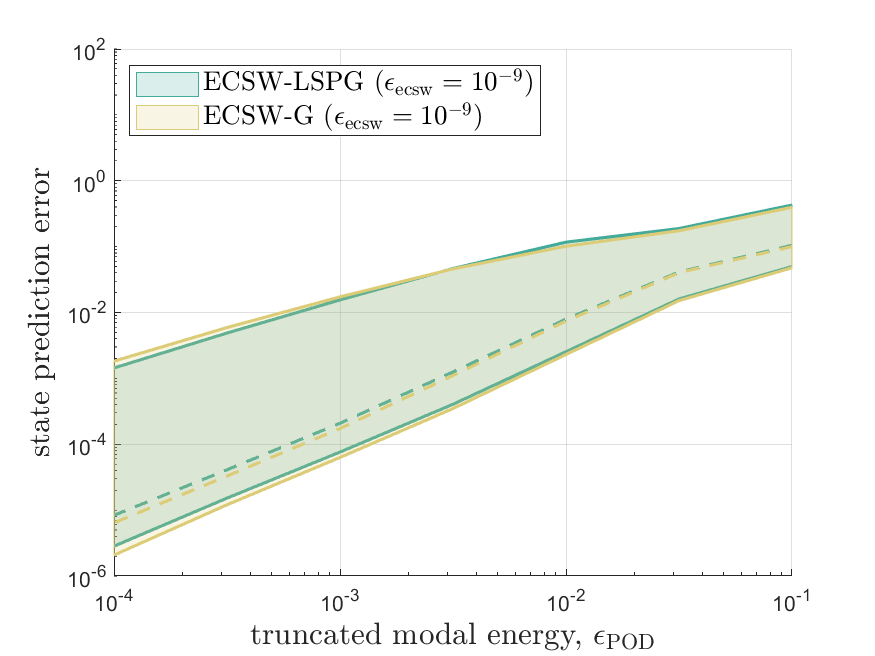}\\
        \raisebox{2.5em}{\rotatebox[origin=lb]{90}{\parbox{2cm}{\centering \footnotesize{test set}}}}&
        \includegraphics[height=0.25\textwidth]{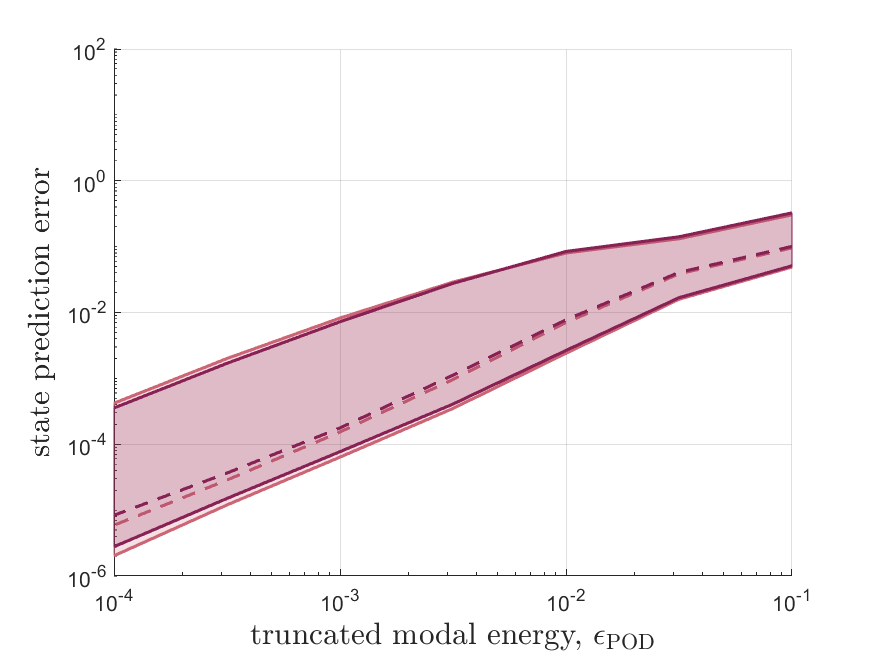}&\hspace{-.5cm}
        \includegraphics[height=0.25\textwidth]{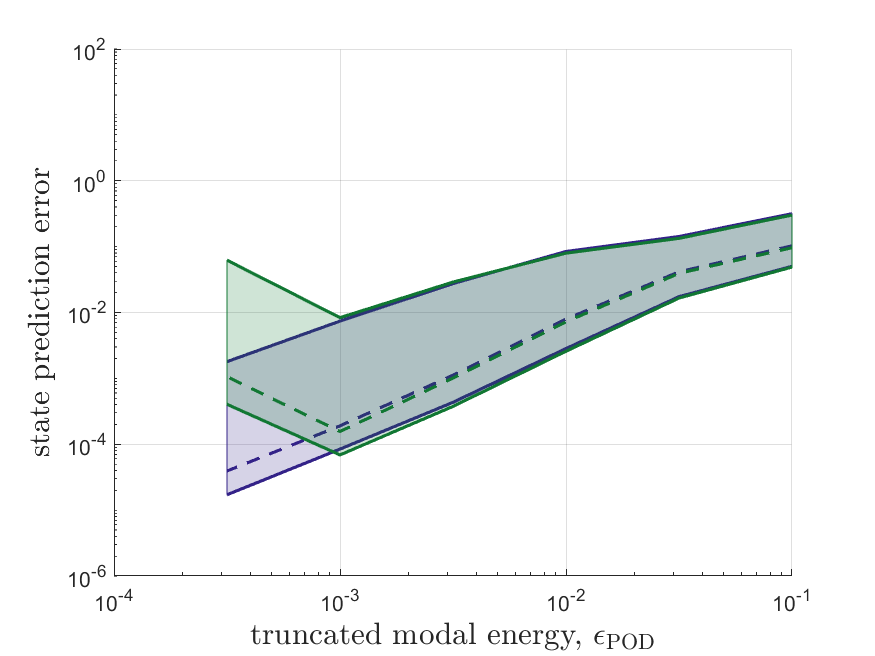}&\hspace{-.5cm}
        \includegraphics[height=0.25\textwidth]{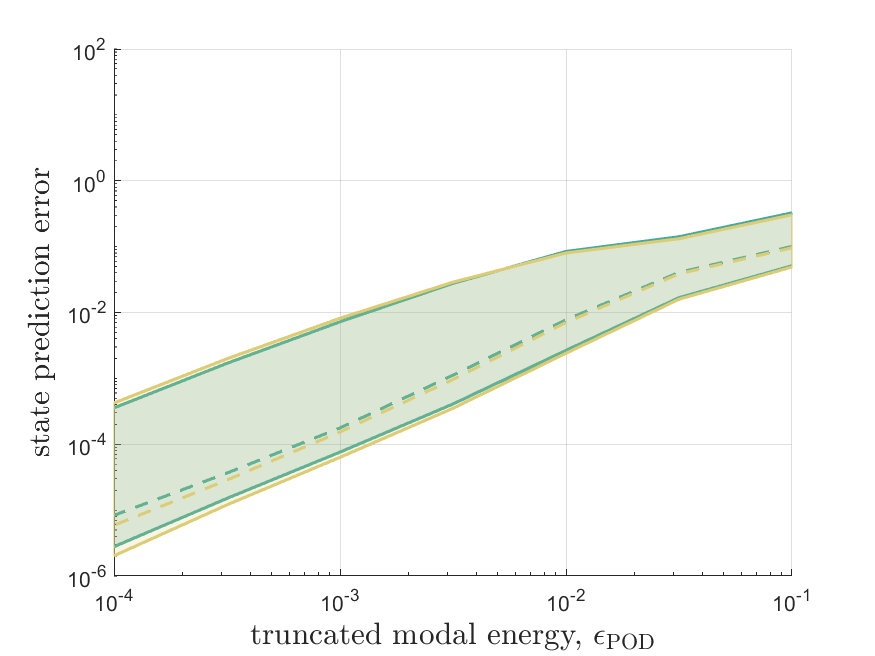}
    \end{tabular}
    \captionsetup{justification=centering}
    \caption{Range of state prediction errors for HRF schemes (left panel), and ECSW schemes for tolerance values of $\epsilon_{\mathrm{ecsw}}=10^{-5}$ (middle panel) and $10^{-9}$ (right panel) plotted with respect to the truncated modal energy, $\epsilon_{\mathrm{POD}}$. The top and bottom rows demonstrate the range of errors for the training set $\boldsymbol{\mu}=(\mu_1=1.0+0.25\mathrm{i}, \mu_2=0.01+0.01\mathrm{j})$, $\mathrm{i},\mathrm{j}=0,\ldots,9$, and for test parameter set $\boldsymbol{\mu}=(\mu_1=1.125+0.25\mathrm{i}, \mu_2=0.015+0.01\mathrm{j})$, $\mathrm{i},\mathrm{j}=0,\ldots,8$, respectively. In each figure, solid lines indicate the maximum and minimum values of error, while the dashed line indicates the median error. Note that for $\epsilon_{\mathrm{ecsw}}=10^{-5}$, all ECSW solutions failed to converge for $\epsilon_{\mathrm{POD}} = 10^{-4}$. The legend for the bottom row figures is identical to that of the corresponding top row figures. (Online version in color.)}
    \label{fig:burgers_fullRange}
\end{figure}

Finally, a more comprehensive assessment of the performance of HRF and ECSW ROMs is conducted over the parametric domain. Figure \ref{fig:burgers_fullRange} presents the range of state prediction errors for HRF and ECSW schemes, and for both Galerkin and LSPG projections over training parameter sets $\boldsymbol{\mu}=(\mu_1=1.0+0.25\mathrm{i}, \mu_2=0.01+0.01\mathrm{j})$, with $\mathrm{i},\mathrm{j}=0,\ldots,9$, and test parameter sets $\boldsymbol{\mu}=(\mu_1=1.125+0.25\mathrm{i}, \mu_2=0.015+0.01\mathrm{j})$, with $\mathrm{i},\mathrm{j}=0,\ldots,8$, plotted with respect to $\epsilon_{\mathrm{POD}}$.  ECSW results are provided for $\epsilon_{\mathrm{ecsw}}=10^{-5}$ and $10^{-9}$. Across all training and test sets, the HRF schemes exhibit state prediction errors that decrease with decreasing truncated modal energy, $\epsilon_{\mathrm{POD}}$. The same trend is observed for ECSW schemes with a tolerance of $\epsilon_{\mathrm{ecsw}}=10^{-9}$. ECSW schemes using $\epsilon_{\mathrm{ecsw}}=10^{-5}$ and $\epsilon_{\mathrm{POD}}<10^{-3}$, however, exhibit error ranges that are generally much higher than those of the HRF schemes and ECSW schemes with $\epsilon_{\mathrm{ecsw}}=10^{-9}$, and even fail to converge for $\epsilon_{\mathrm{POD}}=10^{-4}$ and $\epsilon_{\mathrm{ecsw}}=10^{-5}$.

\subsection{One-dimensional unsteady heat equation with cubic reaction term}

As the second numerical experiment, a 1D unsteady heat equation with a cubic reaction term is considered, following the model introduced in \cite{mcquarrie2025bayesian}. The objectives for this example are threefold: (i) to compare the temporal and parametric predictive capabilities of HRF approaches to those of ECSW, (ii) to assess the performance of HRF schemes when a lifting transformation is employed to expose the desired quadratic 
structure in \eqref{eq:quadraticForm}, and (iii) to demonstrate how HRF schemes are extended to a system of PDEs. 

The dynamics of the 1D unsteady heat equation are governed by the following parameterized first-order PDE and the corresponding initial and boundary conditions:

\begingroup
\begin{align} \label{eq:heatdiffCubic}
    \begin{cases}
        \dfrac{\partial q(x,t;\boldsymbol{\mu})}{\partial t} = (0.005) \dfrac{\partial^2 (q(x,t;\boldsymbol{\mu}))}{\partial x^2} - (q(x,t;\boldsymbol{\mu}))^3 + u(x,t;\boldsymbol{\mu}), & \forall x\in (0,L), \forall t \in (0,T_f],\\[6pt]
        q(0,t;\boldsymbol{\mu}) = 0, \;\; q(L,t;\boldsymbol{\mu}) = 1, & \forall t \in (0,T_f],\\[3pt]
        q(x,0; \boldsymbol{\mu})  = q_0(x)= \dfrac{x}{L}\left(1-\dfrac{x}{L}\right)\left(6\left(1-{\dfrac{x}{L}}\right)^2 e^{-x/L} - 10e^{x/L}\mathrm{sin}\left(\dfrac{x}{6L}\right) \right) + \dfrac{x}{L}, & \forall x \in (0,L),
    \end{cases}
\end{align}
\endgroup
where $L=1$, $T_f = 10$, and the diffusion coefficient is set to 0.005. The system is parameterized through the input term $u(x,t;\boldsymbol{\mu})$, with $\boldsymbol{\mu} = (a,b)$ and $a,\,b\in\mathbb{R}$,
\begin{equation}
    u(x,t;\boldsymbol{\mu}) = \frac{a \mathrm{sin}(2\pi t)}{1 + 100\left(\frac{x}{L}-{\frac{1}{4}}\right)^2}+ \frac{b \mathrm{sin}(4\pi t)}{1 + 100\left(\frac{x}{L}-{\frac{3}{4}}\right)^2}.
\end{equation}

To expose the desired quadratic structure in \eqref{eq:quadraticForm}, a lifting transformation is introduced, which reduces the cubic system \eqref{eq:heatdiffCubic} to a quadratic system through the introduction of the auxiliary variable $w=q^2$, 
\begin{align} 
    \begin{cases}\label{eq:heatdiffQuad}
        \dfrac{\partial q(x,t;\boldsymbol{\mu})}{\partial t} = (0.005) \dfrac{\partial^2 (q(x,t;\boldsymbol{\mu}))}{\partial x^2} - q(x,t;\boldsymbol{\mu})w(x,t;\boldsymbol{\mu}) +u(x,t;\boldsymbol{\mu}), & \forall x\in (0,L), \forall t \in (0,T_f], \\[6pt]
        \dfrac{\partial w(x,t;\boldsymbol{\mu})}{\partial t} = 2 (0.005) q(x,t;\boldsymbol{\mu}) \dfrac{\partial^2 (q(x,t;\boldsymbol{\mu}))}{\partial x^2} - 2(w(x,t;\boldsymbol{\mu}))^2 + 2q(x,t;\boldsymbol{\mu})u(x,t;\boldsymbol{\mu}), & \forall x\in (0,L), \forall t \in (0,T_f], \\[6pt]
        q(0,t;\boldsymbol{\mu}) = 0,\;\; q(L,t;\boldsymbol{\mu} ) = 1, \;\;w(0,t;\boldsymbol{\mu}) = 0, \;\; w(L,t;\boldsymbol{\mu}) = 1, & \forall t \in (0,T_f], \\[3pt]
        q(x,0; \boldsymbol{\mu})=q_0(x) ,\;\; w(x,0; \boldsymbol{\mu}) = (q_0(x))^2, & \forall x \in (0,L), 
    \end{cases}
\end{align}
where $q_0(x)$ is defined in \eqref{eq:heatdiffCubic}.

The lifted formulation of \eqref{eq:heatdiffQuad} transforms a single PDE of \eqref{eq:heatdiffCubic} to a system of two PDEs. The second-order central difference method is used to spatially discretize \eqref{eq:heatdiffQuad},
\begin{align} \label{eq:semidisc_2heatquad}
    &\begin{cases}
        \dfrac{\mathrm{d}}{\mathrm{d}t}\mathbf{x}_1(t;\boldsymbol{\mu}) = \mathbf{C}_1 + \mathbf{A}_{1,1}\mathbf{x}_1(t;\boldsymbol{\mu}) + \mathbf{F}_{1,12} (\mathbf{x}_1(t;\boldsymbol{\mu}) \otimes \mathbf{x}_2(t;\boldsymbol{\mu})) + \mathbf{B}_1 \mathbf{u}(t;\boldsymbol{\mu}), \\[6pt]
        \begin{aligned}
            \dfrac{\mathrm{d}}{\mathrm{d}t}\mathbf{x}_2(t;\boldsymbol{\mu}) = &\mathbf{A}_{2,1}\mathbf{x}_1(t;\boldsymbol{\mu}) + \mathbf{F}_{2,11} (\mathbf{x}_1(t;\boldsymbol{\mu}) \otimes \mathbf{x}_1(t;\boldsymbol{\mu})) \\ &+ \mathbf{F}_{2,22} (\mathbf{x}_2(t;\boldsymbol{\mu}) \otimes \mathbf{x}_2(t;\boldsymbol{\mu})) + \mathbf{N}_{2,1}(\mathbf{u}(t;\boldsymbol{\mu}) \otimes \mathbf{x}_1(t;\boldsymbol{\mu}) ),
        \end{aligned}
    \end{cases}
\end{align}
where index $i$ ($i=1,2$) in $\mathbf{A}_{i,j}$ denotes the state variable whose evolution is being studied, and index $j$ ($j=1,2$) indicates the state variable on which the linear operator $\mathbf{A}_{i,j}$ acts. The same indexing convention applies to $\mathbf{C}_i$, $\mathbf{F}_{i,jk}$, $\mathbf{B}_i$, and $\mathbf{N}_{i,j}$. \ref{appendix:fd_heat} summarizes the construction of the semidiscretized form of \eqref{eq:semidisc_2heatquad} from lifted formulation of \eqref{eq:heatdiffQuad}.

A unified equation is defined using the block-structured representation for the operators and concatenating state variables $\mathbf{x}_1$ and $\mathbf{x}_2$, i.e., $\mathbf{x}(t;\boldsymbol{\mu}) =[(\mathbf{x}_1(t;\boldsymbol{\mu}))^\top, \, (\mathbf{x}_2(t; \boldsymbol{\mu}))^\top]^\top =  [(\mathbf{q}(t;\boldsymbol{\mu}))^\top, \,(\mathbf{w}(t; \boldsymbol{\mu}))^\top]^\top$,
\begin{align}\label{eq:semidisc_1heatquad}
    &\frac{\mathrm{d}}{\mathrm{d}t}\mathbf{x}(t;\boldsymbol{\mu}) = \mathbf{C}+\mathbf{A}\mathbf{x}(t;\boldsymbol{\mu}) + \mathbf{F} (\mathbf{x}(t;\boldsymbol{\mu}) \otimes \mathbf{x}(t;\boldsymbol{\mu}))  + \mathbf{B} \mathbf{u}(t;\boldsymbol{\mu}) + \mathbf{N}(\mathbf{u}(t;\boldsymbol{\mu}) \otimes \mathbf{x}(t;\boldsymbol{\mu}) ),
\end{align}
where the operators are also defined in a block-structured form as follows
\begin{align*}
    \mathbf{C}=\left[\begin{array}{c}
        \mathbf{C}_1 \\
        \mathbf{0}
    \end{array}\right],\;\; \mathbf{A}=\left[\begin{array}{cc}
        \mathbf{A}_{1,1} & \mathbf{0} \\
        \mathbf{A}_{2,1} & \mathbf{0}
    \end{array}\right], \;\; \mathbf{F} = \left[\begin{array}{cccc}
        \mathbf{0} & \mathbf{F}_{1,12} & \mathbf{0} & \mathbf{0}\\
        \mathbf{F}_{2,11} & \mathbf{0} & \mathbf{0} &\mathbf{F}_{2,22} 
    \end{array}\right], \;\; \mathbf{B} =\left[\begin{array}{c}
        \mathbf{B}_1 \\
        \mathbf{0}
    \end{array} \right],\;\; \mathbf{N} = \left[ \begin{array}{cc}
        \mathbf{0} & \mathbf{0} \\
        \mathbf{N}_{2,1} & \mathbf{0}
    \end{array}\right]
\end{align*}
and the Kronecker products $(\mathbf{x}(t;\boldsymbol{\mu}) \otimes \mathbf{x}(t;\boldsymbol{\mu}))$ and $(\mathbf{u}(t;\boldsymbol{\mu}) \otimes \mathbf{x}(t;\boldsymbol{\mu}) )$ maintain the block-structured form of 
\begin{align*}
    \mathbf{x}(t;\boldsymbol{\mu}) \otimes \mathbf{x}(t;\boldsymbol{\mu}) = &[\left(\mathbf{x}_1(t;\boldsymbol{\mu}) \otimes \mathbf{x}_1(t;\boldsymbol{\mu})\right)^\top, \left(\mathbf{x}_1(t;\boldsymbol{\mu}) \otimes \mathbf{x}_2(t;\boldsymbol{\mu})\right)^\top, \left(\mathbf{x}_2(t;\boldsymbol{\mu}) \otimes \mathbf{x}_1(t;\boldsymbol{\mu})\right)^\top, \left(\mathbf{x}_2(t;\boldsymbol{\mu}) \otimes \mathbf{x}_2(t;\boldsymbol{\mu})\right)^\top]^\top,\\
    \mathbf{u}(t;\boldsymbol{\mu}) \otimes \mathbf{x}(t;\boldsymbol{\mu}) = &\left[\left(\mathbf{u}(t;\boldsymbol{\mu}) \otimes \mathbf{x}_1(t;\boldsymbol{\mu})\right)^\top,\, \left(\mathbf{u}(t;\boldsymbol{\mu}) \otimes \mathbf{x}_2(t;\boldsymbol{\mu})\right)^\top\right]^\top.
\end{align*}
In this example, $\mathbf{C}$, $\mathbf{A}$, $\mathbf{F}$, $\mathbf{B}$, and $\mathbf{N}$ are not parameter-dependent and are therefore constant operators that can be precomputed once in the offline phase.

In this numerical example, the spatial domain is discretized using $1024$ grid points, resulting in a lifted semi-discrete system of dimension $2N=2048$. Time integration is carried out using the backward Euler scheme with constants $\tau=1$, $\alpha_0=1$, $\alpha_1=-1$, $\beta_0=1$, and $\beta_1=0$. The time-predictive performance of the HRF schemes is evaluated by generating training data over a time horizon of $T_h=2$ with time step $\Delta t= 0.001$, corresponding to $N_h = 2001$ time steps (including initial conditions). In the online phase, solutions are computed over an extended horizon $T_f = 10$ using the same time step, resulting in $N_t=10001$ time steps including initial conditions. Consistent with \cite{mcquarrie2025bayesian}, the training parameter sets are chosen as $\boldsymbol{\mu}=(a,b)\in \left\{(-2,0),\, (-1,-2),\, (0,1),\, (1,-1), \,(2,2)\right\}$, while the test parameter set is $\boldsymbol{\mu} =(1.5, 0.5)$.

For the lifted formulation of \eqref{eq:heatdiffQuad} with two PDEs, a block-diagonal trial basis is defined,
\begin{equation}\label{eq:trialbasis_system}
    \boldsymbol{\Phi} = \begin{bmatrix}
        \mathbf{V}_1 & \mathbf{0} \\
        \mathbf{0} & \mathbf{V}_2
        \end{bmatrix} \in \mathbb{R}^{2N \times n},
\end{equation}
where $n=n_1+n_2$, and $\mathbf{V}_1\in \mathbb{R}^{N\times n_1}$ and $\mathbf{V}_2\in \mathbb{R}^{N\times n_2}$ are obtained by separately performing the method of snapshots \cite{sirovich1987turbulence} on the trajectories of $q$ and $w$ for $n_s=5$ training parameter sets. To illustrate, a snapshot matrix for $q$ is constructed by concatenating the FOM solution trajectories,
\begin{equation}\label{eq:Q_snapshot}
    \mathbf{Q} = \begin{bmatrix}
        \mathbf{Q}_{(-2,0)} & \mathbf{Q}_{(-1,-2)} & \mathbf{Q}_{(0,1)} & \mathbf{Q}_{(1,-1)} & \mathbf{Q}_{(2,2)}
    \end{bmatrix} \in \mathbb{R}^{N \times 5N_h},
\end{equation}
where $\mathbf{Q}_{(a,b)} \in \mathbb{R}^{N \times N_h}$ denotes the high-dimensional state solution trajectory obtained from the FOM for the parameter set $\boldsymbol{\mu} = (a,b)$. The corresponding solution trajectories for the auxiliary variable, $w$, are not computed by solving an additional PDE, but are instead obtained from the lifting map. Specifically,
\begin{equation}\label{eq:W_snapshot}
    \mathbf{W} = \begin{bmatrix}
        \mathbf{W}_{(-2,0)} & \mathbf{W}_{(-1,-2)} & \mathbf{W}_{(0,1)} & \mathbf{W}_{(1,-1)} & \mathbf{W}_{(2,2)}
    \end{bmatrix} \in \mathbb{R}^{N \times 5N_h}, 
\end{equation}
where
\begingroup
\begin{align*}
    \mathbf{W}_{(a,b)} = \begin{bmatrix}
        \mathbf{q}(t_0;a,b)\odot \mathbf{q}(t_0;a,b) & \mathbf{q}(t_1;a,b)\odot \mathbf{q}(t_1;a,b)) & \cdots & \mathbf{q}({t_{N_h}};a,b)\odot \mathbf{q}({t_{N_h}};a,b))
    \end{bmatrix} \in \mathbb{R}^{N \times N_h}
\end{align*}
\endgroup
denotes the high-dimensional trajectory of the auxiliary variable, $w$, for the parameter set $\boldsymbol{\mu} = (a,b)$, and $\odot$ denotes the Hadamard (element-wise) product. Finally, the POD bases $\mathbf{V}_1$ and $\mathbf{V}_2$ are obtained via SVD of $\mathbf{Q} \in \mathbb{R}^{1024 \times 10005}$ and $\mathbf{W} \in \mathbb{R}^{1024 \times 10005}$, respectively. In this numerical experiment, the wall-clock time for generating the snapshot matrix $\mathbf{Q}$ in \eqref{eq:Q_snapshot} (by computing FOM solution trajectories for all five training parameter sets) is $20.578$ seconds. On the other hand, constructing the snapshot matrix $\mathbf{W}$ in \eqref{eq:W_snapshot} via the lifting map requires only $0.003$ seconds, resulting in a negligible additional offline cost for building the snapshot matrix for the auxiliary variable $w$. Performing SVD on $\mathbf{Q}$ and $\mathbf{W}$ to obtain POD bases $\mathbf{V}_1$ and $\mathbf{V}_2$ takes $0.334$ and $0.333$ seconds, respectively. Consequently, the offline cost associated with POD basis generation is effectively doubled for the lifted formulation. However, this increase remains insignificant for the present example, which involves only one introduced auxiliary variable.

In the Galerkin projection, the test basis $\boldsymbol{\Psi}^{\text{Galerkin}}$ is identical to the trial basis $\boldsymbol{\Phi}$ in \eqref{eq:trialbasis_system}. As shown in  \cite{mcquarrie2023popinf}, for systems of PDEs, the Galerkin projection preserves the polynomial structure of the FOM when a block diagonal basis of \eqref{eq:trialbasis_system} is employed to separate the state variables. For the LSPG projection, the test basis will depend on the residuals of both equations in \eqref{eq:semidisc_2heatquad}, and is defined in a block-structured manner as follows
\begin{align}\label{eq:test_LSPG_system}
    &\boldsymbol{\Psi}^{\text{LSPG}} = \left[ \begin{array}{cc}
        \left.\dfrac{\partial \tilde{\mathbf{r}}_1}{\partial \hat{\mathbf{x}}_1}\right|_{\hat{\mathbf{x}}_1^{m(k)}} & \left.\dfrac{\partial \tilde{\mathbf{r}}_1}{\partial \hat{\mathbf{x}}_2}\right|_{\hat{\mathbf{x}}_2^{m(k)}} \\[6pt]
        \left.\dfrac{\partial \tilde{\mathbf{r}}_2}{\partial \hat{\mathbf{x}}_1}\right|_{\hat{\mathbf{x}}_1^{m(k)}} & \left.\dfrac{\partial \tilde{\mathbf{r}}_2}{\partial \hat{\mathbf{x}}_2}\right|_{\hat{\mathbf{x}}_2^{m(k)}}
    \end{array}\right],
\end{align}
where $\tilde{\mathbf{r}}_i$ denotes the approximate residual for the $i^{\text{th}}$ equation in \eqref{eq:semidisc_2heatquad}, with $i=1,2$. Systems of PDEs with more than two equations (including those arising from lifting transformations) are treated analogously. A block-diagonal trial basis is constructed as in \eqref{eq:trialbasis_system}, where the diagonal block $i$ is obtained via the method of snapshots applied to the snapshot matrix of state variable $i$. The test basis for the LSPG scheme is defined in a similar manner to \eqref{eq:test_LSPG_system}, consisting of $k\times k$ blocks for a system of $k$ PDEs with the $(i,j)$ block given by the partial derivative of the $i^{\text{th}}$ approximate residual $\tilde{\mathbf{r}}_i$ with respect to $\hat{\mathbf{x}}_j^{m(k)}$.

In this example, the HRF solutions constructed from ROMs based on the quadratic lifted formulation \eqref{eq:heatdiffQuad} are compared against the non-lifted FOM solution of \eqref{eq:heatdiffCubic}. Additionally, ECSW results derived for the non-lifted FOM formulation are reported for two tolerance values, $\epsilon_{\mathrm{ecsw}}=10^{-5}$ and $10^{-9}$, along with HRF-G solutions obtained directly from the non-lifted cubic formulation. As discussed in Section \ref{sec:HRF}, the cubic HRF-G projection is enabled by rewriting the Kronecker terms on the left-hand side of \eqref{eq:genProjRes} as a summation over indices of $\hat{\mathbf{x}}\otimes \hat{\mathbf{x}}$, rather than over the indices of $\hat{\mathbf{x}}$ as in the quadratic case. Further details on the efficient evaluation of low-dimensional operators associated with cubic terms can be found in \cite{khodabakhshi2022opinf}. State prediction errors are computed with respect to the original state variable $q(x,t;\boldsymbol{\mu})$. \ref{appendix:ECSW} provides a breakdown of the number of sampled points in the ECSW schemes for each truncated modal energy, $\epsilon_{\text{POD}}$, and ECSW tolerance value, $\epsilon_{\text{ecsw}}$.

\begin{figure}[ht!]
    \centering
     \begin{tabular}{cc|cc}
        {\footnotesize{ground truth (FOM)}} & & {\footnotesize{\;\,HRF schemes (lifted)}} & {\footnotesize{\, ECSW schemes ($\epsilon_{\mathrm{ecsw}}=10^{-9}$)}} \\[-1pt] 
        \includegraphics[height=0.225\textwidth]{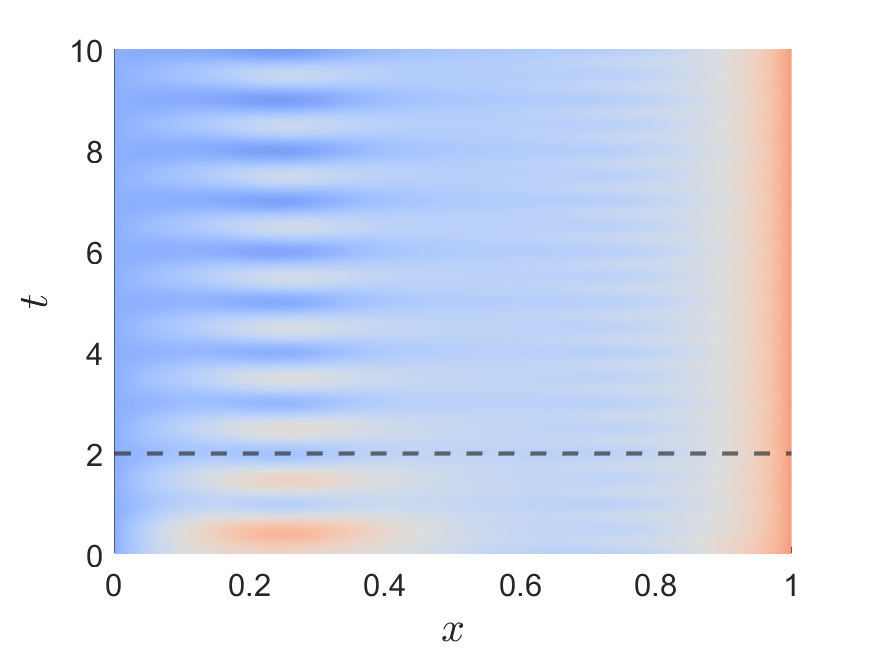}
        &
        \raisebox{1.2em}{\rotatebox[origin=lb]{90}{\parbox{2.5cm}{\centering \footnotesize{Galerkin projection}}}}&
        \includegraphics[height=0.225\textwidth]{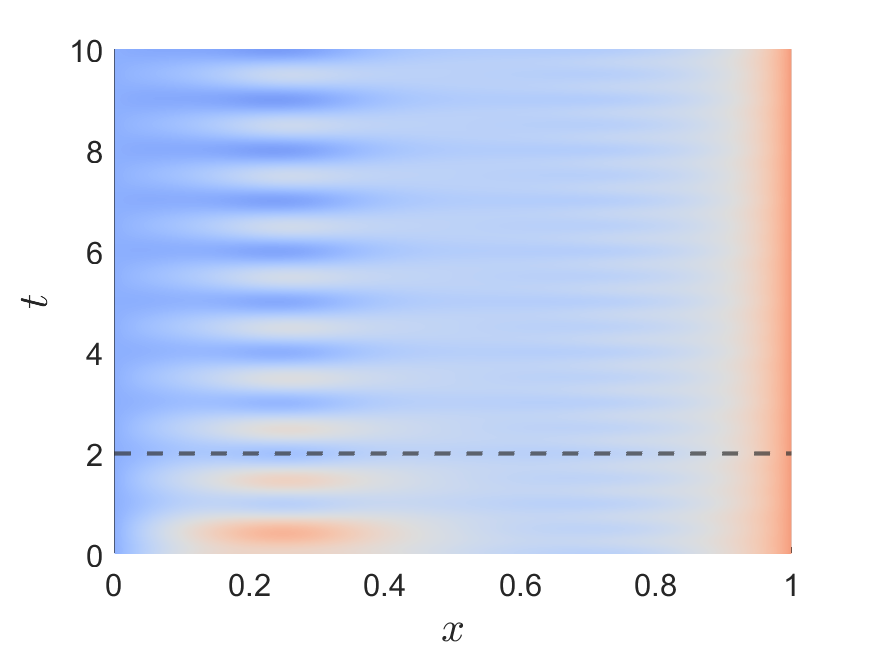}&
        \includegraphics[height=0.225\textwidth]{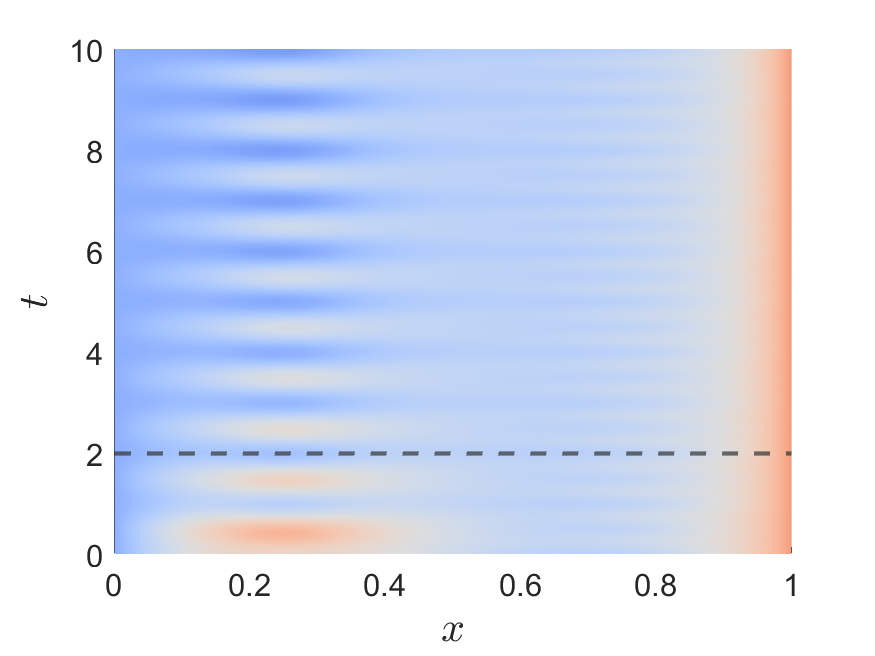}\\ \includegraphics[height=0.225\textwidth]{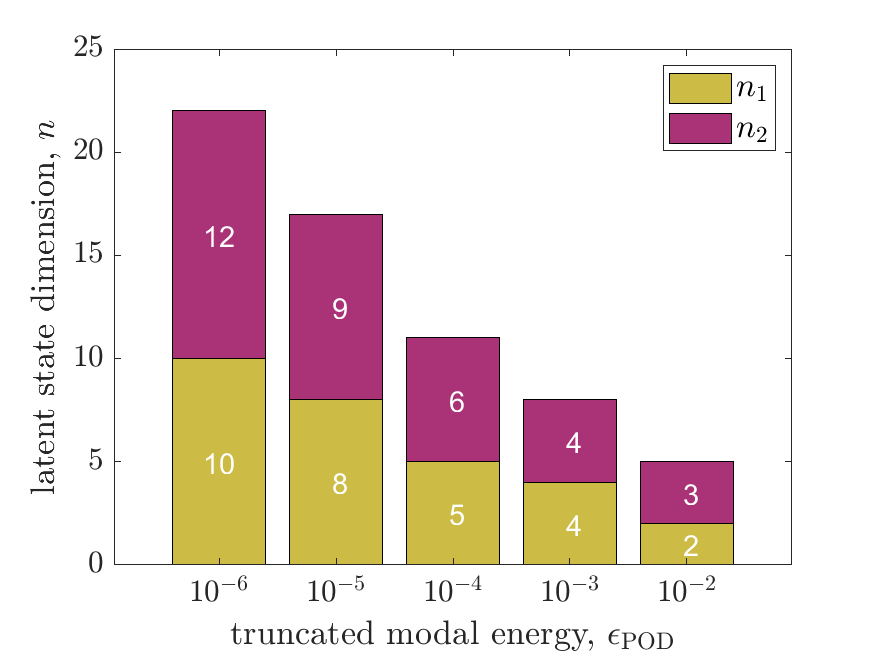}&
        \raisebox{2.0em}{\rotatebox[origin=lb]{90}{\parbox{2cm}{\centering \footnotesize{LSPG projection}}}} &
        \includegraphics[height=0.225\textwidth]{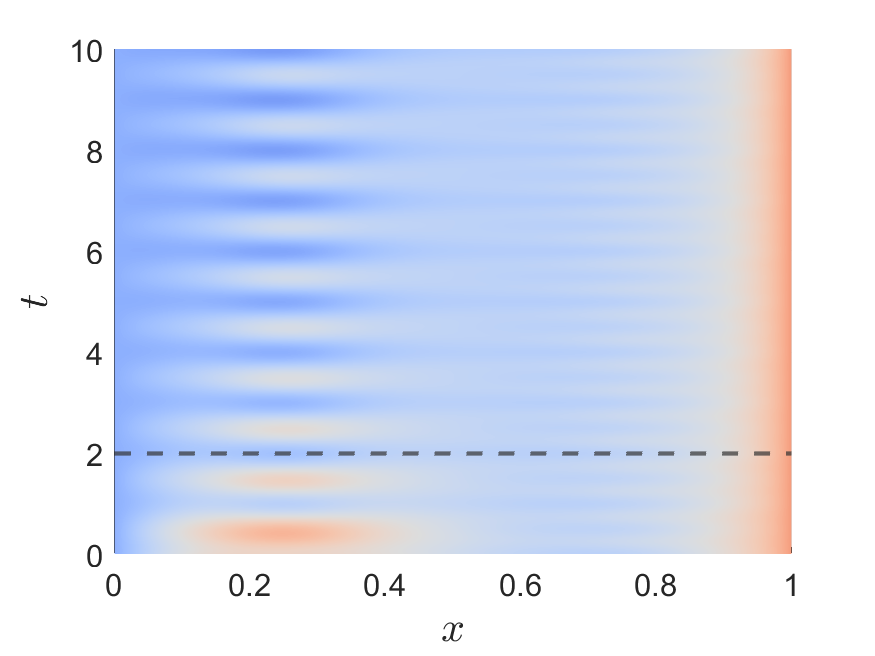}&
        \includegraphics[height=0.225\textwidth]{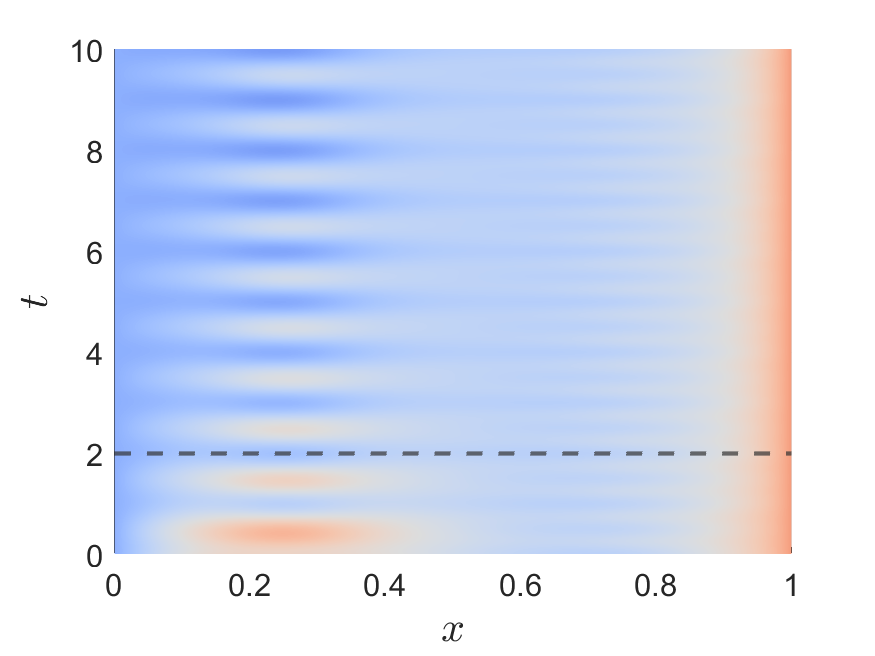}\\
        \multicolumn{4}{c}{\includegraphics[scale=.32]{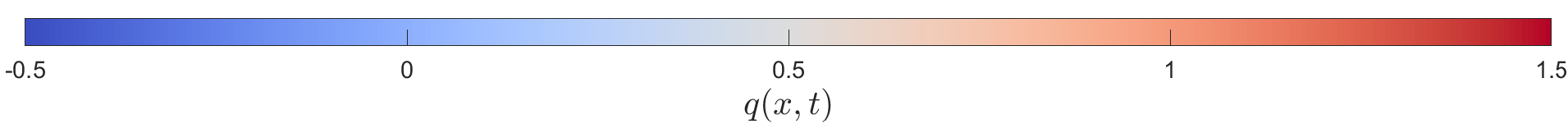}}
    \end{tabular}
    \captionsetup{justification=centering}
    \caption{State solutions for variable $q$ corresponding to the test parameter set $\boldsymbol{\mu}=(1.5,0.5)$. The top left figure shows the ground truth solution generated by the FOM. The middle column presents the solution produced by the lifted HRF-G and lifted HRF-LSPG schemes, while the right column presents the solutions from the ECSW scheme using a tolerance $\epsilon_{\mathrm{ecsw}}=10^{-9}$. In ROM solutions, the top and bottom rows correspond to Galerkin and LSPG projection schemes, respectively. All ROM solutions, using a truncated modal energy of $\epsilon_{\mathrm{POD}}=10^{-3}$, accurately predict the time evolution of state variable $q$ well beyond the training data, which is limited to the interval $t=[0,T_h=2]$ (denoted by the dashed horizontal line). The bottom left figure demonstrates how the latent space dimension of the state variables $q$ and $w$, denoted by $n_1$ and $n_2$, respectively, varies with the truncated modal energy $\epsilon_{\text{POD}}$. (Online version in color.)}
    \label{fig:heatdiff_XT}
\end{figure}

Figure \ref{fig:heatdiff_XT} presents the solutions generated by the FOM (i.e., ground truth), lifted HRF-G, lifted HRF-LSPG, ECSW-G, and ECSW-LSPG for the test parameter set $\boldsymbol{\mu} = (a,b)=(1.5, 0.5)$, using a truncated modal energy $\epsilon_{\mathrm{POD}}=10^{-3}$. ECSW results are reported for the tolerance $\epsilon_{\mathrm{ecsw}}=10^{-9}$; however, for the chosen $\epsilon_{\mathrm{POD}}$, the corresponding ECSW results obtained with $\epsilon_{\mathrm{ecsw}}=10^{-5}$ are very similar. Qualitatively, all presented ROMs accurately predict the solution for the test parameter set over the extended prediction horizon $T_f=10$ beyond the training interval $T_h=2$ over which training snapshots were acquired. In addition, Figure \ref{fig:heatdiff_XT} reports the latent space dimensions $n_1$ and $n_2$ associated with the state variables $q$ and $w$, respectively, as functions of the truncated modal energy $\epsilon_{\text{POD}}$. As expected, the number of POD modes for both state variables increases as $\epsilon_{\text{POD}}$ decreases. Moreover, the lifted ROMs require approximately twice as many POD modes with $n=n_1+n_2$ compared to ECSW and non-lifted HRF-G, requiring a basis only for the original variable $q$ (i.e., $n_1$). For the solutions shown in Figure \ref{fig:heatdiff_XT}, the latent space dimensions for $q$ and $w$ are $n_1=4$ and $n_2=4$, respectively.

\begin{figure}[ht!]
    \centering
     \begin{tabular}{cc}
        {\footnotesize Galerkin projection} & {\footnotesize LSPG projection} \\
        \includegraphics[height=0.3\textwidth]{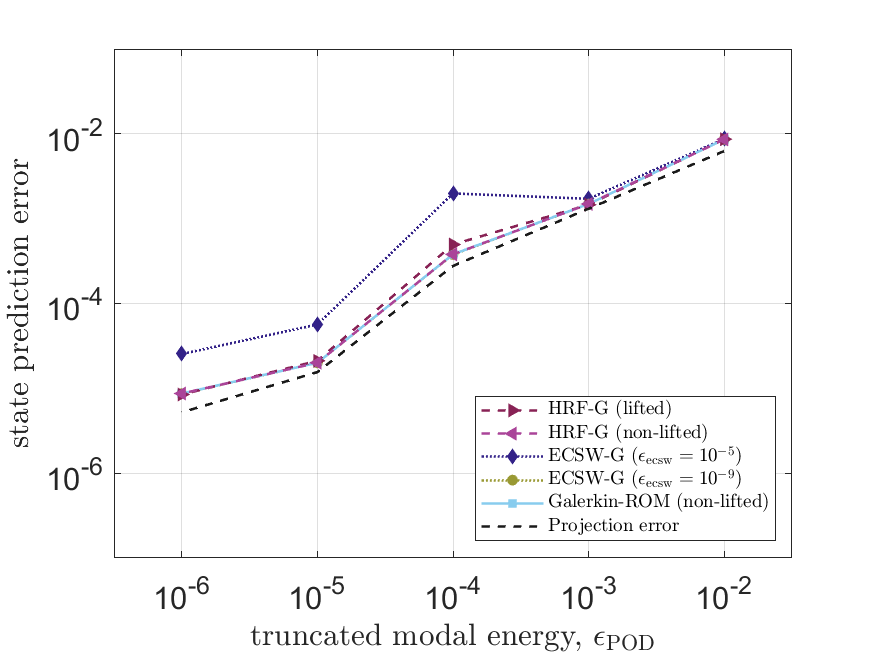}
        & \includegraphics[height=0.3\textwidth]{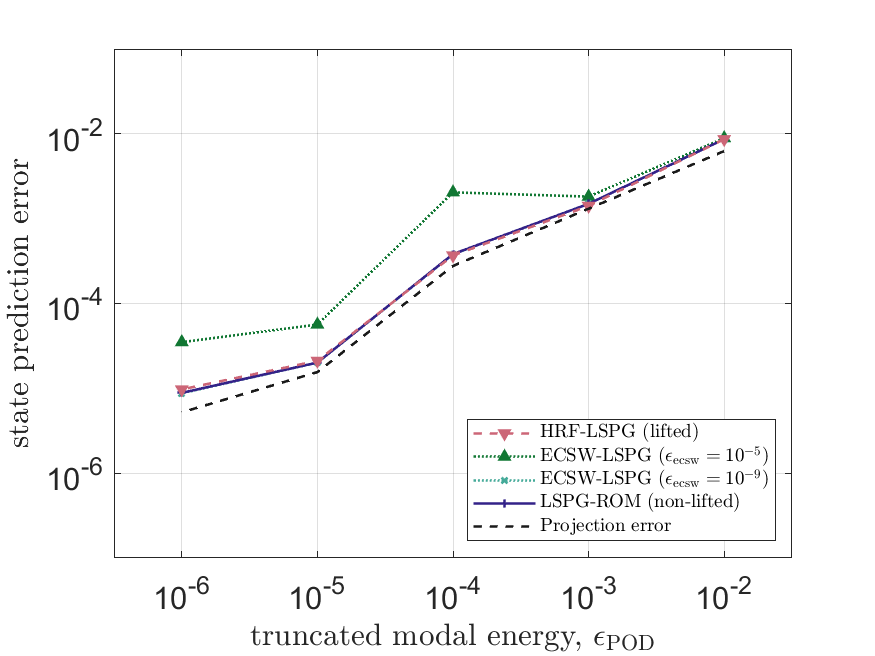}\\
        \includegraphics[height=0.3\textwidth]{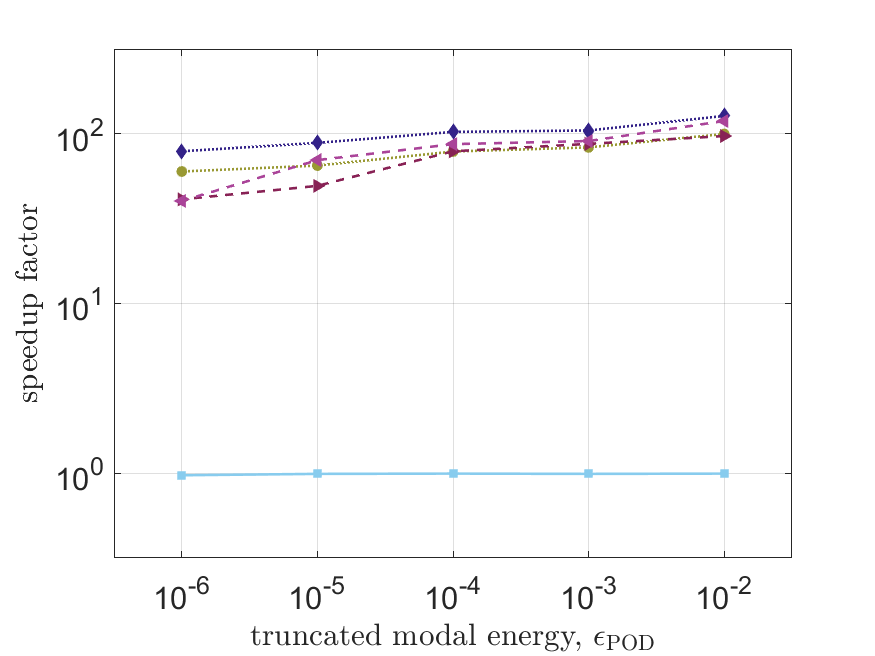}
        & \includegraphics[height=0.3\textwidth]{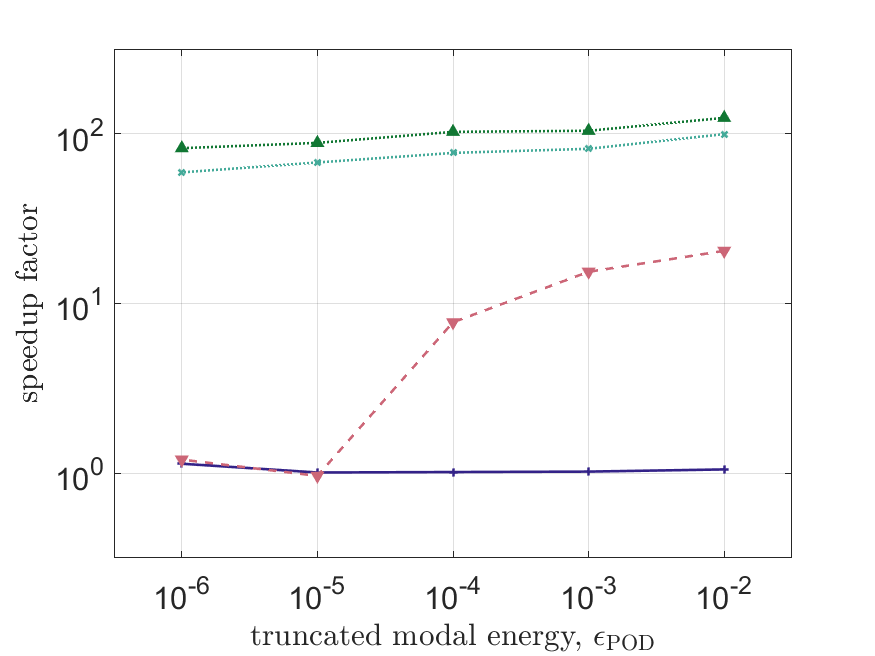}
    \end{tabular}
    \captionsetup{justification=centering}
    \caption{The top row presents state prediction errors, along with the corresponding projection errors,  plotted against the truncated modal energy, $\epsilon_{\mathrm{POD}}$, while the bottom row presents speedup factors plotted against the truncated modal energy, $\epsilon_{\mathrm{POD}}$. In the left column, results are presented for lifted HRF-G, non-lifted HRF-G, ECSW-G at $\epsilon_{\mathrm{escw}}=10^{-5}$ and $10^{-9}$, and non-lifted Galerkin-ROM. The right column shows the results for lifted HRF-LSPG, non-lifted ECSW-LSPG, and non-lifted LSPG-ROM at the same tolerance values. All results correspond to the test parameter set $\boldsymbol{\mu}=(a,b)=(1.5,0.5)$. The state prediction errors highlight the close agreement between HRF predictions and ECSW results at the tighter tolerance. The wall-clock time to run the FOM was 18.55 seconds. Note that the legends in the bottom figures are the same as in the corresponding top figures. (Online version in color.)}
    \label{fig:heatdiff_latTolStatePred_speedup}
\end{figure}

Figure \ref{fig:heatdiff_latTolStatePred_speedup} reports the state prediction errors (along with the projection error) and speedup factors plotted against the truncated modal energy, $\epsilon_{\mathrm{POD}}$, for the lifted HRF-G, lifted HRF-LSPG, non-lifted HRF-G, ECSW-G, ECSW-LSPG, Galerkin-ROM, and LSPG-ROM schemes. Note that, with the exception of the lifted HRF-G and lifted HRF-LSPG schemes, all other reported methods use the non-lifted formulation of \eqref{eq:heatdiffCubic}. The state prediction errors decrease with decreasing  $\epsilon_{\mathrm{POD}}$ for ROMs based on both projection schemes. The predictions obtained using lifted HRF-G, lifted HRF-LSPG, non-lifted HRF-G, and both ECSW schemes using the tolerance $\epsilon_{\mathrm{ecsw}}=10^{-9}$ are in close agreement with those of the Galerkin-ROM and LSPG-ROM. Moreover, the alignment between the lifted and non-lifted HRF-G solutions indicates that the lifting transformation introduces no major sources of error and can be reliably used to transform systems with general nonlinearities into a system amenable to \eqref{eq:quadraticForm} for subsequent deployment in the HRF-G framework. The state prediction errors for both ECSW schemes, with tolerance $\epsilon_{\mathrm{ecsw}}=10^{-5}$ and truncated modal energy $\epsilon_{\mathrm{POD}}\leq10^{-4}$, are roughly half an order of magnitude larger due to the looser tolerance. Lifted HRF-G and non-lifted HRF-G schemes achieve speedup factors comparable to both ECSW-G and ECSW-LSPG. This indicates that the larger latent space dimension associated with the lifting transformation can be justified by the computational savings achieved by the absence of the cubic term. In contrast, the HRF-LSPG method is once again the slowest among the studied methods. Additionally, the Galerkin-ROM and LSPG-ROM results once again fail to achieve any meaningful cost savings, due to the lack of an offline-online decomposition. All ROM methods achieve a higher speedup factor than Galerkin-ROM and LSPG-ROM, while LSPG-ROM attains comparable speedup to HRF-LSPG for $\epsilon_{\mathrm{POD}}\leq 10^{-5}$.

\begin{figure}[ht!]
    \centering
     \begin{tabular}{cc}
        {\footnotesize Galerkin projection} & {\footnotesize LSPG projection} \\
        \includegraphics[height=0.3\textwidth]{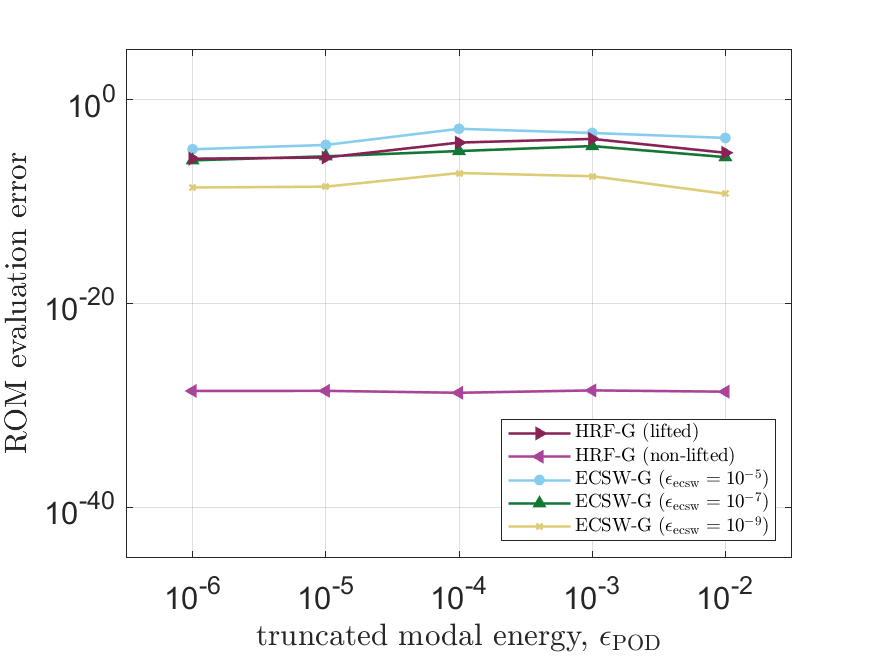}
        & \includegraphics[height=0.3\textwidth]{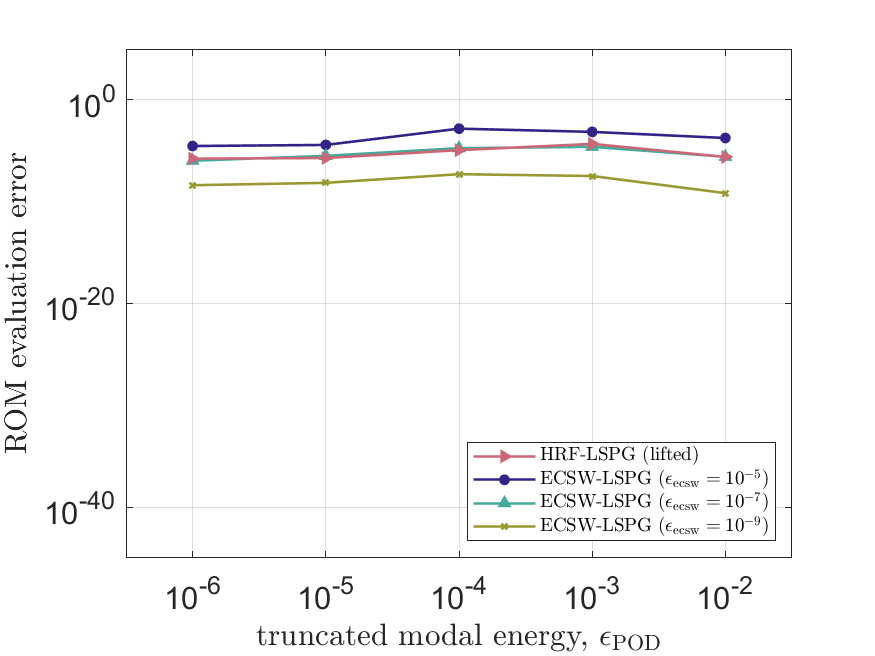}
    \end{tabular}
    \captionsetup{justification=centering}
    \caption{ROM evaluation errors plotted with respect to truncated modal energy, $\epsilon_{\mathrm{POD}}$. Solutions are generated for the test parameter set $\boldsymbol{\mu} = (a,b) = (1.5, 0.5)$. The left and right figures present the errors for ROMs using Galerkin and LSPG projection schemes, respectively. ECSW results are provided for tolerances $\epsilon_{\mathrm{ecsw}}=10^{-5}$, $10^{-7}$, and $10^{-9}$. All ROM evaluation errors are evaluated relative to their corresponding Galerkin-ROM or LSPG-ROM solutions. (Online version in color.)}
    \label{fig:heatdiff_latTolROMeval}
\end{figure}

Figure \ref{fig:heatdiff_latTolROMeval} reports the ROM evaluation errors plotted with respect to truncated modal energy, $\epsilon_{\mathrm{POD}}$, for the lifted HRF-G and HRF-LSPG schemes, as well as the non-lifted HRF-G, ECSW-G, and ECSW-LSPG schemes. ECSW results are provided for three tolerance values of $\epsilon_{\mathrm{ecsw}}=10^{-5}$, $10^{-7}$, and $10^{-9}$. Note that for each ROM, the ROM evaluation error is measured relative to its respective Galerkin-ROM or LSPG-ROM. As expected, the non-lifted HRF-G scheme achieves very low ROM evaluation errors, whereas, the lifted HRF schemes and ECSW solutions exhibit comparable ROM evaluation errors larger than that of the non-lifted HRF-G scheme. This indicates that the lifting transformation introduces a small additional error relative to the Galerkin-ROM and LSPG-ROM; however, as shown in Figure \ref{fig:heatdiff_latTolStatePred_speedup}, its impact on the state prediction error is minimal.

\begin{figure}[ht!]
    \centering
     \begin{tabular}{c}
        \includegraphics[height=0.3\textwidth]{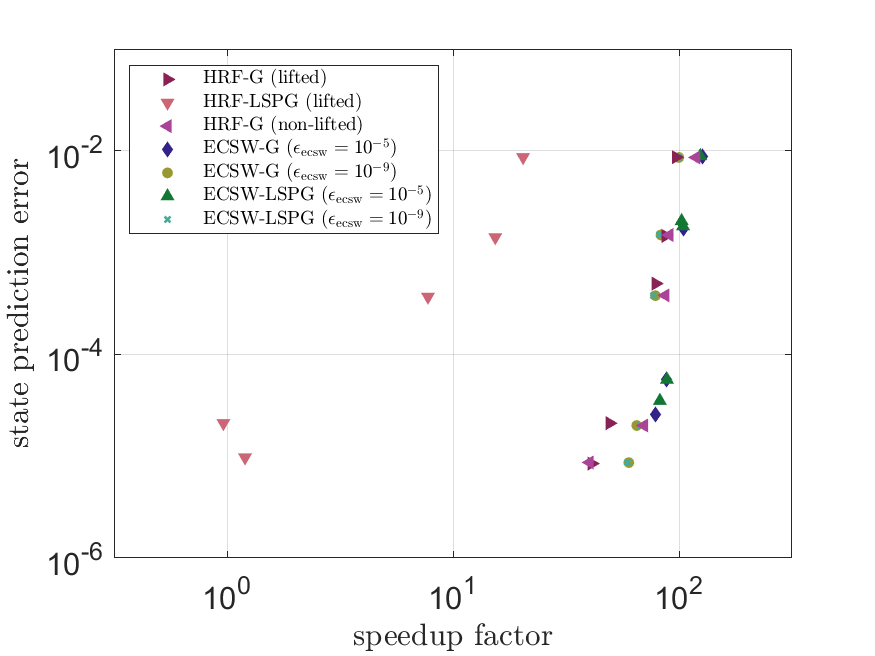}
    \end{tabular}
    \captionsetup{justification=centering}
    \caption{State prediction error for lifted HRF-G, lifted HRF-LSPG, non-lifted HRF-G, ECSW-G, and ECSW-G with respect to the speedup factor.  ECSW results are provided for tolerances $\epsilon_{\mathrm{ecsw}}=10^{-5}$ and $10^{-9}$. Each data point in each scatter plot corresponds to a unique truncated modal energy, $\epsilon_{\mathrm{POD}}$. All solutions are for the test parameter set $\boldsymbol{\mu}=(a=1.5,b=0.5)$. The lifted HRF-G and non-lifted HRF-G solutions consistently achieve speedup factors comparable to ECSW-G and ECSW-LSPG for a desired state prediction accuracy. Alternatively, the HRF-LSPG solution consistently leads to the lowest speedup factor, but still achieves substantial cost savings for several solutions. The wall-clock time to run the FOM was 18.55 seconds. (Online version in color.)}
    \label{fig:heatdiff_speedupFactorStatePred}
\end{figure}

To further investigate the trade-off between accuracy and computational cost, Figure \ref{fig:heatdiff_speedupFactorStatePred} presents a scatter plot of the state prediction error with respect to the speedup factor. Lifted and non-lifted HRF-G achieve speedup factors comparable to those of the ECSW ROMs for a given state prediction error. In contrast, while HRF-LSPG can achieve accurate state predictions, it incurs substantially higher computational cost.

\section{Conclusions and future work} \label{sec:conclusions}

In this paper, HRF schemes are developed as a class of intrusive PMOR methods for implementing Newton solvers in a reduced space. The proposed framework is readily applicable to nonlinear systems exhibiting the polynomial structure of \eqref{eq:quadraticForm}. For such problems, it was shown that the operators appearing in the projected approximate residual and its Jacobian can be precomputed in the offline phase. The resulting ROM operators depend only on the reduced dimension, enabling offline-online decomposition. This framework, presented for both Galerkin  and LSPG projections, allows for efficient implementation of reduced Newton solvers while eliminating the need for hyper-reduction and the associated additional approximation errors. 

The performance of the HRF methods was evaluated in terms of accuracy and computational speedup using two numerical examples. In both cases, the HRF results are compared with those obtained using ECSW (commonly used hyper-reduction scheme for reduced Newton solvers)  for two tolerance values. In the first example, a parameterized 1D Burgers' equation is considered to assess the predictive capabilities of the studied ROMs for previously unseen parameter sets. In the second example, a parameterized 1D unsteady heat equation with a cubic reaction term is studied, for which a lifting transformation is employed to expose the polynomial structure in \eqref{eq:quadraticForm}. The objective of this example is to evaluate both the temporal and parametric predictive capabilities of the resulting ROMs. In addition, the effectiveness of the HRF schemes is investigated for problems where nonlinearities do not naturally conform to \eqref{eq:quadraticForm}, but can be rewritten in the form of \eqref{eq:quadraticForm} through the use of lifting transformations. In both examples, it was shown that HRF-G and HRF-LSPG yield state predictions comparable to their ECSW counterparts. However, while HRF-G is computationally comparable to or faster than both ECSW-G  and ECSW-LSPG, HRF-LSPG generally provides a smaller computational speedup relative to the other studied methods. It was also demonstrated that the accuracy of the HRF schemes does not depend on an additional hyperparameter, such as the tolerance required in the ECSW scheme. In the first numerical example, the ECSW schemes failed to converge when the larger tolerance value and the smallest truncated modal energy were considered. The second numerical example further demonstrated that lifting transformations can be used to recast a nonlinear problem into the desired form \eqref{eq:quadraticForm}, enabling the successful application of HRF schemes to construct accurate ROMs that faithfully capture the dynamics of the original nonlinear model.

The intrusive nature of the HRF schemes restricts their applicability to problems of the form \eqref{eq:quadraticForm} for which the full-order operators are accessible. As a result, HRF methods cannot be directly applied to models with unknown or inaccessible operators, such as those from legacy codes or commercial software. One avenue for future work is to extend ideas from OpInf \cite{peherstorfer2016data,qian2020lift} to perform non-intrusive model-order reduction using the HRF schemes upon systems where the full-order operators are not accessible. Toward this end, the framework outlined in \cite{gkimisis2025nonintrusive} could be used to infer full-space operators from snapshot data using knowledge of mesh connectivity, thereby enabling the deployment of HRF schemes in scenarios where only snapshot data are available. Another potential area of study is to extend HRF solvers to other classes of Petrov-Galerkin projections, such as the adjoint Petrov-Galerkin scheme \cite{parish2020adjoint}.

\paragraph{Acknowledgements}
This version of the article has been accepted for publication, after peer review (when applicable) but is not the Version of Record and does not reflect post-acceptance improvements, or any corrections. The Version of Record is available online at: \url{http://dx.doi.org/10.1007/s00366-026-02358-6}.

\paragraph{Funding Statement}
L. K. Magargal is supported by the Department of Defense (DoD) through the National Defense Science \& Engineering Graduate (NDSEG) Fellowship Program. This material is based upon work supported by the Air Force Office of Scientific Research under award number FA9550-23-F-0014. P. Khodabakhshi acknowledges the support of the National Science Foundation under award 2450804. S. N. Rodriguez acknowledges OUSD R\&E for funding through the Laboratory and University Collaborative Initiative (LUCI). Portions of this research were conducted on Lehigh University's Research Computing infrastructure, partially supported by NSF Award 2019035.

\paragraph{Data Availability Statement}
The data that support the findings of this study are openly available at \url{https://github.com/liam-magargal/HRF-Newton}.

\bibliographystyle{elsarticle-num} 
\biboptions{sort&compress}
\bibliography{main}

\appendix
\setcounter{table}{0}

\section{Hyper-reduction-free least-squares Petrov-Galerkin reduced-order Newton solver}\label{appendix:HRFlspg}
This appendix summarizes the expressions for the operators contributing to the left- and right-hand sides of \eqref{eq:genProjRes} for the LSPG projection scheme, which serve as the counterparts of \eqref{eq:HFGalerkinLHS} and \eqref{eq:HFGalerkinRHS} in the Galerkin projection. Specifically, the LSPG projection \cite{carlberg2011lspg,Barnett2023NNAugmentedPROM} is obtained by taking the test basis as $\boldsymbol{\Psi} = \frac{\partial{\tilde{\mathbf{r}}}}{\partial {\hat{\mathbf{x}}}} \Big \vert_{\hat{\mathbf{x}}^{m(k)}}$. In what follows, \eqref{eq:HRFLSPGLHS3} and \eqref{eq:HRFLSPGRHS2} provide the corresponding operators contributing to the left- and right-hand sides of \eqref{eq:genProjRes}, respectively. For notational simplicity, the superscript $(k)$ is dropped on the superscript of $\hat{\mathbf{x}}^{m}$ term appearing in $\hat{\mathbf{x}}^{m-j}$ in \eqref{eq:HRFLSPGRHS2}. Additionally, for brevity purposes, the function arguments denoting the affine parametric dependence of the operators $\mathbf{C}$, $\mathbf{A}$, $\mathbf{F}$, $\mathbf{B}$, and $\mathbf{N}$ are omitted from both \eqref{eq:HRFLSPGLHS3} and \eqref{eq:HRFLSPGRHS2}, although the operators are assumed to have affine parametric dependence. As with the Galerkin projection, the low-dimensional operators (marked with underbraces in \eqref{eq:HRFLSPGLHS3} and \eqref{eq:HRFLSPGRHS2}), can be precomputed in the offline phase, enabling offline-online decomposition for the ROM. However, unlike the Galerkin formulation, the left- and right-hand sides of the HRF-LSPG projection in \eqref{eq:HRFLSPGLHS3} and \eqref{eq:HRFLSPGRHS2} involve two and one nested loops over the reduced dimension $n$, respectively, compared with one and zero nested loops in the corresponding HRF-G equations \eqref{eq:HFGalerkinLHS} and \eqref{eq:HFGalerkinRHS}. As demonstrated in Section \ref{sec:experiments}, this additional loop directly impacts the computational efficiency of the HRF-LSPG solution.

\begingroup
\begin{align} \label{eq:HRFLSPGLHS3}
    &\left(\frac{\partial \tilde{\mathbf{r}}}{\partial \hat{\mathbf{x}}^{m(k)}}\right)^\top \left(\frac{\partial \tilde{\mathbf{r}}}{\partial \hat{\mathbf{x}}^{m(k)}}\right) = \alpha_0^2 {\underbrace{\boldsymbol{\Phi}^\top \boldsymbol{\Phi}}_{\in \mathbb{R}^{n\times n}}}  - \Delta t \alpha_0 \beta_0 \bigg\{\underbrace{\boldsymbol{\Phi}^\top \mathbf{A} \boldsymbol{\Phi}}_{\in \mathbb{R}^{n\times n}} + \sum_{\ell=1}^n \Big[\underbrace{\boldsymbol{\Phi}^\top\mathbf{F}\left( \boldsymbol{\Phi} \otimes \boldsymbol{\Phi} \right) \mathbf{G}^{\ell}}_{\in \mathbb{R}^{n\times n}} + \underbrace{\boldsymbol{\Phi}^\top \mathbf{F} \left(\boldsymbol{\Phi} \otimes \boldsymbol{\Phi} \right) \mathbf{H}^{\ell}}_{\in \mathbb{R}^{n\times n}} \Big]\hat{x}_{{\ell}}^{m(k)} \notag \\
    & \qquad + \underbrace{\boldsymbol{\Phi}^\top \mathbf{N} \left( \mathbf{I}_{N_u} \otimes \boldsymbol{\Phi} \right)}_{\in \mathbb{R}^{n\times N_u n}} \left( \mathbf{u}^m \otimes \mathbf{I}_n \right)\bigg\} - \Delta t \alpha_0 \beta_0 \underbrace{\boldsymbol{\Phi}^\top \mathbf{A}^\top \boldsymbol{\Phi}}_{\in \mathbb{R}^{n\times n}} + \left(\Delta t \beta_0\right)^2 \bigg\{\underbrace{\boldsymbol{\Phi}^\top \mathbf{A}^\top \mathbf{A} \boldsymbol{\Phi}}_{\in \mathbb{R}^{n\times n}} \notag\\
    &\qquad  +  \sum_{\ell=1}^n \Big[\underbrace{\boldsymbol{\Phi}^\top \mathbf{A}^\top \mathbf{F} \left( \boldsymbol{\Phi} \otimes \boldsymbol{\Phi} \right) \mathbf{G}^{\ell}}_{\in \mathbb{R}^{n\times n}} + \underbrace{\boldsymbol{\Phi}^\top \mathbf{A}^\top \mathbf{F} \left( \boldsymbol{\Phi} \otimes \boldsymbol{\Phi} \right) \mathbf{H}^{\ell}}_{\in \mathbb{R}^{n\times n}}\Big]\hat{x}^{m(k)}_{\ell} + \underbrace{\boldsymbol{\Phi}^\top \mathbf{A}^\top \mathbf{N}\left( \mathbf{I}_{N_u} \otimes \boldsymbol{\Phi} \right)}_{\in \mathbb{R}^{n\times N_u n}} \left(\mathbf{u}^m \otimes \mathbf{I}_n \right) \bigg\} \notag\\
    & \qquad - \Delta t \alpha_0 \beta_0 \sum_{\ell=1}^n \underbrace{(\mathbf{G}^{\ell})^\top \left( \boldsymbol{\Phi} \otimes \boldsymbol{\Phi} \right)^\top \mathbf{F}^\top \boldsymbol{\Phi}}_{\in \mathbb{R}^{n\times n}} \hat{x}^{m(k)}_{\ell} + \left(\Delta t \beta_0\right)^2 \bigg\{\sum_{\ell=1}^n \underbrace{(\mathbf{G}^{\ell})^\top \left( \boldsymbol{\Phi} \otimes \boldsymbol{\Phi} \right)^\top \mathbf{F}^\top \mathbf{A} \boldsymbol{\Phi}}_{\in \mathbb{R}^{n\times n}} \hat{x}_{\ell}^{m(k)} \notag\\
    & \qquad + \sum_{\ell_1=1}^n \sum_{\ell_2=1}^n \Big[\underbrace{(\mathbf{G}^{\ell_1})^\top \left( \boldsymbol{\Phi} \otimes \boldsymbol{\Phi} \right)^\top \mathbf{F}^\top \mathbf{F} \left( \boldsymbol{\Phi} \otimes \boldsymbol{\Phi} \right) \mathbf{G}^{\ell_2}}_{\in \mathbb{R}^{n\times n}} + \underbrace{(\mathbf{G}^{\ell_1})^\top \left( \boldsymbol{\Phi} \otimes \boldsymbol{\Phi} \right)^\top \mathbf{F}^\top \mathbf{F} \left( \boldsymbol{\Phi} \otimes \boldsymbol{\Phi} \right) \mathbf{H}^{\ell_2}}_{\in \mathbb{R}^{n\times n}} \Big] \hat{x}^{m(k)}_{\ell_1} \hat{x}^{m(k)}_{\ell_2} \notag\\
    & \qquad + \sum_{\ell=1}^n \underbrace{(\mathbf{G}^{\ell})^\top \left( \boldsymbol{\Phi} \otimes \boldsymbol{\Phi} \right)^\top \mathbf{F}^\top \mathbf{N} \left( \mathbf{I}_{N_u} \otimes \boldsymbol{\Phi} \right)}_{\in \mathbb{R}^{{n\times N_u n}}} \left( \mathbf{u}^m \otimes \mathbf{I}_n \right) \hat{x}^{m(k)}_{\ell} \bigg\}  - \Delta t \alpha_0 \beta_0 \sum_{\ell=1}^n \underbrace{(\mathbf{H}^{\ell})^\top \left( \boldsymbol{\Phi} \otimes \boldsymbol{\Phi} \right)^\top \mathbf{F}^\top \boldsymbol{\Phi}}_{\in \mathbb{R}^{n\times n}} \hat{x}^{m(k)}_{\ell} \notag\\
    &\qquad + \left(\Delta t \beta_0\right)^2 \bigg\{ \sum_{\ell=1}^n \underbrace{(\mathbf{H}^{\ell})^\top \left( \boldsymbol{\Phi} \otimes \boldsymbol{\Phi} \right)^\top \mathbf{F}^\top \mathbf{A} \boldsymbol{\Phi}}_{\in \mathbb{R}^{n\times n}} \hat{x}_{\ell}^{m(k)}  + \sum_{\ell_1=1}^n \sum_{\ell_2=1}^n 
    \Big[\underbrace{(\mathbf{H}^{\ell_1})^\top \left( \boldsymbol{\Phi} \otimes \boldsymbol{\Phi} \right)^\top \mathbf{F}^\top \mathbf{F} \left( \boldsymbol{\Phi} \otimes \boldsymbol{\Phi} \right) \mathbf{G}^{\ell_2}}_{\in \mathbb{R}^{n\times n}} \notag\\
    & \qquad  + \underbrace{(\mathbf{H}^{\ell_1})^\top \left( \boldsymbol{\Phi} \otimes \boldsymbol{\Phi} \right)^\top \mathbf{F}^\top \mathbf{F} \left( \boldsymbol{\Phi} \otimes \boldsymbol{\Phi} \right) \mathbf{H}^{\ell_2}}_{\in \mathbb{R}^{n\times n}} \Big]\hat{x}^{m(k)}_{\ell_1} \hat{x}^{m(k)}_{\ell_2}  + \sum_{\ell=1}^n \underbrace{(\mathbf{H}^{\ell})^\top \left( \boldsymbol{\Phi} \otimes \boldsymbol{\Phi} \right)^\top \mathbf{F}^\top \mathbf{N} \left( \mathbf{I}_{N_u} \otimes \boldsymbol{\Phi} \right)}_{\in \mathbb{R}^{n\times N_u n}} \left( \mathbf{u}^m \otimes \mathbf{I}_n \right) \hat{x}^{m(k)}_{\ell} \bigg\} \notag\\
    & \qquad - \Delta t \alpha_0 \beta_0 \left( \mathbf{u}^m \otimes \mathbf{I}_n \right)^\top \underbrace{\left( \mathbf{I}_{N_u} \otimes \boldsymbol{\Phi} \right)^\top \mathbf{N}^\top \mathbf{\Phi}}_{\in \mathbb{R}^{N_u n\times n}} + \left(\Delta t \beta_0\right)^2 \left( \mathbf{u}^m \otimes \mathbf{I}_n \right)^\top \bigg\{ \underbrace{\left( \mathbf{I}_{N_u} \otimes \boldsymbol{\Phi} \right)^\top \mathbf{N}^\top \mathbf{A} \boldsymbol{\Phi}}_{\in \mathbb{R}^{N_u n\times n}} \notag\\
    & \qquad + \sum_{\ell=1}^n \Big[\underbrace{\left( \mathbf{I}_{N_u} \otimes \boldsymbol{\Phi} \right)^\top \mathbf{N}^\top \mathbf{F} \left( \boldsymbol{\Phi} \otimes \boldsymbol{\Phi} \right) \mathbf{G}^{\ell}}_{\in \mathbb{R}^{N_u n\times n}} + \underbrace{\left( \mathbf{I}_{N_u} \otimes \boldsymbol{\Phi} \right)^\top \mathbf{N}^\top \mathbf{F} \left( \boldsymbol{\Phi} \otimes \boldsymbol{\Phi} \right) \mathbf{H}^{\ell}}_{\in \mathbb{R}^{N_u n\times n}} \Big]\hat{x}^{m(k)}_{\ell} \notag\\
    & \qquad + \underbrace{\left( \mathbf{I}_{N_u} \otimes \boldsymbol{\Phi} \right)^\top \mathbf{N}^\top \mathbf{N} \left( \mathbf{I}_{N_u} \otimes \boldsymbol{\Phi} \right)}_{\in \mathbb{R}^{N_u n\times N_u n}} \left( \mathbf{u}^m \otimes \mathbf{I}_n \right) \bigg\},
\end{align}

\begin{align} \label{eq:HRFLSPGRHS2}
    & -\left(\frac{\partial \tilde{\mathbf{r}}}{\partial \hat{\mathbf{x}}^{m(k)}}\right)^\top \tilde{\mathbf{r}}^{m(k)} = -\alpha_0 \sum_{j=0}^{\tau} \alpha_j {\underbrace{\boldsymbol{\Phi}^\top \boldsymbol{\Phi}}_{\in \mathbb{R}^{n\times n}}} \hat{\mathbf{x}}^{m-j} + \alpha_0 \Delta t \Bigg\{\sum_{j=0}^{\tau} \beta_j \Big[ \underbrace{\boldsymbol{\Phi}^\top \mathbf{C}}_{\in \mathbb{R}^n} + \underbrace{\boldsymbol{\Phi}^\top \mathbf{A} \boldsymbol{\Phi}}_{\in \mathbb{R}^{n\times n}} \hat{\mathbf{x}}^{m-j} + \underbrace{\boldsymbol{\Phi}^\top \mathbf{F} \left(\boldsymbol{\Phi} \otimes \boldsymbol{\Phi} \right)}_{\in \mathbb{R}^{n\times n^2}} \left( \hat{\mathbf{x}}^{m-j} \otimes \hat{\mathbf{x}}^{m-j} \right) \notag\\
    &\qquad  + \underbrace{\boldsymbol{\Phi}^\top \mathbf{B}}_{\in \mathbb{R}^{n \times N_u}} \mathbf{u}^{m-j} + \underbrace{\boldsymbol{\Phi}^\top\mathbf{N}\left( \mathbf{I}_{N_u} \otimes \boldsymbol{\Phi} \right)}_{\in \mathbb{R}^{n \times N_u n}} \left( \mathbf{u}^{m-j} \otimes \hat{\mathbf{x}}^{m-j} \right) \Big]\Bigg\}  + \Delta t \beta_0 \sum_{{j=0}}^{\tau} \alpha_j \underbrace{\boldsymbol{\Phi}^\top \mathbf{A}^\top \boldsymbol{\Phi}}_{\in \mathbb{R}^{n \times n}} \hat{\mathbf{x}}^{m-j} \notag\\
    & \qquad - \left(\Delta t\right)^2 \beta_0 \Bigg\{\sum_{{j=0}}^{\tau} \beta_j \Big[ \underbrace{\boldsymbol{\Phi}^\top \mathbf{A}^\top \mathbf{C}}_{\in \mathbb{R}^{n}} + \underbrace{\boldsymbol{\Phi}^\top \mathbf{A}^\top \mathbf{A} \boldsymbol{\Phi}}_{\in \mathbb{R}^{n \times n}} \hat{\mathbf{x}}^{m-j} + \underbrace{\boldsymbol{\Phi}^\top \mathbf{A}^\top \mathbf{F} \left( \boldsymbol{\Phi} \otimes \boldsymbol{\Phi} \right)}_{\in \mathbb{R}^{n \times n^2}} \left( \hat{\mathbf{x}}^{m-j} \otimes \hat{\mathbf{x}}^{m-j} \right) + \underbrace{\boldsymbol{\Phi}^\top \mathbf{A}^\top \mathbf{B}}_{\in \mathbb{R}^{n \times N_u}} \mathbf{u}^{m-j} \notag\\
    & \qquad  + \underbrace{\boldsymbol{\Phi}^\top \mathbf{A}^\top \mathbf{N} \left( \mathbf{I}_{N_u} \otimes \boldsymbol{\Phi} \right)}_{\in \mathbb{R}^{n \times N_u n}} \left( \mathbf{u}^{m-j} \otimes \hat{\mathbf{x}}^{m-j} \right) \Big]\Bigg\} + \Delta t \beta_0 \sum_{{j=0}}^{\tau} \sum_{\ell=1}^n \alpha_j \hat{x}^{m(k)}_{\ell}(\underbrace{\mathbf{G}^{\ell})^\top \left( \boldsymbol{\Phi} \otimes \boldsymbol{\Phi} \right)^\top \mathbf{F}^\top \boldsymbol{\Phi}}_{\in \mathbb{R}^{n \times n}} \hat{\mathbf{x}}^{m-j} \notag\\
    & \qquad  - \left(\Delta t\right)^2 \beta_0 \Bigg\{ \sum_{{j=0}}^{\tau}\sum_{\ell=1}^n \beta_j \hat{x}^{m(k)}_{\ell}\Big[  \underbrace{(\mathbf{G}^{\ell})^\top \left( \boldsymbol{\Phi} \otimes \boldsymbol{\Phi} \right)^\top \mathbf{F}^\top \mathbf{C}}_{\in \mathbb{R}^{n}} + \underbrace{(\mathbf{G}^{\ell})^\top \left( \boldsymbol{\Phi} \otimes \boldsymbol{\Phi} \right)^\top \mathbf{F}^\top \mathbf{A} \boldsymbol{\Phi} }_{\in \mathbb{R}^{n \times n}} \hat{\mathbf{x}}^{m-j} \notag\\ 
    & \qquad   + \underbrace{(\mathbf{G}^{\ell})^\top \left( \boldsymbol{\Phi} \otimes \boldsymbol{\Phi} \right)^\top \mathbf{F}^\top \mathbf{F} \left( \boldsymbol{\Phi} \otimes \boldsymbol{\Phi} \right)}_{\in \mathbb{R}^{n\times n^2}} \left( \hat{\mathbf{x}}^{m-j} \otimes \hat{\mathbf{x}}^{m-j} \right) + \underbrace{(\mathbf{G}^{\ell})^\top \left( \boldsymbol{\Phi} \otimes \boldsymbol{\Phi} \right)^\top \mathbf{F}^\top \mathbf{B}}_{\in \mathbb{R}^{{n\times N_u}}} \mathbf{u}^{m-j} \notag\\ 
    & \qquad  + \underbrace{(\mathbf{G}^{\ell})^\top \left( \boldsymbol{\Phi} \otimes \boldsymbol{\Phi} \right)^\top \mathbf{F}^\top \mathbf{N} \left( \mathbf{I}_{N_u} \otimes \boldsymbol{\Phi} \right)}_{\in \mathbb{R}^{n \times N_u n}} \left( \mathbf{u}^{m-j} \otimes \hat{\mathbf{x}}^{m-j} \right) \Big]\Bigg\} + \Delta t \beta_0 \sum_{{j=0}}^{\tau} \sum_{\ell=1}^n \alpha_j \hat{x}^{m(k)}_{\ell}\underbrace{(\mathbf{H}^{\ell})^\top \left( \boldsymbol{\Phi} \otimes \boldsymbol{\Phi} \right)^\top \mathbf{F}^\top \boldsymbol{\Phi}}_{\in \mathbb{R}^{n \times n}} \hat{\mathbf{x}}^{m-j} \notag\\
    & \qquad - \left(\Delta t\right)^2 \beta_0 \Bigg\{ \sum_{{j=0}}^{\tau} \sum_{\ell=1}^n \beta_j \hat{x}^{m(k)}_{\ell} \Big[  \underbrace{(\mathbf{H}^{\ell})^\top \left( \boldsymbol{\Phi} \otimes \boldsymbol{\Phi} \right)^\top \mathbf{F}^\top \mathbf{C}}_{\in \mathbb{R}^{n}} + \underbrace{(\mathbf{H}^{\ell})^\top \left( \boldsymbol{\Phi} \otimes \boldsymbol{\Phi} \right)^\top \mathbf{F}^\top \mathbf{A} \boldsymbol{\Phi}}_{\in \mathbb{R}^{n\times n}} \hat{\mathbf{x}}^{m-j} \notag\\ 
    & \qquad  + \underbrace{(\mathbf{H}^{\ell})^\top \left( \boldsymbol{\Phi} \otimes \boldsymbol{\Phi} \right)^\top \mathbf{F}^\top \mathbf{F} \left( \boldsymbol{\Phi} \otimes \boldsymbol{\Phi} \right)}_{\in \mathbb{R}^{n \times n^2}} \left( \hat{\mathbf{x}}^{m-j} \otimes \hat{\mathbf{x}}^{m-j} \right) + \underbrace{(\mathbf{H}^{\ell})^\top \left( \boldsymbol{\Phi} \otimes \boldsymbol{\Phi} \right)^\top \mathbf{F}^\top \mathbf{B}}_{\in \mathbb{R}^{{n\times N_u}}} \mathbf{u}^{m-j} \notag\\
    & \qquad  + \underbrace{(\mathbf{H}^{\ell})^\top \left( \boldsymbol{\Phi} \otimes \boldsymbol{\Phi} \right)^\top \mathbf{F}^\top \mathbf{N} \left( \mathbf{I}_{N_u} \otimes \boldsymbol{\Phi} \right)}_{\in \mathbb{R}^{n \times N_u n}} \left( \mathbf{u}^{m-j} \otimes \hat{\mathbf{x}}^{m-j} \right) \Big] \Bigg\} + \Delta t \beta_0 \sum_{{j=0}}^{\tau} \alpha_j \left( \mathbf{u}^m \otimes \mathbf{I}_{n} \right)^\top \underbrace{\left( \mathbf{I}_{N_u} \otimes \boldsymbol{\Phi} \right)^\top \mathbf{N}^\top \boldsymbol{\Phi}}_{\in \mathbb{R}^{N_u n \times n}} \hat{\mathbf{x}}^{m-j} \notag\\
    & \qquad  - \left(\Delta t\right)^2 \beta_0 \Bigg\{ \sum_{{j=0}}^{\tau} \beta_j \left( \mathbf{u}^m \otimes \mathbf{I}_{n} \right)^\top \Big[ \underbrace{\left( \mathbf{I}_{N_u} \otimes \boldsymbol{\Phi} \right)^\top \mathbf{N}^\top \mathbf{C}}_{\in \mathbb{R}^{{N_u n}}} + \underbrace{\left( \mathbf{I}_{N_u} \otimes \boldsymbol{\Phi} \right)^\top \mathbf{N}^\top \mathbf{A} \boldsymbol{\Phi}}_{\in \mathbb{R}^{N_u n\times n}} \hat{\mathbf{x}}^{m-j} \notag\\
    & \qquad + \underbrace{\left( \mathbf{I}_{N_u} \otimes \boldsymbol{\Phi} \right)^\top \mathbf{N}^\top \mathbf{F} \left( \boldsymbol{\Phi} \otimes \boldsymbol{\Phi} \right)}_{\in \mathbb{R}^{N_u n\times n^2}} \left( \hat{\mathbf{x}}^{m-j} \otimes \hat{\mathbf{x}}^{m-j} \right) + \underbrace{\left( \mathbf{I}_{N_u} \otimes \boldsymbol{\Phi} \right)^\top \mathbf{N}^\top \mathbf{B}}_{\in \mathbb{R}^{N_u n\times N_u}} \mathbf{u}^{m-j} \notag\\
    & \qquad  +  \underbrace{\left( \mathbf{I}_{N_u} \otimes \boldsymbol{\Phi} \right)^\top \mathbf{N}^\top \mathbf{N} \left( \mathbf{I}_{N_u} \otimes \boldsymbol{\Phi} \right)}_{\in \mathbb{R}^{N_u n \times N_u n}} \left( \mathbf{u}^{m-j} \otimes \hat{\mathbf{x}}^{m-j} \right) \Big] \Bigg\},
\end{align}

\endgroup

\setcounter{figure}{0}
\setcounter{equation}{0}
\section{Energy-conserving sampling and weighting (ECSW) hyper-reduction scheme }\label{appendix:ECSW}
As presented in \cite{farhat2015ecsw,grimberg2021ecsw}, at each Newton iteration, the projected full-order approximate residual and Jacobian (left- and right-hand sides of \eqref{eq:genProjRes}) can be exactly expressed as a summation over their indices. ECSW aims to approximate the projected quantities as a weighted sum over a \textit{sampled subset of indices}. Therefore, the left- and right-hand sides of \eqref{eq:genProjRes} are approximated as
\begin{equation}\label{eq:ecswLHS}
    \mathbf{\Psi}^\top \left(\frac{\partial{\tilde{\mathbf{r}}}}{\partial {\hat{\mathbf{x}}}} \Big \vert_{\hat{\mathbf{x}}^{m(k)}} \right) = \sum_{s \in \mathcal{S}} \mathbf{\Psi}^\top \mathbf{P}_s^\top \left(\frac{\partial{\tilde{\mathbf{r}}}}{\partial {\mathbf{x}}} \Big \vert_{\mathbf{P}_{s^+}\boldsymbol{\Phi}\hat{\mathbf{x}}^{m(k)}}\right) \mathbf{P}_{s^+} \boldsymbol{\Phi} \approx \sum_{s \in \tilde{\mathcal{S}}} \omega_s \mathbf{\Psi}^\top \mathbf{P}_s^\top \left(\frac{\partial{\tilde{\mathbf{r}}}}{\partial {\mathbf{x}}} \Big \vert_{\mathbf{P}_{s^+}\boldsymbol{\Phi}\hat{\mathbf{x}}^{m(k)}}\right) \mathbf{P}_{s^+} \boldsymbol{\Phi},
\end{equation}
\begin{equation}\label{eq:ecswRHS}
    -\mathbf{\Psi}^\top \tilde{\mathbf{r}}^{m(k)} = -\sum_{s \in \mathcal{S}} \mathbf{\Psi}^\top \mathbf{P}_s^\top \tilde{\mathbf{r}}^{m(k)} (\mathbf{P}_{s^+}\boldsymbol{\Phi}\hat{\mathbf{x}}^{m(k)}) \approx -\sum_{s \in \tilde{\mathcal{S}}} \omega_s \mathbf{\Psi}^\top \mathbf{P}_s^\top \tilde{\mathbf{r}}^{m(k)}(\mathbf{P}_{s^+}\boldsymbol{\Phi}\hat{\mathbf{x}}^{m(k)}),
\end{equation}
where $\mathcal{S}$ denotes the set of all collocation points in the spatial discretization, $\tilde{\mathcal{S}} \subset \mathcal{S}$ denotes the subset of sampled points used in the hyper-reduction scheme (with $\vert \tilde{\mathcal{S}} \vert \ll \vert \mathcal{S} \vert$), and $\boldsymbol{\omega} \in \mathbb{R}^{N}$ is the corresponding sparse weight vector. Here, $\mathbf{P}_s \in \mathbb{R}^{N \times N}$ and $\mathbf{P}_{s^+} \in \mathbb{R}^{N \times N}$ denote the sampling matrices that extract specific rows of a vector or matrix. Specifically, $\mathbf{P}_s$ extracts the $s^{\mathrm{th}}$ row, while $\mathbf{P}_{s^+}$ extracts the $s^{\mathrm{th}}$ row together with the rows corresponding to its neighboring sampling points in the spatial discretization. The sampling matrices allow for sparse evaluations of the full-order residual and its Jacobian at sampled indices in the mesh. As is common in the literature \cite{farhat2015ecsw,grimberg2021ecsw}, the weight vector $\boldsymbol{\omega}$ and the set of sampled points $\tilde{\mathcal{S}}$ are determined by applying a non-negative least-squares solver \cite{lawson1995nnls} to residual snapshot data from the FOM. The user-prescribed hyperparameter $\epsilon_{\mathrm{ecsw}} \in [0,1]$ controls the trade-off between ROM accuracy and computational cost. For further details on ECSW and its implementation, see \cite{farhat2015ecsw,grimberg2021ecsw}.

\begin{figure}[ht!]
    \centering
     \begin{tabular}{ccc}
        & {\footnotesize Galerkin projection} & {\footnotesize LSPG projection} \\
        \raisebox{2.4em}{\rotatebox[origin=lb]{90}{\parbox{3.0cm}{\centering \footnotesize{1D viscous Burgers' equation}}}} & \includegraphics[height=0.3\textwidth]{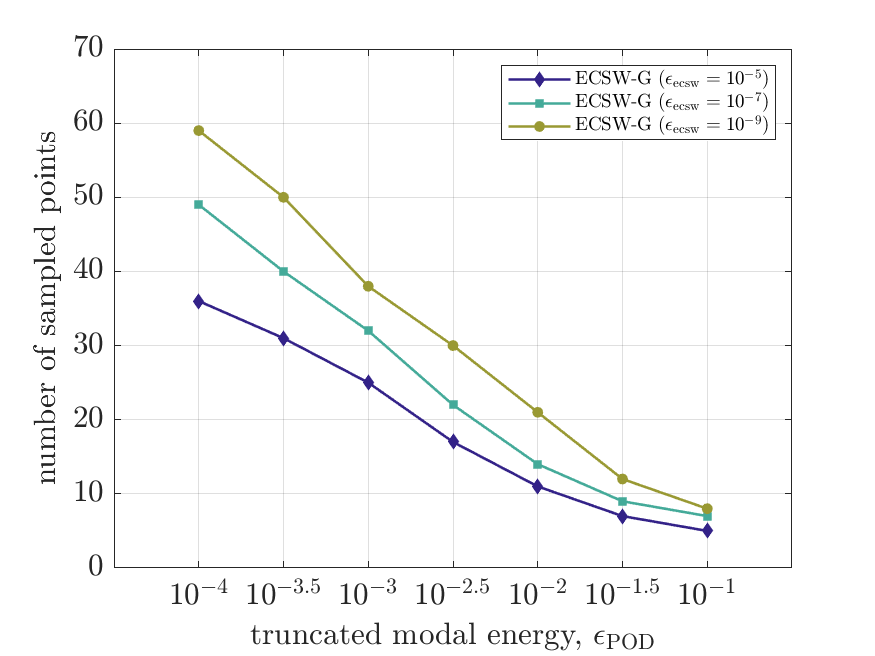}
        & \includegraphics[height=0.3\textwidth]{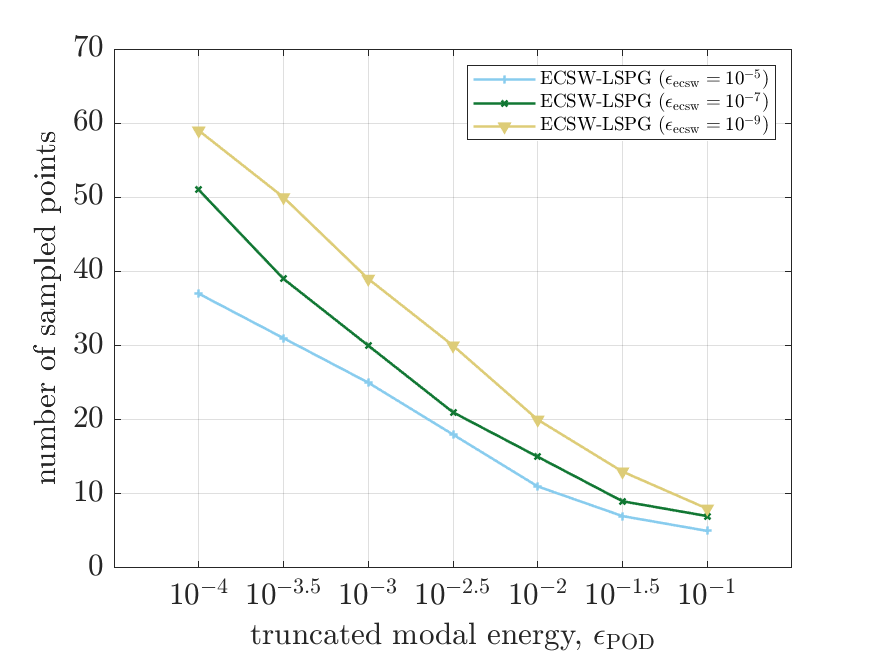}\\
        
        \raisebox{2.8em}{\rotatebox[origin=lb]{90}{\parbox{3.0cm}{\centering \footnotesize{1D unsteady heat equation with cubic reaction term}}}} & \includegraphics[height=0.3\textwidth]{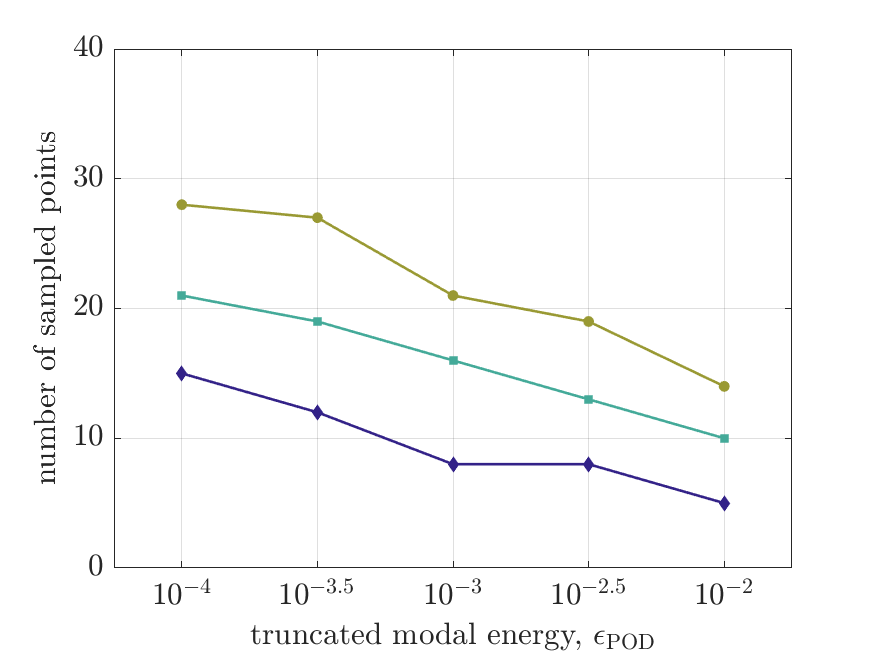}
        & \includegraphics[height=0.3\textwidth]{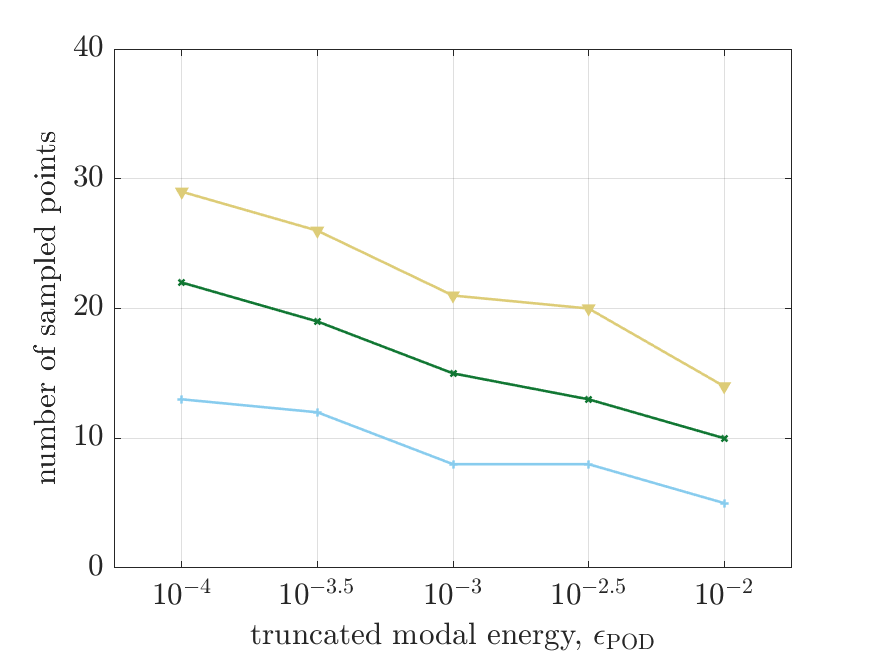}
    \end{tabular}
    \captionsetup{justification=centering}
    \caption{Number of sampled points in ECSW-G (left) and ECSW-LSPG (right) schemes for the 1D Burgers' equation model (top row) and the the 1D unsteady heat equation with a cubic reaction term (bottom row) plotted with respect to truncated modal energy, $\epsilon_{\mathrm{POD}}$, for ECSW tolerances $\epsilon_{\mathrm{ecsw}}=10^{-5}$, $10^{-7}$, and $10^{-9}$. Lower values of $\epsilon_{\mathrm{POD}}$ and $\epsilon_{\mathrm{ecsw}}$ lead to more sampled points in the corresponding ECSW scheme. Note that the legends in the bottom figures are the same as in the corresponding top figures. (Online version in color.)}
    \label{fig:latTolNoSample}
\end{figure}

Figure \ref{fig:latTolNoSample} presents the number of sample points for the ECSW-G and ECSW-LSPG schemes for $\epsilon_{\mathrm{ecsw}}=10^{-5}$, 
$10^{-7},$ and $10^{-9}$ with respect to the truncated modal energy, $\epsilon_{\mathrm{POD}}$ for both numerical examples. For each example, $5000$ residual snapshots randomly selected from the FOM were used. As anticipated, smaller values of $\epsilon_{\mathrm{POD}}$ and $\epsilon_{\mathrm{ecsw}}$ incur a larger number of sampled points.

\setcounter{figure}{0}
\setcounter{equation}{0}
\section{Spatial Semi-discretization} \label{appendix:finitedifference}

This appendix presents the application of the second-order central difference scheme to derive the semi-discrete formulations of the PDEs in both numerical experiments, resulting in the form given by \eqref{eq:quadraticForm}. Section \ref{appendix:fd_burgers} details the derivation for the 1D viscous Burgers' equation, while Section \ref{appendix:fd_heat} provides the corresponding derivation for the 1D heat equation with a cubic reaction term.

\subsection{One-dimensional viscous Burgers' equation}\label{appendix:fd_burgers}
The Dirichlet boundary conditions in \eqref{eq:burgers} are enforced using ``ghost grid points'' at $x=0$ and $x=L$. Since the Dirichlet boundary condition at $x=L$ is homogeneous, it does not require explicit treatment in the semi-discrete form. Alternatively, the inhomogeneous nature of the Dirichlet boundary condition at $x=0$  is accounted for through an input term.

The diffusion and advection terms in \eqref{eq:burgers} are spatially discretized using a second-order central difference scheme, with Dirichlet boundary conditions $w(0,t;\boldsymbol{\mu})=\mu_1$ and $w(L,t;\boldsymbol{\mu})=0$. The diffusion term in \eqref{eq:burgers} is discretized as
\begin{equation}
    \left[\mu_2 \frac{\partial^2 w}{\partial x^2}\right]_i \approx
    \frac{\mu_2}{(\Delta x)^2}
    \begin{cases}
        \mu_1 - 2w_1 + w_2, & i=1,\\[4pt]
        w_{i-1} - 2w_i + w_{i+1}, & 2 \le i \le N-1,\\[4pt]
        w_{N-1} - 2w_N, & i=N,
    \end{cases}
\end{equation}
where $w_i \approx w(x_i,t;\boldsymbol{\mu})$ for nodes $i=1,\cdots,N$, denote the solution values at all grid points (excluding the boundary nodes). Consequently, the semi-discretization of the diffusion term yields a linear term of the form $\mathbf{A}(\mu_2)\mathbf{x}(t;\boldsymbol{\mu})$ at the nodes $i=1,\cdots,N$. In addition, the enforcement of the left Dirichlet boundary condition introduces an input term of the form $\mathbf{B}(\mu_2)\mathbf{u}(t;\boldsymbol{\mu})$ for $i=1$, where $\mathbf{B}(\mu_2)$ is affinely dependent on $\mu_2$. Here, $\mathbf{x}(t;\boldsymbol{\mu}) = \mathbf{w}(t; \boldsymbol{\mu})$ and $\mathbf{u}(t;\boldsymbol{\mu})=[\mu_1]$.

Likewise, the advection term in \eqref{eq:burgers} is discretized as
\begin{equation}
    \left[ w \frac{\partial w}{\partial x} \right]_i
    \approx \frac{1}{2\Delta x}
    \begin{cases}
        w_1 w_2 - w_1 \mu_1, & i=1,\\[4pt]
        w_i w_{i+1} - w_i w_{i-1}, & 2 \le i \le N-1,\\[4pt]
        -\, w_N w_{N-1}, & i=N,
    \end{cases}
\end{equation}
where the Dirichlet boundary condition at the left boundary condition has been incorporated. The semi-discretization of the advection term yields a quadratic terms that appears as $\mathbf{F}(\mathbf{x}(t; \boldsymbol{\mu})\otimes\mathbf{x}(t; \boldsymbol{\mu}))$ in \eqref{eq:burgersQuadratic} for $i=1,\cdots,N$, and a bilinear term $\mathbf{N}(\mathbf{u}(t; \boldsymbol{\mu}) \otimes \mathbf{x}(t; \boldsymbol{\mu}))$ with contribution from the grid point at $i=1$. Note that $\mathbf{C}=\mathbf{0}$, since \eqref{eq:burgers} contains no constant term.

\subsection{One-dimensional unsteady heat equation with cubic reaction term}\label{appendix:fd_heat}
As in the previous example, Dirichlet boundary conditions are enforced using ghost grid points at $x=0$ and $x=L$. The input vector is defined as $\mathbf{u}(t;\boldsymbol{\mu}) = [a \mathrm{sin}(2\pi t),\; b \mathrm{sin}(4\pi t)]^{\top}$. 

The terms arising from the application of a second-order central difference scheme to the spatial discretization of the lifted form of \eqref{eq:heatdiffQuad} are first considered. Imposing homogeneous Dirichlet boundary conditions at the left boundary (i.e., $q(0,t;\boldsymbol{\mu})=w(0,t;\boldsymbol{\mu})=0$) and inhomogeneous Dirichlet boundary conditions at the right (i.e., $q(L,t;\boldsymbol{\mu})=w(L,t;\boldsymbol{\mu})=1$), we obtain 
\begin{align}
    &\left[ 0.005\, \frac{\partial^2 q}{\partial x^2} \right]_i \approx
    \frac{0.005}{(\Delta x)^2}
    \begin{cases}
        -2q_1 + q_2, & i=1,\\[4pt]
        q_{i-1} - 2q_i + q_{i+1}, & 2 \le i \le N-1,\\[4pt]
        q_{N-1} - 2q_N + 1, & i=N,
    \end{cases} \label{eq:semidisc2_1}\\
    & \left[ 2(0.005)\, q\, \frac{\partial^2 q}{\partial x^2} \right]_i \approx \frac{2(0.005)}{(\Delta x)^2}
    \begin{cases}
        -2q_1^2 + q_1 q_2, & i=1,\\[4pt]
        q_i q_{i-1} - 2q_i^2 + q_i q_{i+1}, & 2 \le i \le N-1,\\[4pt]
        q_N q_{N-1} - 2q_N^2 + q_N, & i=N,
    \end{cases} \label{eq:semidisc2_2}
\end{align}
where $q_i \approx q(x_i,t;\boldsymbol{\mu})$ for nodes $i=1,\cdots,N$ denote the solution values for $q$ at the interior grid points not including the boundary nodes, and $\mathbf{x}_1(t;\boldsymbol{\mu}) = \mathbf{q}(t; \boldsymbol{\mu})$. The semi-discrete form of \eqref{eq:semidisc2_1} yields a linear term of the form $\mathbf{A}_{1,1}\mathbf{x}_1(t;\boldsymbol{\mu})$ over all nodes $i=1,\cdots,N$, together with an additional constant term $\mathbf{C}_1$ with non-zero contribution in its last row ($i=N$).  Similarly, \eqref{eq:semidisc2_2} yields a quadratic term of the form $\mathbf{F}_{2,11} (\mathbf{x}_1(t;\boldsymbol{\mu}) \otimes \mathbf{x}_1(t;\boldsymbol{\mu}))$ for $i=1,\cdots,N$ and a linear term of the form $\mathbf{A}_{2,1}\mathbf{x}_1(t;\boldsymbol{\mu})$ arising due to the enforcement of the boundary condition at $i=N$.

The remaining terms in \eqref{eq:heatdiffQuad} are discretized at all nodes $i=1,\cdots,N$ as
\begingroup
\begin{align}
    &\left[q w\right]_i = q_i w_i \quad \text{yielding the quadratic term}\;\mathbf{F}_{1,12}(\mathbf{x}_1(t;\boldsymbol{\mu}) \otimes \mathbf{x}_2(t;\boldsymbol{\mu})), \\
    &\left[u\right]_i = \frac{a \mathrm{sin}(2\pi t)}{1 + 100\left(\frac{x_i}{L}-{\frac{1}{4}}\right)^2}+ \frac{b \mathrm{sin}(4\pi t)}{1 + 100\left(\frac{x_i}{L}-{\frac{3}{4}}\right)^2} \quad  \text{yielding the input term}\;\mathbf{B}_{1}\mathbf{u}(t;\boldsymbol{\mu}),\\
    &\left[2w^2\right]_i = 2w_i^2  \quad\text{yielding the quadratic term}\;\mathbf{F}_{2,22}(\mathbf{x}_2(t;\boldsymbol{\mu}) \otimes \mathbf{x}_2(t;\boldsymbol{\mu})),\\
    & [2q u]_i = \frac{2 q_i a \mathrm{sin}(2\pi t)}{1 + 100\left(\frac{x_i}{L}-{\frac{1}{4}}\right)^2}+ \frac{2 q_i b \mathrm{sin}(4\pi t)}{1 + 100\left(\frac{x_i}{L}-{\frac{3}{4}}\right)^2} \quad \text{yielding the bilinear term}\; \mathbf{N}_{2,1}(\mathbf{u}(t;\boldsymbol{\mu}) \otimes \mathbf{x}_1(t;\boldsymbol{\mu})),
\end{align}
\endgroup
where $w_i \approx w(x_i,t;\boldsymbol{\mu})$ denotes nodal solutions for $w$ for nodes $i=1,\cdots,N$ excluding the boundary nodes, and $\mathbf{x}_2(t;\boldsymbol{\mu}) = \mathbf{w}(t; \boldsymbol{\mu})$.

\end{document}